\documentclass[12pt]{iopart}
\usepackage[utf8]{inputenc}
\usepackage{graphicx}
\expandafter\let\csname equation*\endcsname\relax
\expandafter\let\csname endequation*\endcsname\relax
\usepackage{amsmath,amssymb,amsfonts}
\usepackage{algorithmic}
\usepackage{textcomp}
\usepackage{xcolor}
\usepackage{booktabs}
\usepackage[normalem]{ulem}
\usepackage{cite}
\usepackage{url}
\usepackage{hyperref}
\hypersetup{
    colorlinks=true,       
    linkcolor=blue,          
    citecolor=blue,        
    urlcolor=magenta        
}
\usepackage{caption,subcaption}
\usepackage{physics}
\usepackage{algorithmic}
\usepackage{algorithm}
\usepackage{bm}
\usepackage{bbm}
\usepackage{fontawesome}
\usepackage{multirow}

\graphicspath{{../}}

\DeclareMathOperator*{\argmin}{arg\,min}

\newcommand{\gitrepo}{https://github.com/tobias-liaudat/wf-psf}
\newcommand{\nblink}[1]{\href{https://github.com/tobias-liaudat/wf-psf/tree/main/papers/article_IOP/notebooks/#1.ipynb}{\faFileCodeO}}
\newcommand{\github}{\href{https://github.com/tobias-liaudat/wf-psf}{\faGithub}}

\begin{document}

\title[Rethinking data-driven PSF modeling with a differentiable optical model]{Rethinking data-driven point spread function modeling with a differentiable optical model}
\author{Tobias Liaudat$^1$, Jean-Luc Starck$^1$, Martin Kilbinger$^{1}$, Pierre-Antoine Frugier$^1$}

\address{$^1$AIM, CEA, CNRS, Université Paris-Saclay, Université de Paris, F-91191 Gif-sur-Yvette, France}
\ead{tobiasliaudat@gmail.com}
\vspace{10pt}
\begin{indented}
\item[]November 2022
\end{indented}

\begin{abstract}

In astronomy, upcoming space telescopes with wide-field optical instruments have a spatially varying point spread function (PSF). Specific scientific goals require a high-fidelity estimation of the PSF at target positions where no direct measurement of the PSF is provided. Even though observations of the PSF are available at some positions of the field of view (FOV), they are undersampled, noisy, and integrated into wavelength in the instrument's passband. PSF modeling represents a challenging ill-posed problem, as it requires building a model from these observations that can infer a super-resolved PSF at any wavelength and position in the FOV. Current data-driven PSF models can tackle spatial variations and super-resolution. However, they are not capable of capturing PSF chromatic variations.

Our model, coined WaveDiff, proposes a paradigm shift in the data-driven modeling of the point spread function field of telescopes. We change the data-driven modeling space from the pixels to the wavefront by adding a differentiable optical forward model into the modeling framework. This change allows the transfer of a great deal of complexity from the instrumental response into the forward model. The proposed model relies on efficient automatic differentiation technology and modern stochastic first-order optimization techniques recently developed by the thriving machine-learning community. Our framework paves the way to building powerful, physically motivated models that do not require special calibration data.

This paper demonstrates the WaveDiff model in a simplified setting of a space telescope. The proposed framework represents a performance breakthrough with respect to the existing state-of-the-art data-driven approach. The pixel reconstruction errors decrease $6$-fold at observation resolution and $44$-fold for a $3$x super-resolution. The ellipticity errors are reduced at least $20$ times, and the size error is reduced more than $250$ times. By only using noisy broad-band in-focus observations, we successfully capture the PSF chromatic variations due to diffraction. WaveDiff source code and examples associated with this paper are available at this link \github.
\end{abstract}
%

\section{Introduction}

\subsection{Context}

Each time an imaging instrument observes, the image taken is affected by instrument-related distortions. We can encompass most distortions into the instrumental response or point spread function (PSF). Many science applications need to consider the PSF's effects for quantitative analysis of images. The characterization of the instrument's response, also known as PSF modeling, is a subject of study in many fields, ranging from astronomical imaging to biomedical imaging. Cosmology provides an exciting application that poses very stringent requirements on the quality of the PSF model. This application is weak gravitational lensing (see \cite{kilbinger2015}, or \cite{mandelbaum2018_bis3} for a review).

The light emitted from observed objects travels through the Universe towards our imaging instrument and is deflected by the gravitational potential of massive objects along the line of sight. The effect is similar to that of a magnifying glass and is therefore given the term gravitational lensing. In the weak lensing regime, this effect represents a minimal change of shape of observed extended objects, usually galaxies. The precise measurement of this change in shape can help cosmologists answer questions that have puzzled them since the 1980s, for example, the nature of dark matter and its spatial distribution.
Weak gravitational lensing is based on measuring galaxy shapes with high accuracy. The instrument's PSF produces a systematic deformation of the images, thus obscuring the lensing-induced shape deformations.
It is, therefore, \textit{essential} to correct these images from the effects of the PSF.

The next generation of powerful imaging instruments is currently being built, such as the \textit{Euclid} space telescope \cite{laureijs2011}, the \textit{Roman} space telescope \cite{wfirst}, and the Vera C.~Rubin Observatory's Legacy Survey of Space and Time (LSST) \cite{LSST2009}. These instruments' wide field-of-view (FOV) will make the PSF vary significantly across the focal plane. The extremely tight instrumental knowledge required by weak-lensing applications requires capturing PSF morphological variations across the FOV into the image correction. The accuracy aimed by these future missions also demands considering the PSF's wavelength variations. We assume that the PSF can be modeled locally by a convolutional kernel. Then, the \textit{PSF field} refers to a continuous function that maps each point in the FOV and each wavelength to the corresponding convolutional kernel. In astronomy, certain observations of unresolved sources like isolated stars provide a measurement of the PSF field at a given FOV position. PSF modeling consists in using these observations to build a model of the PSF field. Then, the PSF model can be used to infer the PSF at the positions of the objects of interest and to correct their images.

This problem has created much attention within the astronomer's community. Over the years, imaging instruments have become more powerful, while the accuracy requirements have become more stringent. This progress has motivated the development of a wide range of methodologies for PSF modeling. These include models based on Moffat functions \cite{moffat1988}, polynomial models \cite{piotrowski2013, bertin2011, miller2013, jarvis2020}, principal component analysis \cite{jee2007, schrabback2010, gentile2013}, sparse matrix factorization \cite{ngole2015,ngole2016,schmitz2020}, specific basis functions \cite{refregier2003a,massey2005}, optimal transport \cite{ngole2017}, neural networks \cite{herbel2018,jia2020a,jia2020b,jia2020c}, and a parametric model of the telescope's optics \cite{krist1993, krist1995, krist2011}.

Future missions, like \textit{Euclid}, impose stringent requirements on the PSF model. The current state-of-the-art data-driven PSF model for \textit{Euclid} \cite{schmitz2020} is not capable of modeling wavelength variations and is far from requirements, at a factor between $100$ and $300$ for the PSF shape and a factor of $10^4$ for the PSF size. These facts triggered us to rethink how we build data-driven PSF models.

\subsection{Challenges in point spread function modeling for space missions}

The main difference compared to a ground-based telescope PSF is the effect of the atmosphere, which acts as a fast-changing stochastic filter. The PSF tends to resemble a low-pass filter as the exposure time increases. Therefore, it strongly affects the observations and changes over time. The PSF stability and quality improvements are the main reasons to send a telescope into space.

The modeling of the PSF in the field of view (FOV) can be seen as an inverse problem. We can consider particular unresolved objects, e.g., stars, as samples of the PSF field in the FOV. These samples can be used to constrain our model, which we will later use to infer the PSF at target positions. PSF field modeling for space missions encompasses several challenges:
\begin{itemize}
    %
    \item[1.] \textit{The PSF varies spatially in the FOV}. 
    The instruments are designed with a wide FOV and are composed of a large array of charge-coupled devices (CCDs). These dimensions make the PSF vary substantially across the focal plane. The lack of atmosphere and the high-quality optics used in telescopes makes the PSF close to the diffraction limit. Consequently, the PSF has more complex shapes with higher spatial frequencies. The model needs to capture the different spatial frequency variations of the PSF from the stars to infer the PSF at target positions. 
    %
    \item[2.] \textit{The observations are undersampled}. 
    This situation is generally the case for space missions, and the model needs to super-resolve the output PSFs. However, this scenario differs from the usual super-resolution (SR) task, as we do not have several low-resolution observations of the same object. In this case, we have several samples of the undersampled PSF field at \textit{different positions} in the FOV.
    %
    \item[3.] \textit{The PSF varies as a function of wavelength}. 
    Also known as chromatic variations, they must be included in the PSF model for most science goals. However, the modeling difficulty resides in some instruments' very broad passbands. Consequently, each star observation is integrated with respect to its wavelength throughout the passband (e.g.~the \textit{Euclid} passband extends from $550$ to $900$nm \cite{cropper2016}). The PSF chromaticity will also be very likely important for the \textit{Roman} space telescope \cite{wfirst} or LSST \cite{LSST2009}.
\end{itemize}

In addition to the three major problems mentioned above, we must also address the fact that the telescope will change over time during its journey in space. Some reasons include strong temperature gradients or degradations of the instrument. Although capturing these variations in the PSF model might seem intricate, in practice, one can build independent PSF models at each point in time. Consequently, we have a different model for each time snapshot and cope with the temporal variations. In the present work, we will analyze a single point in time. The only limitation this approach imposes is the number of observed stars available to constrain the PSF model. The stability of a space mission allows using several exposures to increase the available information when building the PSF model. 

\subsection{Current approaches for PSF modeling}
\label{sc:approaches_PSF_modeling}

There are two main approaches, or families of methods, for PSF modeling, as follows: 

\paragraph{Parametric PSF models}
This approach consists in building a parametric model of the entire optical system that aims to be as close as possible to the actual telescope. Then, a few model parameters are fit to the star observations. This family of methods can handle the PSF's chromatic variations. This approach has been used for the Hubble Space Telescope (HST), with the Tiny Tim method \cite{krist1993, krist1995b, krist2011}. However, it was later shown that a simple data-driven model \cite{HST_PSF2017} outperformed Tiny Tim, exposing some limitations of parametric modeling. Errors will arise if there is a mismatch between the parametric model and the ground truth. Furthermore, even if, ideally, there was no mismatch, the optimization of these models is a degenerate problem. It requires potentially expensive calibration information, usually in the form of out-of-focus observations, to break degeneracies. Some events, e.g., launch vibrations and ice contaminations, introduce significant variations into the model, which prevent a complete ground characterization from being successful.

\paragraph{Data-driven PSF models}
This approach, also known as \textit{non-parametric PSF models}, only relies on the observed stars to build the model in pixel space. It is blind to the physics of the inverse problem. The models assume some regularity in the variation of the PSF field across the FOV. These methods differ in how they exploit this regularity and handle the super-resolution, for example, \cite{liaudat2020, bertin2011, schmitz2020, jarvis2020, miller2013}. Data-driven methods can smoothly adapt to the current state of the optical system. However, they have difficulties in modeling complex PSF shapes. A limitation shared by all the data-driven methods is their sensitivity to the available number of stars to constrain the model. When the number of stars falls below some threshold, the model built is usually considered unusable. This family of methods has been widely used for modeling ground-based telescope PSFs. Nevertheless, they cannot successfully model the chromatic variations in addition to the spatial variations and the super-resolution.

\subsection{Our contribution}

This paper proposes a novel data-driven point spread function model coined WaveDiff. Our model can handle super-resolution and model spatial variations of wide-field telescopes. To the best of our knowledge, this is the first data-driven PSF model that can also capture the diffraction-induced chromatic variations of the PSF.

The WaveDiff model is build in the wavefront space and based on two WFE parts, a parametric and a non-parametric, or data-driven, part. The central idea of WaveDiff is to correct the possible errors of the parametric part with a data-driven contribution that can better adapt to the observations. The parametric part is based on Zernike polynomials \cite{noll1976}, which are convenient to model the WFE of circular apertures as they are orthogonal in the unit disk. The data-driven part is based on a low-rank matrix factorization approach with data-driven features. We show that the WaveDiff model can estimate a useful manifold in the wavefront space that approximates the underlying pixel PSF field with low error through automatic differentiation. We present three variations of the WaveDiff model that we refer to as its flavors.

The proposed WaveDiff model represents a \textit{paradigm shift} with respect to current state-of-the-art \textit{data-driven} PSF models as it builds the model in the wavefront space. Although this is also the case for parametric models, WaveDiff remains data-driven, or non-parametric. We can include the physics behind the PSF modeling inverse problem into the WaveDiff model using the \textit{differentiable optical forward model} that we have designed. The forward model is theoretically based on optics principles and practically implemented in an automatically differentiable framework, \textsc{TensorFlow} \cite{tensorflow2015}, that runs on GPU. The optical forward model represents a simplified version of the telescope's optical system and allows propagating the wavefront from the pupil plane to the focal plane to compute the pixel representation of the PSF finally. This change allows us to build the PSF model directly on the wavefront space and to propagate it to obtain its pixel PSF representation. The forward model is fast, as it avoids the resampling of the wavefront space for each wavelength, and versatile, as it depends on a reduced number of parameters known for optical telescopes. The model could be, therefore, easily adapted to model the PSF field of other telescopes. In addition, we propose a framework for estimating the WaveDiff model based on modern stochastic gradient methods. Even though the WaveDiff model is built in the wavefront space, it is constrained only by noisy in-focus observations without requiring particular calibration observations. 
This paper extends the work in \cite{liaudat2021,liaudat2021_b}.

The rest of the paper is organized as follows: \autoref{sc:psf_field_inv_problem} describes the inverse problem we are targeting, \autoref{sc:related_work} gives an overview of related work, \autoref{sc:wavefront_psf_model} presents the proposed model from the physical motivation to the practical implementation. In \autoref{sc:numerical_experiments}, we describe the numerical experiments we have done to validate the WaveDiff model and benchmark it against state-of-the-art PSF models. We also include a description of the experiment setup and the presentation of the results. We discuss important points in \autoref{sc:discussions}, and finally conclude this study in \autoref{sc:conclusions}.

\subsection{Notation}
We use capital letters, e.g., $H$, for discrete functions where the codomain is $\mathbb{K}^{a \times b}$ for some $a,b \in \mathbb{N}$, and $\mathbb{K}$ is $\mathbb{R}$ or $\mathbb{C}$. We use calligraphic capital letters, e.g., $\mathcal{H}$, for continuous functions where the codomain is $\mathbb{K}$. The variables $(x,y)$ are assigned for the positions in the FOV, $[u,v]$ for positions in the focal plane, and $[\xi,\eta]$ for positions in the pupil plane. For example, $\bar{I}(x_i,y_i)[u,v]$ represents the $[u,v]$ pixel of the observation at the FOV position $(x_i,y_i)$, and $\Phi(x_i,y_i)[\xi,\eta]$ the $[\xi,\eta]$ pixel of the wavefront representation of the object from the same FOV position $(x_i,y_i)$.

\section{The PSF field and the inverse problem}
\label{sc:psf_field_inv_problem}

Let us define the PSF field for a particular image exposure as a function that has as inputs a position in the FOV and a specific wavelength and outputs a monochromatic PSF, i.e., a PSF at a specific wavelength. We denote this function by $I: \mathbb{R}^{2} \times \mathbb{R}_{+} \to \mathbb{R}^{M \times M} $, where $M^2$ is the number of pixels of the desired super-resolved postage stamp. PSF modeling consists in building an estimator, $H$, of the PSF field and then using it to output any monochromatic PSF at a set of target FOV positions $\{(x_j,y_j) \}_{j=1, \ldots, m_{\text{target}}}$. The imaging instrument has a certain passband and several degradations that we need to consider in our model to match our PSF estimation with the observations. 

\subsection{The observation model}
\begin{figure}
    \begin{subfigure}{.24\textwidth}
      \centering
      \includegraphics[width=.9\linewidth]{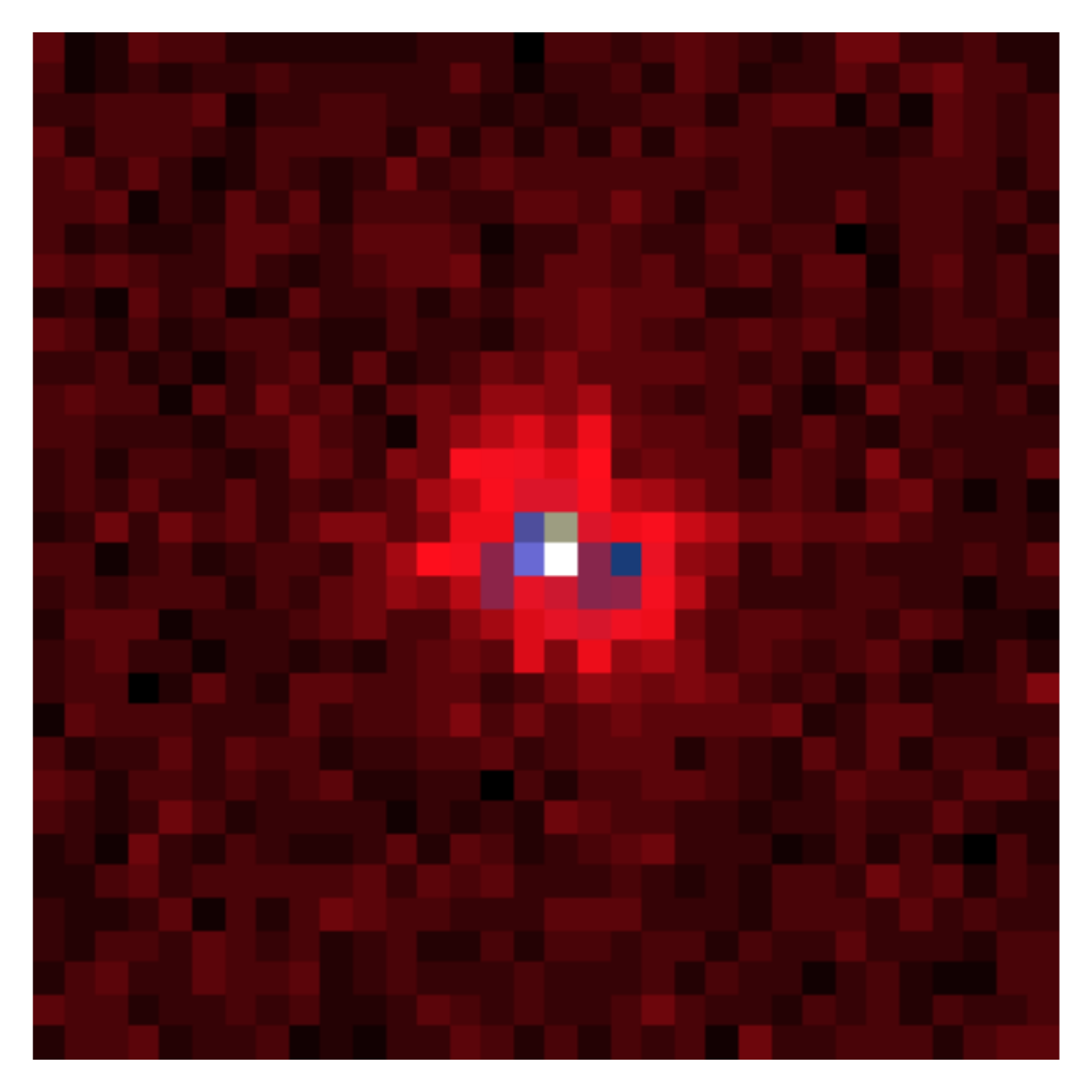}
      \caption{}
      \label{fig:sub_a}
    \end{subfigure}%
    \begin{subfigure}{.24\textwidth}
      \centering
      \includegraphics[width=.9\linewidth]{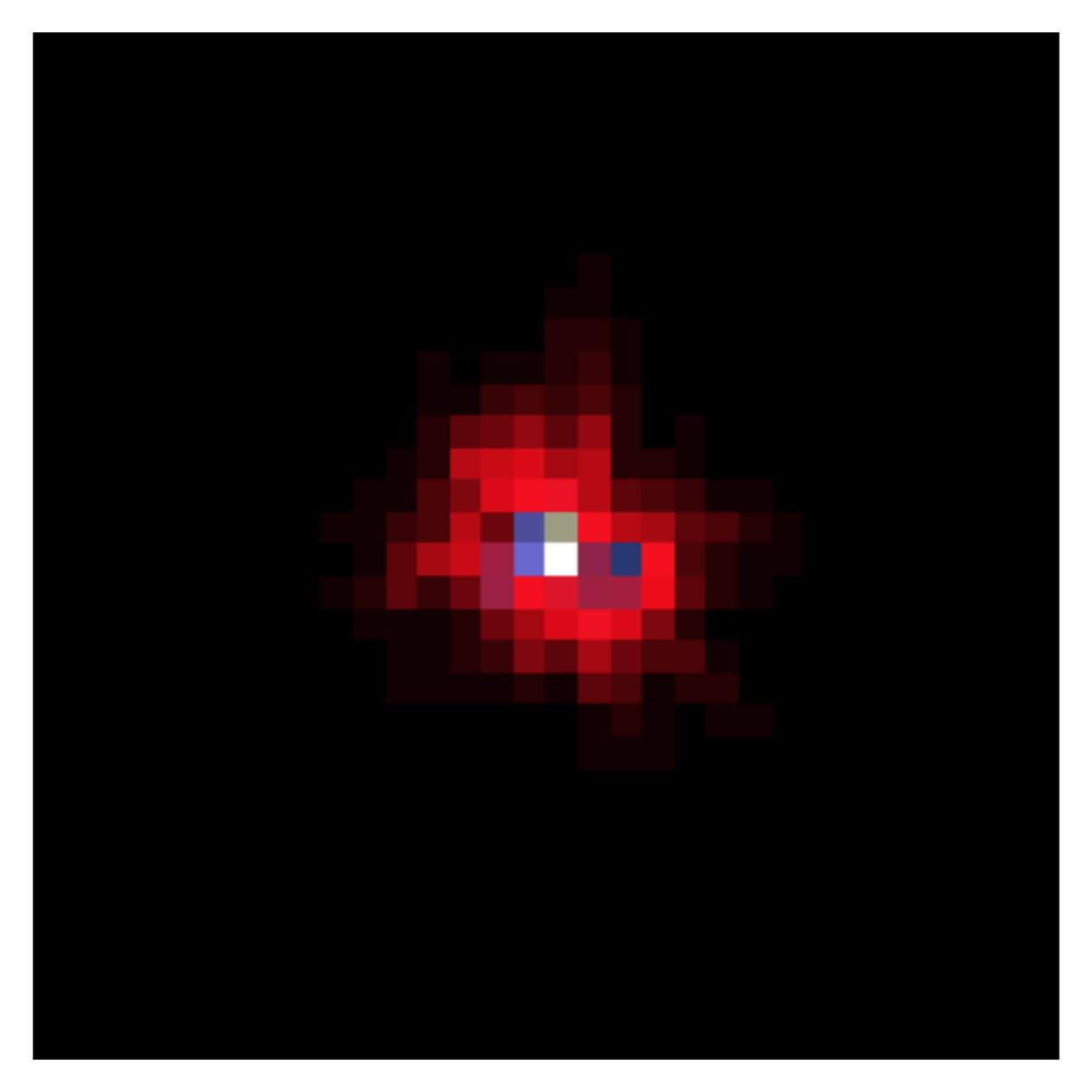}
      \caption{}
      \label{fig:sub_b}
    \end{subfigure}
    \begin{subfigure}{.24\textwidth}
      \centering
      \includegraphics[width=.9\linewidth]{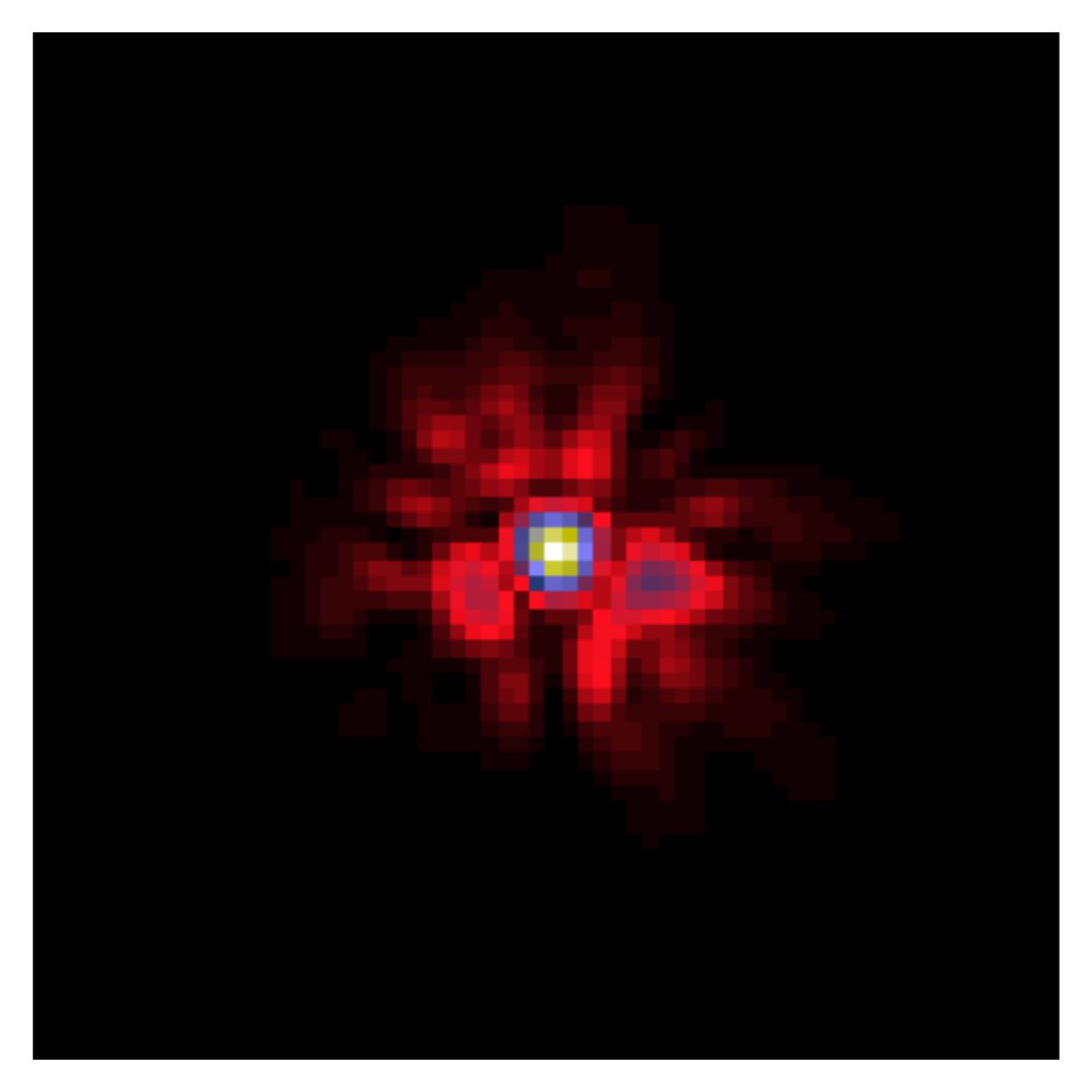}
      \caption{}
      \label{fig:sub_c}
    \end{subfigure}
    \begin{subfigure}{.24\textwidth}
      \centering
      \includegraphics[width=.9\linewidth]{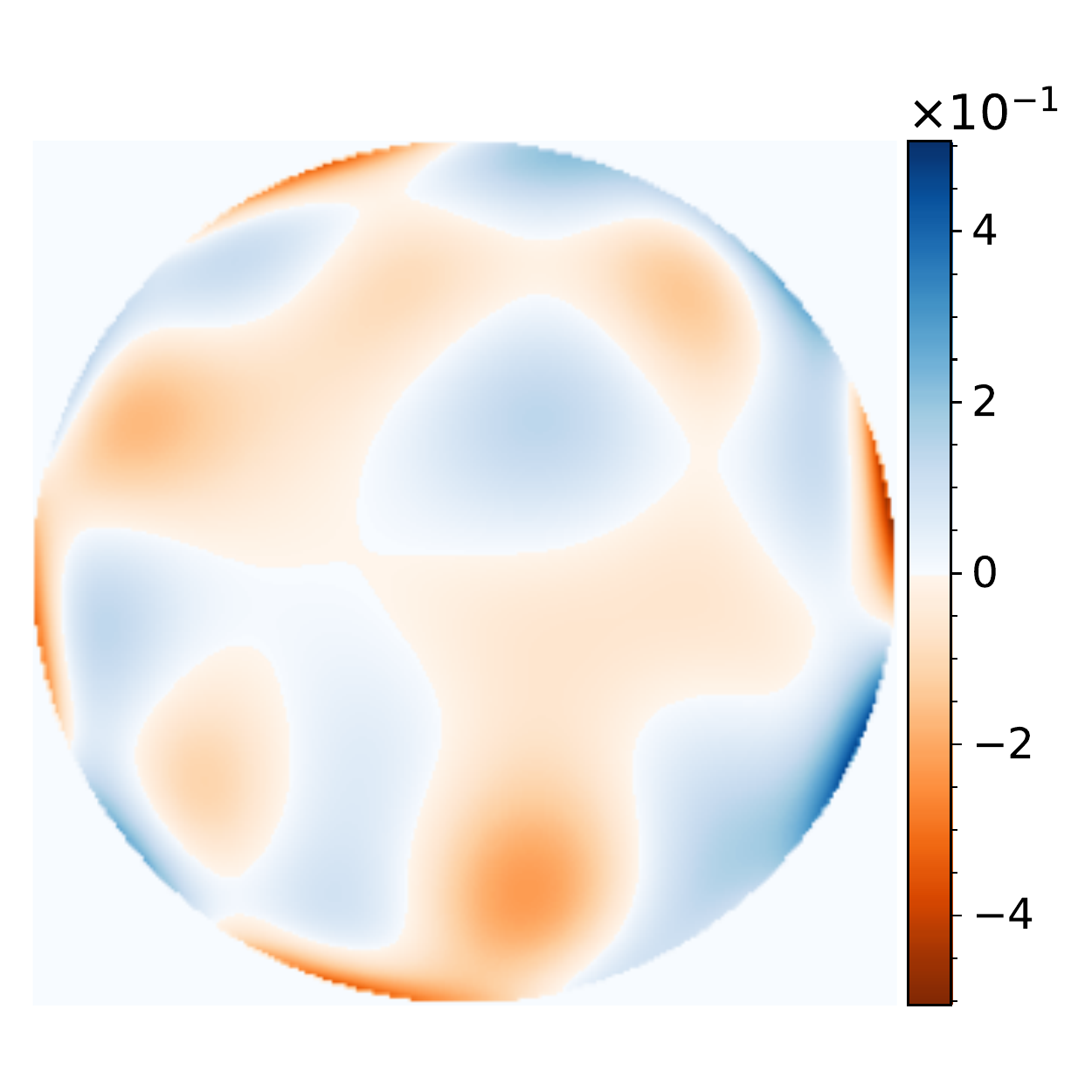}
      \caption{}
      \label{fig:sub_d}
    \end{subfigure}\\
    \begin{subfigure}{\textwidth}
      \centering
      \includegraphics[width=0.95\linewidth]{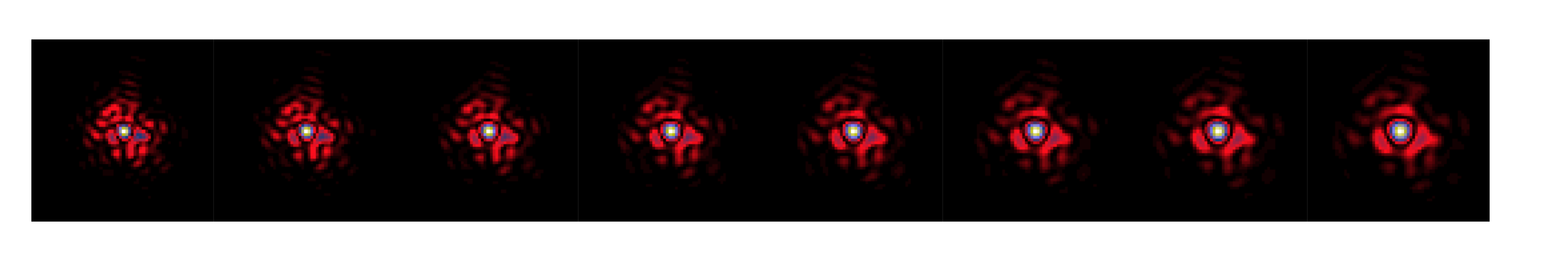}
      \caption{}
      \label{fig:sub_e}
    \end{subfigure}
    \caption{The figure shows different representations of a space-like PSF at a single position in the field of view. The SED used corresponds to one of the templates from the numerical experiments described in \autoref{sc:experiment_set_up}. \textbf{(a)} Noisy polychromatic PSF at observation resolution, \textbf{(b)} Noiseless polychromatic observation at observation resolution, \textbf{(c)} High-resolution noiseless polychromatic observation, \textbf{(d)} Wavefront error map representing the aberrations in the optical system. Units are in $\mu$m, \textbf{(e)} High-resolution chromatic variations of the PSF at equally spaced wavelengths in the passband $[550, 900]$nm.}
    \label{fig:PSF_examples}
\end{figure}

We consider stars as point sources. Therefore, star observations provide samples of the PSF field at their corresponding positions. The spectral energy of the point source at each wavelength is often different, and we have to consider this for our observations. The spectral energy distribution (SED) gives us the normalized energy as a function of the wavelength of a star at a FOV position $(x_i, y_i)$ or $\int_{\text{passband}} \text{SED}(x_i, y_i; \lambda) \text{d}\lambda = 1$. Each star $i$ has its own $\text{SED}(\lambda)$ that writes $\text{SED}(x_i, y_i; \lambda)$, which usually differs for a different star $k$ at $(x_k, y_k)$. From now on, we assume the different SEDs to be known. The following observational model, 
\begin{equation}
    \bar{I}(x_i, y_i) = F_d \left\{ \int_{\text{passband}} \text{SED}(x_i, y_i; \lambda) \; I(x_i, y_i; \lambda) \; \text{d}\lambda \right\} + \mathbf{n}_{i}\;, 
    \label{eq:poly_star}
\end{equation} 
relates the star observation $\bar{I}(x_i, y_i) \in \mathbb{R}^{N \times N}$, where $N^2$ is the number of pixels in the observed postage stamp, at the FOV position $(x_i, y_i) \in \mathbb{R}^{2}$ to the objective PSF field $I(x_i, y_i; \lambda) \in \mathbb{R}^{M \times M}$, which is at a higher resolution with respect to the observations ($M>N$). The desired instrumental response $I$ is integrated into the instrument's passband weighted by the star's normalized spectral energy distribution $\text{SED}(x_i, y_i; \lambda) \in \mathbb{R}_{+}$. Then, it is degraded with the operator $F_d : \mathbb{R}^{M \times M} \to \mathbb{R}^{N \times N}$, which accounts for: 
(i) downsampling that affects the pixels in the stamp by a factor $D$ (meaning that $D N = M$),
(ii) a sub-pixel shift that depends on where the position of the centroid of the object is placed with respect to the pixel grid, and
(iii) other pixel-level degradations such as detector effects or guiding errors.
The observational noise is modeled by $\mathbf{n}_{i}$, which we assume to be white and Gaussian for simplicity, i.e., $\mathbf{n}_{i} \sim \mathcal{N}(0, \sigma_{i}^{2} \mathbf{I}_{N})$. We leave for future work the consideration of Poisson noise known to arise in the CCDs. We make a distinction between a monochromatic PSF $I(x_i, y_i; \lambda)$, that is evaluated at a single wavelength, and its polychromatic counterpart $\bar{I}(x_i, y_i)$, which has been integrated in wavelength over a passband. \autoref{fig:PSF_examples} presents examples of different representations of a PSF at a single FOV position.

\subsection{Inverse problem and regularizations}
\label{sc:inv_problem}

The PSF modeling problem consists in building a model $H: \mathbb{R}^{2} \times \mathbb{R}_{+} \to \mathbb{R}^{M \times M}$ that best approximates $\{I(x_j, y_j; \lambda)\}_{j=1, \ldots, n_{\text{target}}}$ for all the target positions and wavelengths. To do so, we have as input a set of noisy degraded polychromatic observations $\{\bar{I}(x_i, y_i)\}_{i=1, \ldots, n_{\text{obs}}}$. We constrain our model by minimizing a reconstruction error between our model and the observations. For example, $\sum_{i=1}^{n_{\text{obs}}}\| \bar{I}(x_i, y_i) - \bar{H}(x_i, y_i) \|_{F}^{2}$, where $\| \cdot \|_{F}$ denotes the Frobenius norm of a matrix\footnote{The Frobenius norm of a matrix $A$ can be defined as $\| A \|_{F}^{2} = \sum_i \sum_j |A_{ij}|^{2}$, and is related to the quadratic data fidelity term and the Gaussian noise assumption.}. We need to consider the same degradations applied to $I(x_i, y_i; \lambda)$ in \autoref{eq:poly_star} to the model $H$ to match the observation. This inverse problem is ill-posed due to the degradations in $F_d$, the broad passband integration, and the noise. As a consequence, many PSF models can reproduce the observations.

We can inject some prior knowledge about the PSF field to regularize the inverse problem. For example:

\begin{itemize}

    \item \textit{Positivity}: Given the way the PSF physically forms, it should not contain negative values. The presence of negative values in the observations is due to the noise. Therefore, one can then consider the positivity of the PSF to regularize the problem.

    \item \textit{Smoothness of the PSF}: One can consider that the PSFs are structured, so the model should output smooth or piecewise-smooth images. One way to consider this would be to use predefined parametric functions that are already smooth to build the PSF model. For example, the Moffat function \cite{moffat1988, li2016}. This choice can be effective in avoiding noise contamination in the model. However, it has problems modeling PSF fields that are not well represented by the chosen base function. Another approach proposed by \cite{ngole2016} would be to use sparsity to enforce structure in the PSF image. In practice, this is used to denoise the model so that the model does not learn the observational noise. It can be achieved by enforcing a sparse representation of the PSF on some dictionary that can be built using, for example, isotropic undecimated wavelets, also known as starlets \cite{starck2011}. Although this technique proved flexible and adaptable to the observed data, the resulting PSFs are biased toward the properties of the dictionary. In the case of starlets, the resulting PSF would be biased towards isotropic features, smoothing out anisotropies of the PSF.
   
    \item \textit{PSF field regularity}: The most used regularization is to exploit the regularity of the PSF field in the FOV. This regularization imposes that two close-by positions in the FOV should have a close PSF representation. The regularity translates as a correlation in space (and time) of the PSFs. Most current PSF models rely on this property to build a model that can accurately generalize target positions. Each PSF model imposes the regularity differently, e.g., \cite{liaudat2020, ngole2016, bertin2011, jee2007}.


\end{itemize}

Up to now, the properties mentioned were mild assumptions of the PSF modeling problem. This prior knowledge is a consequence of the physics involved in the image formation of the telescope. 
For example, the CCD detectors measure a voltage proportional to the number of photons exciting the CCD electrons and output a positive quantity.
Also, the smoothness and structure of the PSF come from the fact that it is the diffraction pattern of the incoming wavefront. Instead of building our PSF model over the assumptions mentioned above, we can take a step back and inject the physics of the optical system into our model. In practice, we account for the physics of the inverse problem with the addition of a differentiable optical forward model. The PSF modeling space is shifted from the pixels to the wavefront, which is closer to the physics of the problem and a more natural space to model the PSF. With this change, we are not only considering the previously mentioned priors but adding much more physical information to solve our inverse problem. This change is fundamentally different from how \textit{data-driven} PSF models are built.

The PSFs generated using the optical forward model are naturally smooth without negative pixel values. However, as already mentioned, current data-driven models can also account for the positivity and smoothness of the PSF. Then, what would be the advantage of building a wavefront-based model? Current data-driven models only tackle consequences of the physical process involved in the inverse problem, for example, positivity. The proposed wavefront-based model incorporates the physics involved, and the reconstructed pixel PSF is computed as the intensity of the diffracted electric field in the focal plane. Furthermore, the optical forward model cannot reproduce observational noise, which gives us a natural way of avoiding overfitting it without using any denoising technique.

It may seem that we are adding a complete model of the optical system as it is done in the parametric PSF models from \autoref{sc:approaches_PSF_modeling}. Nevertheless, we remain on a general data-driven framework, as we will see in \autoref{sc:wavefront_psf_model}. We include in our model the diffraction phenomena, which have been well studied \cite{goodman2005}, as well as general properties of the telescope, like the focal length, the pupil diameter, the obscurations, and the pixel spacing.

\section{Related work}
\label{sc:related_work}

\paragraph{Data-driven efforts to model chromatic variations}
Previous work attempted to solve the problem of modeling the chromatic variations from the data-driven side. An approach based on optimal transport \cite{schmitz2018} was developed in \cite{schmitz2019}. The chromatic variations were modeled as the displacement interpolation \cite{mccann1997}, a weighted Wasserstein barycenter \cite{agueh2011}, between the monochromatic PSFs at the two extreme wavelengths. However, the results were not encouraging, and the hypothesis taken by this approach might not hold for \textit{Euclid}. The authors in \cite{soulez2016} tried to model the propagation of the incoming wavefront from the stars through the different mirrors in the optical system. They propose to infer the aberrations introduced by each mirror by solving a phase retrieval type of problem. Finally, the proposed formulation is recast into a constrained optimization problem using the star observations in the FOV. However, it is a proof-of-concept article, as many difficulties remained unaddressed, and the results were only qualitative. None of these studies proved to be a satisfying solution.

\paragraph{Phase retrieval with automatic differentiation}
Estimating the wavefront of an intensity observation falls into phase retrieval problems. In this type of problem, we want to estimate a particular complex signal, i.e., the wavefront, from one or several intensity-only measurements of the original signal, i.e., camera observations. Current detectors only provide intensity measurements and cannot measure the incoming phase information, thus giving rise to the phase retrieval problem. Recent works \cite{Wang2020, Wong2021}, based on a framework established by \cite{Jurling2014}, are tackling the phase retrieval problem \cite{schechtman2015} using automatic differentiation \cite{baydin2017}. These works rely on using first-order optimization methods over their differentiable forward model, some prior knowledge about the complex signal, and the information available in the intensity measurement. The images used by \cite{Wang2020} are natural images with noticeable patterns, while \cite{Wong2021} focuses on the \textsc{toliman} space telescope's PSF \cite{tuthill2018} that was designed to detect exoplanets. In both cases, the intensity image has $256 \times 256$ pixels, and the target phase image is $256 \times 256$ or $128 \times 128$ pixels, for \cite{Wang2020} and \cite{Wong2021}, respectively. Our PSF modeling problem deals with undersampled noisy intensity PSF images of $32 \times 32$ pixels, e.g., \autoref{fig:sub_a}. These PSF images have poor image diversity as most light rays converge into a single point in the focal plane, making the phase retrieval problem even more ill-posed. Even though the input conditions make the problem harder, the wavefront representation we require still has $256 \times 256$ pixels, e.g., \autoref{fig:sub_d}. In addition, in our PSF problem, we do not have access to the PSF intensity image at the FOV positions of interest. Therefore, we need to build a  PSF field model to recover the PSF at such positions, which none of these methods consider in their formulations. The aforementioned conditions make using the phase retrieval methods mentioned above unpractical. However, we can still take inspiration from these works and use an automatic differentiation framework to build our optical forward model.

\paragraph{State-of-the-art data-driven PSF models}
Several state-of-the-art data-driven PSF models exist. Here we briefly review the three most relevant methods, which we will later use for comparison. Although none can model spectral variations, all can model spatial variations and handle super-resolution. The three methods are as follows: 

\begin{itemize}
    \item \texttt{PSFEx}\footnote{\url{https://github.com/astromatic/psfex}} \cite{bertin2011} has been widely used in astronomy for weak-lensing surveys, e.g., \cite{zuntz2018}. It was designed to work together with the \texttt{SExtractor}\cite{bertin1996} software which builds catalogs from astronomical images and measures several properties of the observed stars. \texttt{PSFEx} models the variability of the PSF in the FOV as a function of these measured properties. It builds independent models for each CCD in the focal plane and works with polychromatic observations. It is not able to model the chromatic variations of the PSF field. The model is based on a matrix factorization scheme, where one matrix represents PSF features and the other matrix the feature weights. Each observed PSF is represented as a linear combination of PSF features. The feature weights are defined as a polynomial law of the selected measured properties. This choice allows having an easy interpolation framework for target positions. In practice, the properties that are generally used are both components of the PSF's FOV position. The PSF features are shared by all the observed PSFs and are learned in an optimization problem. The PSF reconstruction in the model's framework at a FOV position $(x_i,y_i)$ can be written as
    \begin{equation}
    \bar{H}^{\text{PSFEx}}(x_i,y_i) = F^{\text{PSFEx}} \left\{ \sum_{\substack{p,q > 0 \\ p+q \leq d}} x_{i}^{p} \, y_{i}^{q} \, S_{p,q} \,+\, S_{0} \right\} \,,
    \label{eq:psfex_model}
    \end{equation}
    where $S_{p,q}$ represents the learned PSF features, $S_{0}$ represents a first guess of the PSF, the polynomial law is defined to be of degree $d$, and $F^{\text{PSFEx}}$ represents the degradations considered to match the model with the observations. The first guess can be computed as a function of the median of all the observations. The PSF features are estimated in an optimization problem that aims to minimize the reconstruction error between the \texttt{PSFEx} model and the observations which read
    \begin{equation}
    \min_{\substack{S_{p,q} \\ \forall p,q > 0 \,,\, p+q \leq d}} \left\{ \sum_{i=1}^{n_{\text{obs}}} \left\| \frac{\bar{I}(x_i,y_i) - \bar{H}^{\text{PSFEx}}(x_i,y_i)}{\hat{\sigma}_{i}^{2}} \right\|_{F}^{2} + \sum_{\substack{p,q > 0 \\ p+q \leq d}} \left\| T_{p,q} S_{p,q} \right\|_{F}^{2} \right\} \,,
    \label{eq:psfex_optim}
    \end{equation}
    where $\hat{\sigma}_{i}^{2}$ represent the estimated per-pixel variances, $\bar{I}$ represents the noisy observations, and $\| \cdot \|_{F}$ the Frobenius norm of a matrix. The second term in \autoref{eq:psfex_optim} corresponds to a Tikhonov regularization, where $T_{p,q}$ represents some regularization weights to favor smoother PSF models. The PSF recovery at target positions is straightforward. One needs to introduce new positions in the \autoref{eq:psfex_model} after estimating the PSF features $S_{p,q}$. 

    \item \texttt{RCA}\footnote{\url{https://github.com/CosmoStat/rca}} \cite{ngole2016} is the state-of-the-art data-driven method designed for the \textit{Euclid} mission \cite{schmitz2020}. The model builds an independent model for each CCD and, like \texttt{PSFEx}, is based on a matrix factorization scheme. However, there are two fundamental changes with respect to \texttt{PSFEx}. The first difference is that, in \texttt{RCA}, the feature weights are defined as a further matrix factorization. The feature weights are also learned from the data and are constrained to be part of a dictionary built with different spatial variations that are based on the harmonics of a fully connected graph. The graph is built using the star positions as the nodes. The edge weights are computed as a function of the inverse of the node's distance.
    The second difference corresponds to the regularizations used in the loss function, and the optimization algorithms \cite{beck2009, condat2013}. \texttt{RCA} uses a positivity constraint, a denoising based on a sparsity constraint in the starlet \cite{starck2015} domain, which is a wavelet representation basis, and a constraint to learn the useful spatial variations from the graph-harmonics-based dictionary. See \cite{ngole2016} for more details.
   
    \item \texttt{MCCD}\footnote{\url{https://github.com/cosmostat/mccd}} \cite{liaudat2020} is a state-of-the-art data-driven method originally designed for the ground-based Canada-France Imaging Survey (CFIS)\footnote{\url{https://www.cfht.hawaii.edu/Science/CFIS/}} on the Canada-France-Hawaii Telescope (CFHT). The \texttt{MCCD} model extends the notions from the two methods described above and builds a single PSF model for the entire CCD mosaic in the focal plane. This choice allows a more robust capture of spatial variations extending over several CCDs. It is composed of global components that are shared between all CCDs and local CCD-specific components that aim to correct for the imperfections of the global component. Both components are based on a matrix factorization framework. The feature weights are defined as a combination of the polynomial strategy used in \texttt{PSFEx} and the graph-based strategy used in \texttt{RCA}. The model's optimization is more complex than the previous models, but it is also based on iterative schemes. The constraints used in \texttt{MCCD} are motivated by the ones used in \texttt{RCA}. See \cite{liaudat2020} for more details.
\end{itemize}

\section{Data-driven wavefront PSF model}
\label{sc:wavefront_psf_model}

\subsection{Physical motivation and the wavefront error}
\label{sc:physical_motivation}

\begin{figure}
    \centering
    \begin{subfigure}[b]{0.32\textwidth}
        \centering
        \includegraphics[width=\textwidth]{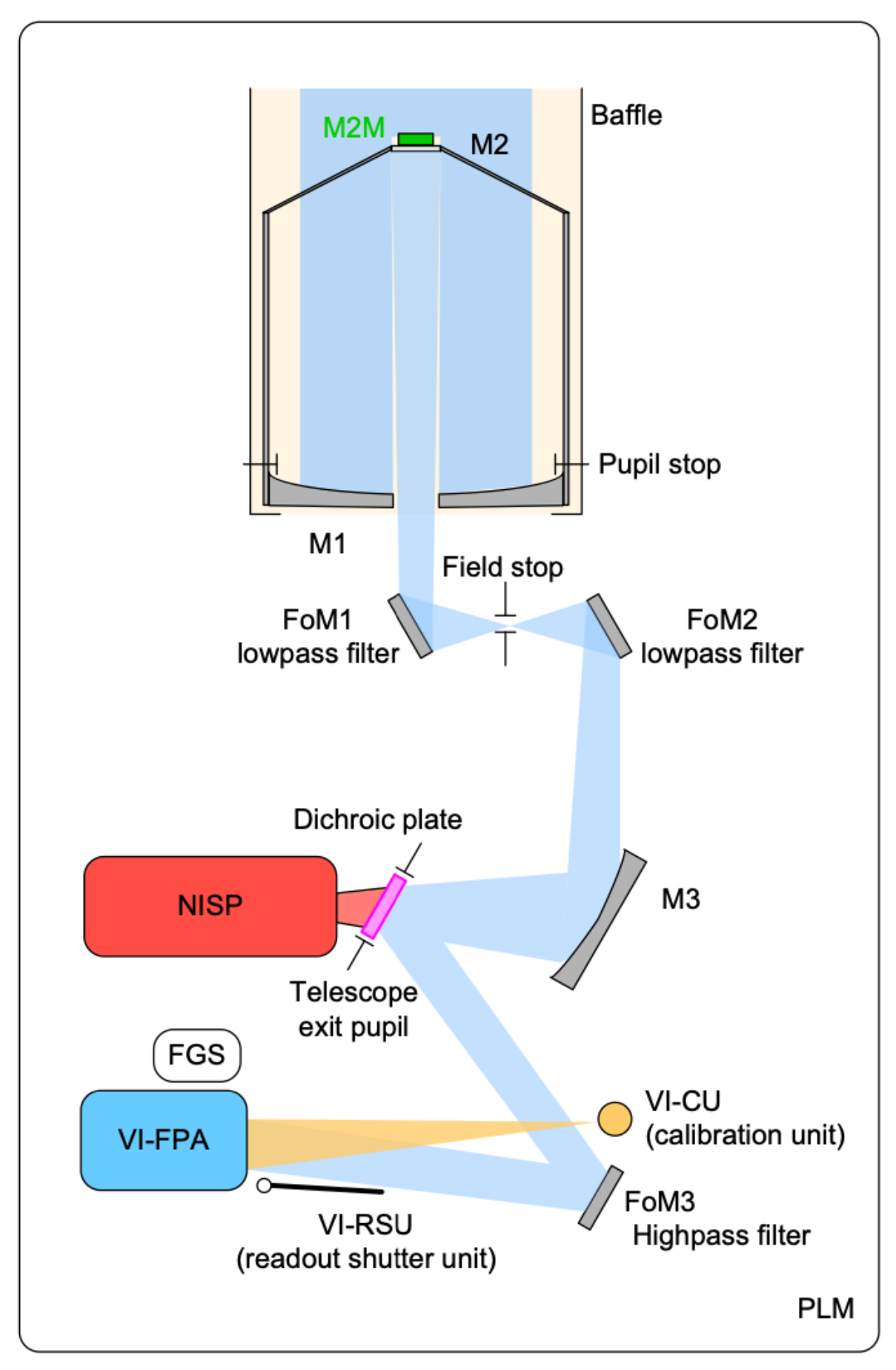}
        \caption{\label{fi:optics_a}}
    \end{subfigure}
    \begin{subfigure}[b]{0.65\textwidth}
        \centering
        \includegraphics[width=\textwidth]{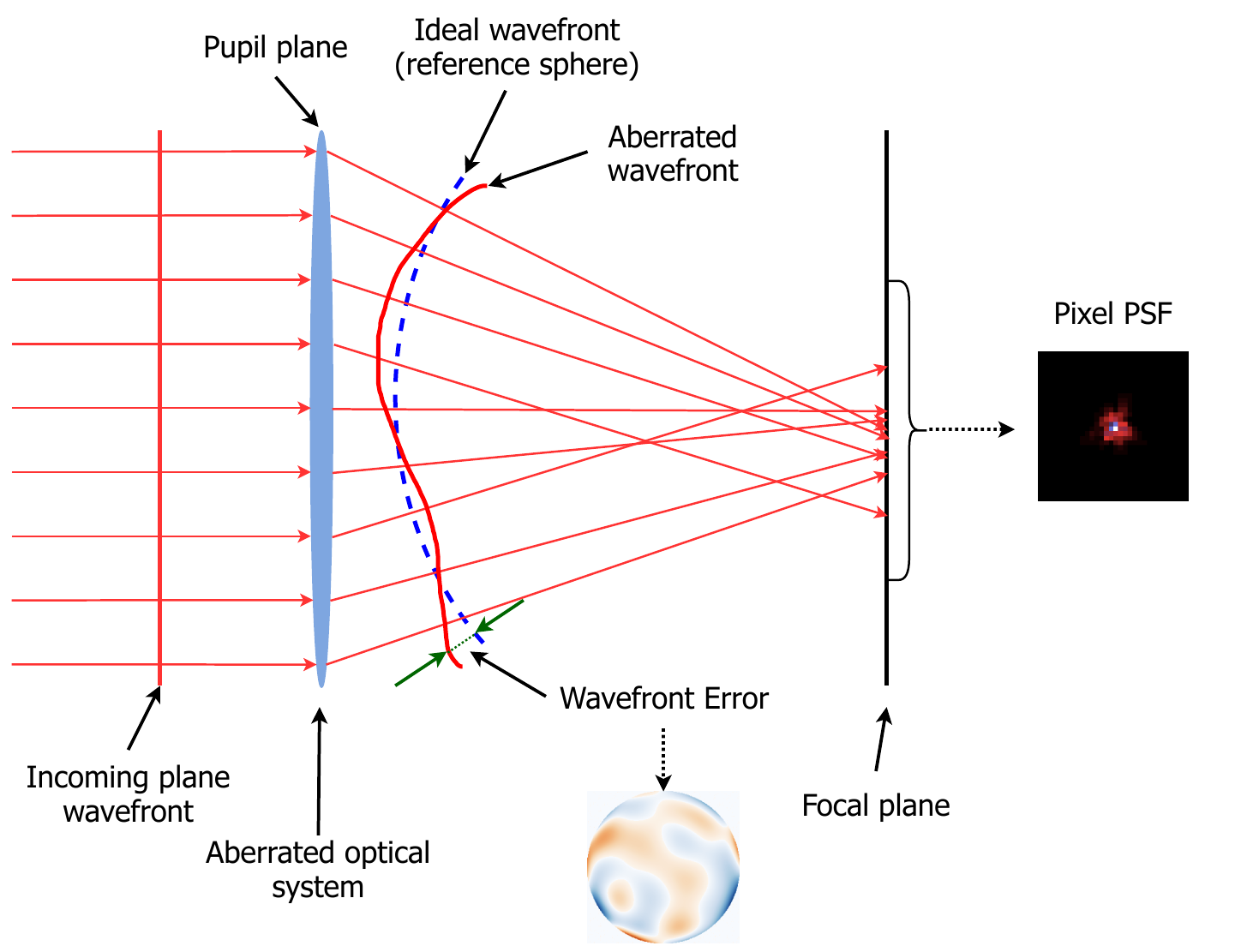}
        \caption{\label{fi:optics_b}}
    \end{subfigure}
    \caption{The figure on the left-hand side (\subref{fi:optics_a}) shows a schematic functional view of the \textit{Euclid}'s payload module taken from \cite{racca2016}. The figure on the right-hand side (\subref{fi:optics_b}) shows a one-dimensional illustration of the single-lens equivalent system where the wavefront error is visible.}
    \label{fi:wfe_explained}
\end{figure}

Current and future telescope optical systems require high optical performance, thus imposing complex designs. To illustrate this, \autoref{fi:optics_a} presents an example of the optical system of the \textit{Euclid} space telescope from \cite{racca2016}. The PSF can be modeled with high accuracy and complexity using powerful commercial optic simulators, e.g., \textsc{Zemax}. They depend on ray-tracing techniques to account for each optical surface. Although very accurate, these simulators are not practical for our problem due to their high complexity. They are not suited for massive PSF computations as are required in survey-type missions.

Nevertheless, we can consider a simpler optical system that approximates the actual system well. It should reproduce the existing system's dominant effects while being more accessible to model. We use some ideas from parametric models \cite{krist2011} and adopt a single converging lens system. Stars are well approximated by point sources at infinity and will be considered plane waves when entering the system. An ideal optical system turns the plane waves into spherical waves, and the rays converge at a single point at the instrument's focal plane. The exact point at which the rays converge depends on the angle of incidence of the incoming rays with respect to the optical axis.

Due to imperfections and aberrations of the optical system, the outgoing wavefront is not precisely spherical, and the rays do not converge in a single point. The difference between the actual aberrated wavefront and the ideal spherical wavefront is called the \textit{wavefront error} (WFE). \autoref{fi:optics_b} presents a simplified sketch of the WFE. The advantage of this simpler optical system is that we can take advantage of Fraunhofer's approximation of diffraction \cite[Chapter~4.3]{goodman2005} and its effect on the wavefront \cite[Chapter~5.2]{goodman2005}. These approximations allow us to relate the electric field's propagation and diffraction between the pupil plane and the focal plane, which can be seen in \autoref{fi:optics_b}. The field at the focal plane is the two-dimensional Fourier transform of the field at the pupil plane. The sensors measure the intensity of the electric field at the focal plane. Therefore, we can relate the electric field at the pupil plane and the observed pixel PSF by taking the squared modulus of its Fourier transform. Let us consider $(x_i,y_i) \in \mathbb{R}^{2}$ to be a FOV position that corresponds to a determined set of sky angular coordinates. For that specific FOV position and a given wavelength, we have the electric field at the pupil plane $\mathcal{U}_p(x_i, y_i; \lambda) : \mathbb{R}^{2} \to \mathbb{C}$. Then, the monochromatic PSF at the focal plane can be written as
\begin{equation}
    \hat{\mathcal{H}}(x_i, y_i; \lambda)[u,v] \propto \left| \iint_{\mathbb{R}^{2}} \mathcal{U}_p(x_i, y_i; \lambda)[\xi, \eta] \exp\left[ \frac{- 2 \pi \text{i}}{\lambda f_{L}} (u \xi + v \eta) \right] \dd \xi \dd \eta \right|^{2} \;,
    \label{eq:mono_PSF_model_theory}
\end{equation}
where $f_L$ is the optical system's focal length. 
The two coordinates of the pupil plane we integrate over are $[\xi, \eta]$, while the coordinates at the focal plane are $[u,v]$, where we observe our intensity PSF. We can rewrite the previous equation using the Fourier transform (FT) \cite{fourier1978} as
\begin{equation}
    \hat{\mathcal{H}}(x_i, y_i; \lambda)[u,v] \propto \left| \text{FT}\left\{\mathcal{U}_p(x_i, y_i; \lambda) \right\} \left[ \frac{u}{\lambda f_L}, \frac{v}{\lambda f_L} \right] \right|^2 \;.
    \label{eq:mono_PSF_model_theory_2}
\end{equation}
Then, we need to specify the electric field at the pupil plane $\mathcal{U}_p$ that characterizes the aberrations of the system. The electric field writes
\begin{equation}
    \mathcal{U}_p(x_i, y_i; \lambda)[\xi, \eta] = \mathcal{P}[\xi, \eta] \exp \left[ \frac{2 \pi \text{i}}{\lambda} \phi_{\theta}(x_i, y_i)[\xi, \eta] \right] \;,
    \label{eq:elec_field_theory}
\end{equation}
where $\mathcal{P}: \mathbb{R}^2 \to [0,1]$ represents the obscuration encountered at the focal plane, and $\phi_{\theta}(x_i, y_i): \mathbb{R}^2 \to \mathbb{R}$ represents the WFE or the aberrations of our optical system. Obscurations at the pupil arise commonly due to the lens' shapes and the superposing of mirrors and supports. It is especially the case for telescopes due to occultation from a secondary mirror and its supporting structures. The obscurations are known and unlikely to change during the telescope's lifetime, although their figure varies with the FOV position. In this work, we assume, for simplicity, that the obscuration patterns do not change with the FOV position. We consider an all-reflective-telescope design and neglect any wavelength dependence of the optical system's WFE, $\phi_{\theta}$. This assumption resumes considering that the dominant source of chromaticity in the PSF is due to the diffraction phenomena encoded in the $1/\lambda$ dependence in \autoref{eq:elec_field_theory}. We note that such wavelength dependence could be added to our model, for example, to model a telescope with a diffractive element in the optical system, e.g., the dichroic filter in \textit{Euclid} \cite{baron2022}.

To summarize, the WFE map is transformed into a complex electric field via \autoref{eq:elec_field_theory}. This field is propagated to the focal plane, and its intensity is computed to provide the pixel PSF with \autoref{eq:mono_PSF_model_theory_2}. We refer to the process of obtaining the pixel images from the WFE map as the \textit{optical forward model}.

\subsection{Wavefront error PSF model}

\begin{figure}
    \centering
    \includegraphics[width=\textwidth]{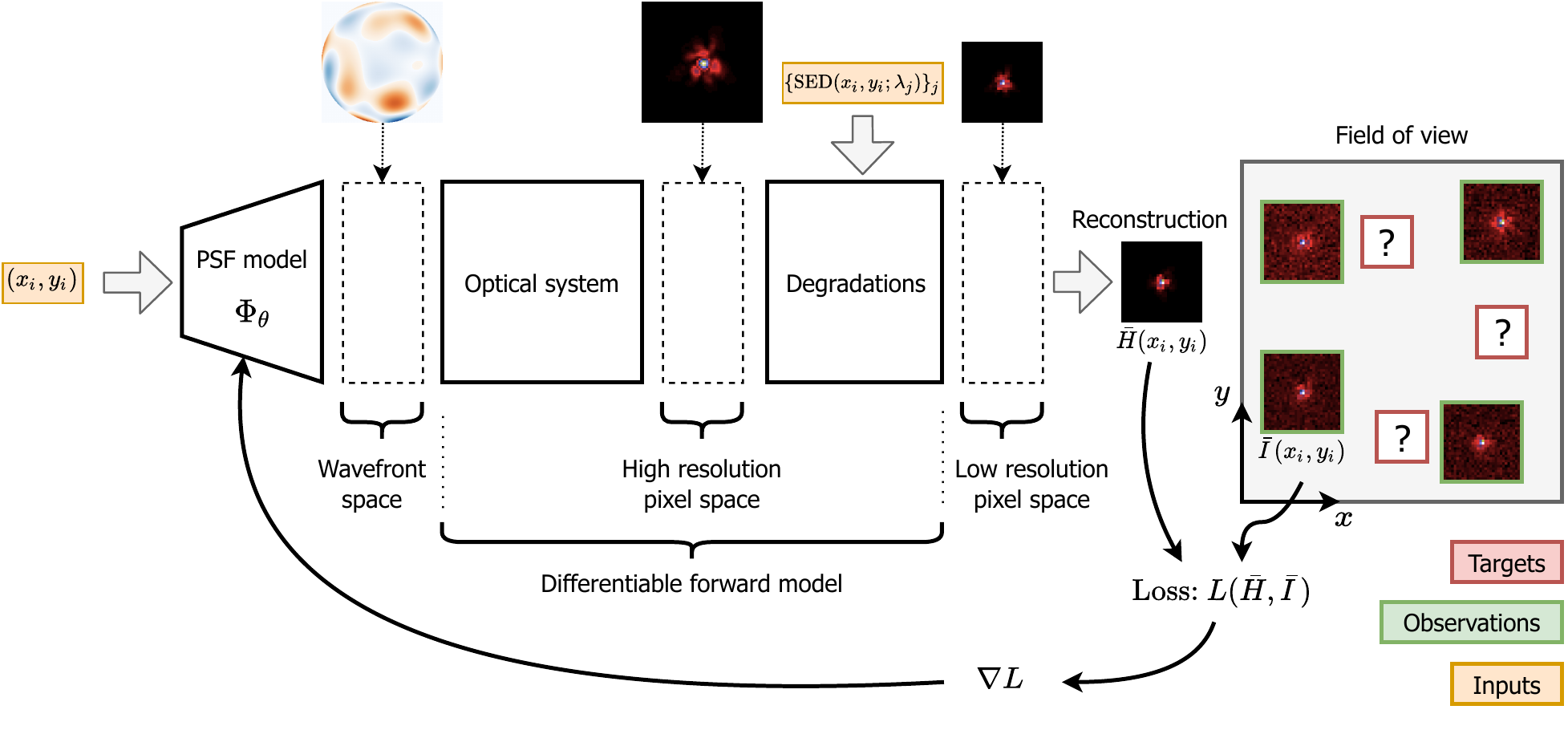}
    \caption{A schematic of the proposed framework for data-driven wavefront PSF modeling.}
    \label{fi:psf_model_diagram}
\end{figure}

In the following, we focus on how to use the WFE to model the PSF.
The core of the proposed framework is the WFE model, or WFE PSF model, $\Phi_{\theta}$, which is given by the following equation of the electric field at the pupil plane
\begin{equation}
    U_p(x_i, y_i; \lambda) = P \odot \exp\left[ \frac{2 \pi \text{i}}{\lambda} \; \Phi_{\theta}(x_i, y_i) \right] \;,
    \label{eq:U_PSF_model}
\end{equation}
where $\odot$ is the Hadamard or element-wise product, $U_p(x_i, y_i; \lambda) \in \mathbb{C}^{K \times K}$, and $P, \Phi_{\theta}(x_i, y_i) \in \mathbb{R}^{K \times K}$. 
The WFE PSF model depends on some parameters, represented with $\theta$, and its goal is to compute a WFE map at any position in the FOV. To conduct this task, we have to estimate, or \textit{learn}, the parameters $\theta$ of our model from the observed images using the full differentiable forward model that includes the degradations from the observational model in \autoref{eq:poly_star}. A schematic of the proposed framework is presented in \autoref{fi:psf_model_diagram}. We provide more detail of the forward model in \autoref{sc:optical_forward_model}.

We propose to base the WFE model on a weighted sum of wavefront features (or \textit{eigenWFE} if we draw a parallel with the notion of \textit{eigenPSF} \cite{liaudat2020, schmitz2020, ngole2016}).
A wavefront feature is distributed across the FOV and, therefore, shared by all the PSFs at any FOV position. Then, the weight that goes with that wavefront feature will change depending on the FOV position. 
We will consider two classes of wavefront features: fixed and data-driven. The former should provide a good representation basis of our WFE. In contrast, the latter should correct for mismatches of the previous ones, be learned from the data, and better adapt to the observations. 
We can write our WFE PSF model as follows
\begin{equation}
    \Phi_{\theta}(x_i,y_i) = \underbrace{\sum_{l=1}^{n_{\text{Z}}} f^{\text{Z}}_l(x_i,y_i) \; S^{\text{Z}}_l}_{\Phi^{\text{Z}}(x_i,y_i)} + \underbrace{\sum_{k=1}^{n_{\text{DD}}} f^{\text{DD}}_k(x_i,y_i) \; S^{\text{DD}}_k}_{\Phi^{\text{DD}}(x_i,y_i)} \;,
    \label{eq:wfe_psf_model}
\end{equation}
where $f^{\text{Z}}_l, f^{\text{DD}}_k : \mathbb{R}^{2} \to \mathbb{R}$ and $S^{\text{Z}}_l, S^{\text{DD}}_{k} \in \mathbb{R}^{K \times K}$. The functions $f^{\text{Z}}_l, f^{\text{DD}}_k$ give the weights for the wavefront features $S^{\text{Z}}_l, S^{\text{DD}}_{k}$ that correspond to the fixed and the data-driven (DD) parts respectively. A fixed number of features is set for both parts, $n_{\text{Z}}$ and $n_{\text{DD}}$, for the fixed and the DD, respectively. These hyperparameters allow controlling the desired complexity of the model.

\subsubsection{Zernike polynomials as fixed features} 
%
Zernike polynomials \cite{noll1976} have been widely used in the optics community. These polynomials provide a sound basis for modeling circular apertures as they are orthogonal on the unit disk. They are well suited to model the low-frequency variations of the WFE. We follow the definition from \cite{noll1976} and use polar coordinates $[\rho,\phi]$ instead of the Cartesian WFE coordinates $[\xi, \eta]$. The Zernike polynomials are defined as
\begin{align}
    Z_{n}^{m}[\rho, \phi] &= \sqrt{n+1} \, R^{m}_{n}(\rho) \, \sqrt{2} \,\cos(m \phi) ,\quad\quad \text{Even polynomial and } m \neq 0 \,,\\
    Z_{n}^{-m}[\rho, \phi] &= \sqrt{n+1} \, R^{m}_{n}(\rho) \, \sqrt{2} \,\sin(m \phi) ,\quad\quad \text{Odd polynomial and } m \neq 0 \,,\\
    Z_{0}^{n}[\rho, \phi] &= \sqrt{n+1} \, R^{0}_{n}(\rho) ,\qquad\qquad\qquad\quad\,\,  \text{ for } m = 0 \,,
\end{align}

where 
\begin{equation}
    R^{m}_{n}(\rho) = \sum_{s=0}^{\frac{n-m}{2}} \frac{\left( -1 \right)^{s} \, \left( n - s \right)!}{s! \, \left( \frac{n+m}{2} -s \right)! \, \left( \frac{n-m}{2} -s \right)!} \rho^{n - 2 s} , \quad \text{ with } R^{m}_{n}(1) = 1 ,\, \forall n,m\,,
\end{equation}

where $m$ and $n$ are nonnegative integers that respect $n \geq m \geq 0$ and that $n - m$ is even. The Zernike polynomial coordinates are limited to $0 \leq \phi < 2\pi$ and $0 \leq \rho \leq 1$. The orthogonality relation of the Zernike polynomials can be written as
\begin{equation}
    \int W(\rho) \, Z_{j} \, Z_{j'} \, d^{2}\rho = \delta_{j, j'},\,
\end{equation}

where $j$ and $j'$ represent a way of indexing the $(n,m)$ and $(n',m')$ values, and $W(\rho)$ is a weight function defined as
\begin{equation}
    W(\rho) = \left \{
        \begin{aligned}
            \frac{1}{\pi} \,, \quad \text{ if } \rho \leq 1, \\
            0 \,, \quad \text{ if } \rho > 1, 
        \end{aligned}
        \right. \;.
\end{equation}

\begin{figure}
    \centering
    \includegraphics[width=0.9\textwidth]{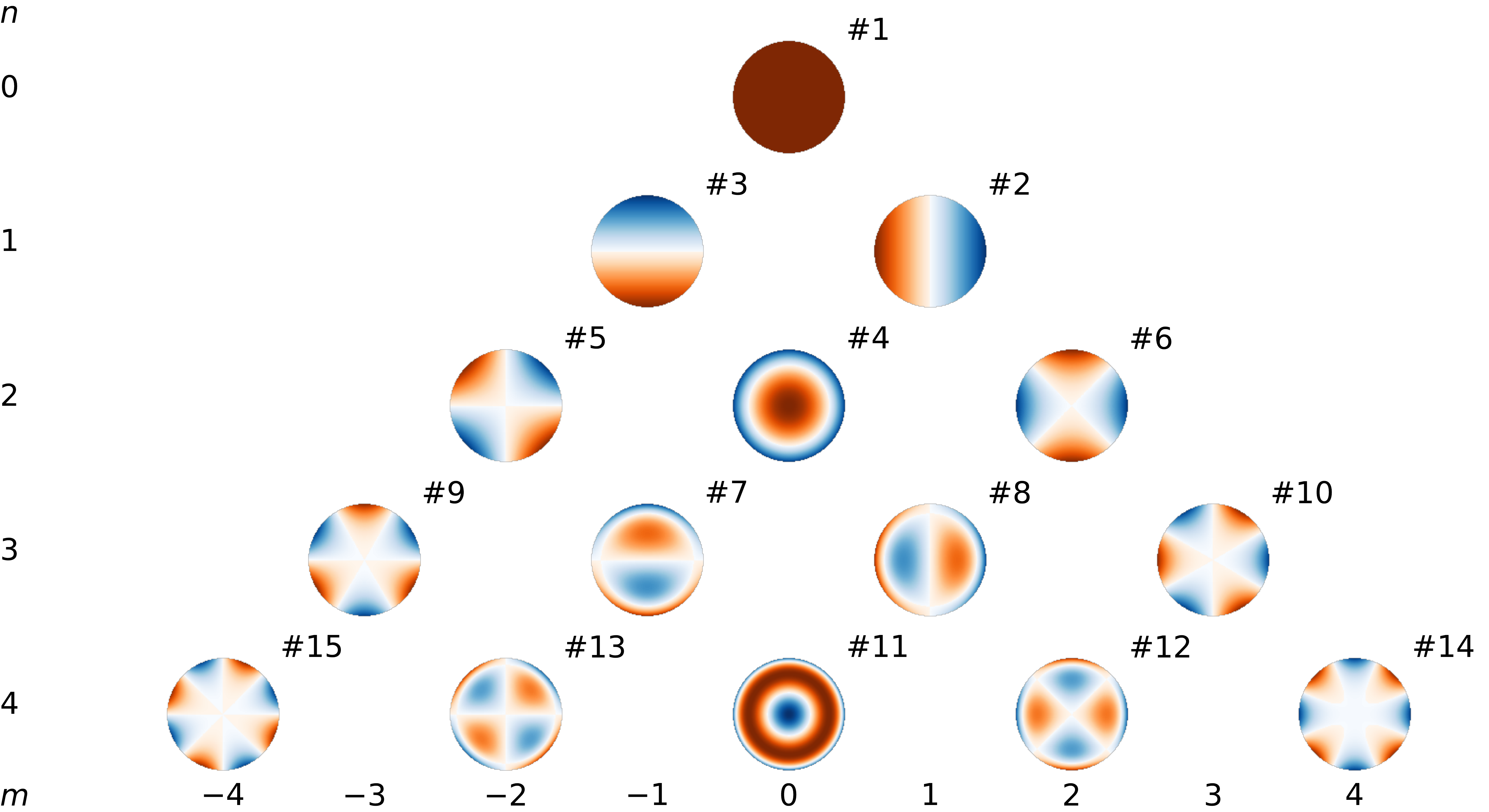}
    \caption{The first $15$ Zernike polynomials with their Noll's single-index prefaced by the hash symbol $\#$, and ordered vertically by their radial degree $n$ and horizontally by their azimuthal degree $m$.  \nblink{Zernike-example-plots}}
    \label{fi:zernike_example}
\end{figure}

\autoref{fi:zernike_example} shows the first $15$ Zernike polynomials that are a function of the pupil coordinates $[\xi, \eta]$ on a circular aperture. The weights $f^{\text{Z}}_l$, for each discretized Zernike polynomial $S^{\text{Z}}_l$, are defined as independent polynomial variations of the FOV position, $(x_i, y_i)$. One of the hyperparameters of the parametric model is the number of Zernike polynomials, $n_{\text{Z}}$. The other hyperparameter is the maximum degree of the FOV position polynomial, $d_{\text{Z}}$, that will accompany each Zernike polynomial. The parametric model can be expressed as
\begin{equation}
    \Phi^{\text{Z}}(x_i, y_i)[\xi, \eta] = \sum_{l=1}^{n_{\text{Z}}} \underbrace{\bm{\pi}^{\text{Z}}_{l}(x_i,y_i)^{\text{T}} \, \mathbbm{1}_{n_{d_{\text{Z}}}}}_{f_{l}^{\text{Z}}(x_i,y_i)} \, S_{l}^{\text{Z}}[\xi, \eta] \,,
    \label{eq_06:parametric_model}
\end{equation}

where $S_{l}^{\text{Z}}$ is the Zernike polynomial, $\mathbbm{1}_{n_{d_{\text{Z}}}} \in \mathbb{R}^{n_{d_{\text{Z}}} \times 1}$ is a vector of ones and the position polynomial vector $\bm{\pi}^{\text{Z}}_{l}$ for the Zernike index $l$ writes
\begin{equation}
    \bm{\pi}^{\text{Z}}_{l}(x,y) = \left[\pi^{\text{Z}}_{l,[0,0]},\, \pi^{\text{Z}}_{l,[1,0]} x,\, \pi^{\text{Z}}_{l,[0,1]} y,\, \cdots \,,\, \pi^{\text{Z}}_{l,[0,d_{\text{Z}}]} y^{d_{\text{Z}}} \right] \,,
\end{equation}

where $\{ \pi^{\text{Z}}_{l,[i,j]} \}_{i+j \leq d_{\text{Z}}}$ are the coefficients for of the position polynomial. The total number of parameters to estimate, if we use $n_{\text{Z}}$ Zernike polynomials, is $n_{\text{Z}} \, n_{d_{\text{Z}}}$, where $n_{d_{\text{Z}}} = (d_{\text{Z}} + 1)(d_{\text{Z}} + 2)/2$ corresponds to the number of monomials (and parameters) in the polynomial. We recall that the Zernike polynomial definition is in pupil plane coordinates $[\xi, \eta]$ (e.g.~the maps seen in \autoref{fi:zernike_example}). Note that in this model, $n_{\text{Z}}$ does not need to be equal to $n_{d_{\text{Z}}}$. On the one hand, $n_{\text{Z}}$ controls the maximum frequency content of the WFE in a given FOV position. On the other hand, $n_{d_{\text{Z}}}$ controls the maximum frequency in the FOV spatial variation of the WFE.

\subsubsection{Data-driven features}
\label{sc:data_driven_features}
The data-driven features, represented by the $S^{\text{DD}}_{k}$ matrices, are non-parametric. We do not impose a specific structure on the features, which are entirely learned from the data. These DD features should adapt to capture variations not well modeled by the Zernike-based features and should also account for unexpected mismatches between our forward model and the ground truth model. 

Maintaining a certain regularity in the PSF field and having a good generalization capability over the entire FOV is essential. 
The underlying hypothesis under the PSF field regularity is a consequence of the optical system that can be seen in \autoref{fi:optics_a}. The system is composed of mirrors of different sizes. The way the mirror's footprints propagate to the instrument's focal plane imposes a regularity in the PSF field. 
The simplified model we propose in \autoref{fi:optics_b} cannot inherently account for this propagation, and it is necessary to impose the regularity in the $f_{k}^{\text{DD}}(x,y)$. 

In the WaveDiff model, the feature weights are based on FOV position polynomials up to a degree $d_{\text{DD}}$. The amount of data-driven features used, $n_{\text{DD}}$, is fixed by the maximum polynomial degree $d_{\text{DD}}$, and is computed as $n_{\text{DD}} = (d_{\text{DD}} + 1)(d_{\text{DD}} + 2)/2$. In order to provide more flexibility to the model without modifying the maximum polynomial degree, we add a mixing matrix $A_{\text{mix}}^{\text{DD}} = [\bm{a}_1 , \bm{a}_2 , \cdots , \bm{a}_{n_{\text{DD}}}]$ that allows the features to contribute to more than one monomial from the FOV position polynomial. The data-driven contribution to the WFE writes
\begin{equation}
    \Phi^{\text{DD}}(x_i, y_i)[\xi, \eta] = \sum_{k=1}^{n_{\text{DD}}} \underbrace{\bm{\pi}^{\text{DD}}(x_i, y_i)^{\text{T}} \, \bm{a}_{k}}_{f_{k}^{\text{DD}}(x_i,y_i)} \, S_{k}^{\text{DD}}[\xi, \eta] \,,
    \label{eq_06:data_driven_model}
\end{equation}
where $\bm{\pi}^{\text{DD}}(x_i, y_i) \in \mathbb{R}^{n_{\text{DD}} \times 1}$ is a vector composed with each monomial of the $n_{\text{DD}}$-dimensional FOV position polynomial, $\bm{a}_{k} \in \mathbb{R}^{n_{\text{DD}} \times 1}$ is a column from the mixing matrix $A_{\text{mix}}^{\text{DD}} \in \mathbb{R}^{n_{\text{DD}} \times n_{\text{DD}} }$, and $S_{k}^{\text{DD}} \in \mathbb{R}^{K \times K}$ is a data-driven feature. The polynomial vector is expressed as follows
\begin{equation}
    \bm{\pi}^{\text{DD}}(x, y) = \left[ \pi^{\text{DD}}_{[0,0]},\, \pi^{\text{DD}}_{[1,0]} x ,\, \pi^{\text{DD}}_{[0,1]} y ,\, \pi^{\text{DD}}_{[2,0]} x^{2} ,\, \pi^{\text{DD}}_{[1,1]} x y ,\, \cdots ,\, \pi^{\text{DD}}_{[0, d_{\text{DD}}]} y^{d_{\text{DD}}} \right]^{\text{T}} ,
\end{equation}
where $\pi^{\text{DD}}_{[i,j]} \in \mathbb{R}$ for $i+j \leq d_{\text{DD}}$ are the polynomial coefficients. In this model, $S_{k}^{\text{DD}}$, $\bm{a}_{k}$ and $\bm{\pi}^{\text{DD}}$ for $k=1,\dots,n_{\text{DD}}$ will be learned from the data. The total number of parameters of the data-driven part is $n_{\text{DD}} + n_{\text{DD}}^{2} + n_{\text{DD}}\,K^{2}$. The proposed model constrains the WFE spatial variations with the degree $d_{\text{DD}}$ of the position polynomial. However, as the data-driven features $S_{k}^{\text{DD}}$ have no imposed structure, they can reproduce high WFE frequencies.

\subsubsection{Additional data-driven models}
We have considered two other WaveDiff data-driven models that we detail in \ref{ap:graph_DD_models}. They are based on a matrix factorization approach similar to \cite{liaudat2020}. The first one considers a graph constraint to capture more localized spatial variations. The second one uses the polynomial variations from  \autoref{eq_06:data_driven_model} in conjunction with the graph-constrained variations, allowing to capture smooth and localized variations. The sutdy of these supplementary models is presented in \ref{ap:wavediff_graph_flavours}.

\subsection{Practical optical forward model}
\label{sc:optical_forward_model}
The choice of the proposed model and Fourier optics allows computing the forward modeling of the wavefront to the pixels using a Fourier Transform, as described in \autoref{sc:physical_motivation}. Given the vast amount of data to process in upcoming missions, it is essential to develop fast methods. This requirement motivates using the Fast Fourier Transform (\texttt{FFT}) algorithm. \autoref{eq:mono_PSF_model_theory_2} shows that the sampling of the wavefront space changes with wavelength ($\lambda$). This observation implies that a resampling of the space is required for each wavelength if we want to work with a fixed space $\mathbb{C}^{K \times K}$. The resampling is impractical as several monochromatic PSFs are required to compute a single polychromatic PSF. Furthermore, we must consider that the input and output matrices of the \texttt{FFT} algorithm are of the same dimension. We want to be Nyquist sampled at all wavelengths to avoid any aliasing problem.

One way to accelerate the calculations is to fix the dimensions of the electric field at the pupil plane from \autoref{eq:U_PSF_model} as $\mathbb{C}^{K \times K}$, and then add a wavelength-dependent zero padding. Let us first define our zero-padding operator as follows,
\begin{equation}
    \underset{K \times K \to p \times p}{\text{pad}} : \mathbb{K}^{K \times K} \to \mathbb{K}^{p \times p}, \text{ where } \underset{K \times K \to p \times p}{\text{pad}}\left\{ A \right\} = P_{\text{p}} \, A \, P_{\text{p}}^{\text{T}}, \text{ with } 
    P_{\text{p}} = \begin{bmatrix}
            0_{\frac{p-K}{2} , K}  \\
            I_{K}  \\
            0_{\frac{p-K}{2} , K}  \\
        \end{bmatrix} ,
\end{equation}
for a matrix $A\in \mathbb{K}^{K \times K}$, where $p>K$, $P_{\text{p}} \in \mathbb{K}^{p \times K}$, $I_{K}$ is the identity matrix of dimension $K$, and $0_{a,b} \in \mathbb{K}^{a \times b}$ is a zero matrix. We have assumed that the padding applied is symmetric; therefore, $(p-K)/2$ will always be an integer.

The addition of zeros into the obscured part and the fixing of the non-obscured dimensions of the wavefront, i.e., $K$, allow us to control the sampling of the electrical field at the pupil plane. This practical method avoids resampling the electrical field for each wavelength and allows it to be Nyquist sampled. We zero-pad the electrical field matrix from $\mathbb{C}^{K \times K}$ to $\mathbb{C}^{p(\lambda) \times p(\lambda)}$. The required dimensions for the input matrix of the \texttt{FFT} operation can be seen in the following equation
\begin{equation}
    p(\lambda) = \frac{K \; \lambda \; f_{\mathrm{L}}}{M_{\mathrm{D}} \; \Delta} \; Q \;,
    \label{eq:wavelength_dep_zero_pad}
\end{equation}
where $K$ is the fixed matrix dimension of the electrical field that is equal to the pupil diameter in pixels, $f_{\mathrm{L}}$ is the focal length, $M_{\mathrm{D}}$ is the length of the pupil diameter, $\Delta$ is the length of a squared-pixel in the focal plane, $\lambda$ is the wavelength, and $Q$ is a dimensionless oversampling factor. See \ref{sc:app_oversamp_formula} for the derivation of the \autoref{eq:wavelength_dep_zero_pad} where we follow \cite{schmidt2010}. In practice, we fix the telescope's parameters, we choose the dimension $K$ we will work with, and then we calculate the required $Q$ so that we respect the Nyquist criterion ($p(\lambda) \geq 2 K$) for every $\lambda$ in our passband. 

Each computed monochromatic PSF will have different dimensions following the variable zero padding. An afterward crop to a fixed size of $\mathbb{R}^{M \times M}$ can solve this issue. We define the crop operation as follows,
\begin{equation}
    \underset{p \times p \to M \times M}{\text{crop}} : \mathbb{K}^{p \times p} \to \mathbb{K}^{M \times M}, \text{ where } \underset{p \times p \to M \times M}{\text{crop}}\left\{ B \right\} = P_{\text{c}} \, B \, P_{\text{c}}^{\text{T}}\;,
\end{equation}
where $p>M$, and $P_{\text{c}} = \begin{bmatrix} 0_{M, \frac{p-M}{2}} & I_{M} & 0_{M, \frac{p-M}{2}} \\ \end{bmatrix} \in \mathbb{K}^{M \times p}$. One can note that composing a zero-padding and crop operation, $\underset{p \times p \to M \times M}{\text{crop}} \circ \underset{M \times M \to p \times p}{\text{pad}} = \text{Id}$, returns the identity operation if the dimensions match. It is safe to assume that the energy contribution of the cropped pixels is negligible, given that $M$ is sufficiently big. The propagation of the wavefront PSF to the oversampled pixel PSF can be resumed as follows
\begin{equation}
    \hat{H}(x_i, y_i; \lambda) = \frac{1}{z_i} \underset{p(\lambda) \times p(\lambda) \to M \times M}{\text{crop}} \underbrace{\Bigg| \texttt{FFT}\bigg\{ \underset{K \times K \to p(\lambda) \times p(\lambda)}{\text{pad}} \Big( \overbrace{P \odot \exp\Big[ \frac{2 \pi i}{\lambda} \underbrace{\Phi_{\theta}(x_i, y_i)}_{\text{WFE PSF model }} }^{\text{Electric field}} \Big) \Big] \bigg\} \Bigg|^{2}}_{\text{Pixel representation}} ,
    \label{eq:mono_PSF_model}
\end{equation}
where $z_i$ is a normalization factor. We need to normalize the PSF so that it has a unit pixel sum, or $\sum\sum_{l,m = 1}^{M} \hat{H}(x_i, y_i; \lambda)[u_l,v_m] = 1$.
We still need to apply the degradations to the previous pixel PSF so that it can match our observations. This action resumes applying the observational model seen in \autoref{eq:poly_star} to the monochromatic oversampled PSF. We can then integrate over the instrument's passband and downsample to match the instrument's sampling, which is resumed as follows
\begin{equation}
    \hat{\bar{H}}(x_i, y_i) = D_\text{D} \left\{ \sum_{k=1}^{n_{\lambda}} \hat{H}(x_i, y_i; \lambda_k) \; \text{SED}(x_i, y_i; \lambda_k) \right\} \; ,
    \label{eq:spectral_approx}
\end{equation}
where we discretized the integral in \autoref{eq:poly_star} in $n_{\lambda}$ evenly spaced bins, and $D_\text{D}$ represents a downsampling by a factor $\text{D}$ that can include a further crop on the postage stamp. We need to use $\text{D} = Q$ to match the observation sampling. The SED in \autoref{eq:spectral_approx} has been normalized so that for a given $n_{\lambda}$ and position $(x,y)$, it verifies $\sum_{k=1}^{n_{\lambda}} \text{SED}(x, y; \lambda_k) = 1$. As one can see from \autoref{eq:spectral_approx}, we are not including other detector effects for the sake of simplicity. We use an oversampling factor $Q \in \mathbb{N}$ as the implementation of the downsampling operator is simplified. 

In a nutshell, \autoref{eq:mono_PSF_model} represents the optical system propagation in \autoref{fi:psf_model_diagram}, which gives us access to the monochromatic oversampled pixel PSF, while \autoref{eq:spectral_approx} represents the degradations from \autoref{fi:psf_model_diagram} that allows us to compute the reconstructed PSF at observing conditions. Finally, using the aforementioned equations, the optical forward model relating our model's parameters and the reconstructed PSF can be resumed as
\begin{equation}
    \hat{\bar{H}}(x_i, y_i) =  \texttt{Fwd}_{\left(\text{SED}(x_i,y_i)\right)} \left\{ \Phi_{\theta}(x_i, y_i) \right\} \; .
\end{equation}

\subsection{Model optimization}
\label{sc:training}

The PSF model's parameters we aim for will be the solution to the following optimization problem
\begin{equation}
    \argmin_{\substack{\bm{\pi}^{\text{Z}} \in\, \mathbb{R}^{n_{\text{Z}} \times 1} \\ A_{\text{mix}}^{\text{DD}} \in\, \mathbb{R}^{n_{\text{DD}} \times n_{\text{DD}}} \\ \bm{\pi}^{\text{DD}} \in\, \mathbb{R}^{n_{\text{DD}} \times 1} \\ S_{k}^{\text{DD}} \in\, \mathbb{R}^{K \times K} \,\forall\, k=1,\dots,n_{\text{DD}}}} \frac{1}{n_{\text{stars}}} \sum_{i=1}^{n_{\text{stars}}} \frac{1}{\hat{\sigma}_i} \left\| \hat{\bar{H}}(x_i, y_i) - \bar{I}(x_i, y_i) \right\|^{2}_{F} \; ,
    \label{eq:optim_problem}
\end{equation}
%
where
\begin{equation}
    \hat{\bar{H}}(x_i, y_i) = \texttt{Fwd}_{\left(\text{SED}(x_i,y_i)\right)} \left\{\sum_{l=1}^{n_{\text{Z}}} \bm{\pi}^{\text{Z}}_{l}(x_i,y_i)^{\text{T}} \, \mathbbm{1}_{n_{d_{\text{Z}}}} \, S_{l}^{\text{Z}} + \sum_{k=1}^{n_{\text{DD}}} \bm{\pi}^{\text{DD}}(x_i, y_i)^{\text{T}} \, \bm{a}_{k} \, S_{k}^{\text{DD}} \right\} \;,
\end{equation}
and where $\hat{\sigma}$ estimates the observed star noise standard deviation. The estimation is carried out using the median absolute deviation (MAD) \cite{starck2015}. We use the pixels from the outer region of the squared postage stamp, where we have masked the circular central region that contains most of the star's energy. 

Estimating the model's parameters from \autoref{eq:optim_problem} is not simple, as it involves solving a non-convex and possibly non-smooth optimization problem. If we do not consider the optical forward model $\texttt{Fwd}()$ for a moment, we would be facing a similar problem as in \cite{liaudat2020}. The non-convexity comes from the matrix factorization, and the non-smoothness from a regularization term added to \autoref{eq:optim_problem} as in the models in \ref{ap:wavediff_graph_flavours}. The use of proximal algorithms \cite{parikh2014, condat2013,beck2009} with a block coordinate descend approach \cite{xu2013} to exploit the multi-convex property of the matrix factorization problem would amount to solve the optimization problem. However, the optical forward model is non-convex and non-smooth as it introduces complex nonlinearities. Furthermore, the observed star images contain little information as they are in-focus PSFs, i.e., the incoming rays are concentrated in a small region. The observations have also suffered several degradations, i.e., noise, downsampling, and spectral integration.

Optimizing our PSF model can resemble the training of a neural network, which is implemented in an automatic differentiation framework and has an overparametrized parameter space, nonlinearities, and non-smoothnesses. The WaveDiff PSF model is automatically differentiable thanks to its implementation in the \textsc{TensorFlow} framework \cite{tensorflow2015}. In addition, the model's wavefront dimension $K$ can be several times bigger than the observing pixel dimension $M$, e.g., $K=256$ and $M=32$, resuming in an overparametrized space. One of the great tools that helped develop neural networks is the adoption of stochastic gradient descent methods. The noise introduced in the gradient estimation by a limited batch number acts as a regulariser to our model and helps to escape spurious local minima with bad performance. 
The optimization algorithm used in practice is Rectified Adam \cite{liu2020}, a stochastic gradient-based method widely used in the machine learning community, which, in our case, proved to be more stable than regular Adam \cite{kingma2014}. 

The data-driven features and the feature weight parameters are initialized using a uniform distribution with zero mean. The bounds of the distribution are kept close to zero, $-1$nm, and $1$nm. This choice means that the starting PSF model contains a reduced amount of aberrations, close to a perfect system, and has almost no spatial variations. This choice corresponds to starting with an almost perfect system, close to a vanishing WFE. However, the small non-zero aberration values help to escape local minima at the start of the optimization algorithm. The optimization adds aberrations to the system until the observed PSF field is well represented. We have found that the best working optimization strategy is to iterate the optimization between the parameters of the parametric and the data-driven parts. 
We use different learning rates and number of epochs for each optimization sequence. Algorithm \ref{al:wavediff_training} presents a summary of the training procedure.


The proposed PSF modeling framework, exploiting automatic differentiation, is versatile. If prior information is available, it can be included as regularizations at the wavefront or pixel levels. The framework can be adapted for other imaging applications, and other task-specific regularizers can be added to \autoref{eq:optim_problem}. The model was developed in the TensorFlow framework \cite{tensorflow2015}, which makes these adaptations straightforward.

\begin{algorithm}
    \caption{WaveDiff training procedure}
    \begin{algorithmic}[1]
    \label{al:wavediff_training}
    
    \STATE \underline{Hyperparameters}:
    \\ Learning rates $\{\eta_{m}^{\text{DD}} \}_{m=1, \ldots, m_{\max}}$, $\{\eta_{m}^{\text{Z}} \}_{m=1, \ldots, m_{\max}}$ 
    \\ Number of epochs $\{N_{\text{ep}, m}^{\text{DD}} \}_{m=1, \ldots, m_{\max}}$, $\{N_{\text{ep}, m}^{\text{Z}} \}_{m=1, \ldots, m_{\max}}$ 
    \bigskip
    
    \STATE \underline{Initialization}:  
    \\ Data-driven features: $S_{k}^{\text{DD}} \sim \mathcal{U}[-10^{-3}, 10^{-3}]^{K \times K}, \forall k$
    \\ Data-driven mixing matrix: $A_{\text{mix}} = I_{n_{\text{DD}}}$
    \\ Data-driven feature weights: $\pi^{\text{DD}}_{[l,m]} \sim \mathcal{U}[-10^{-2}, 10^{-2}], \forall \, l+m \leq d_{\text{DD}}$ 
    \\ Parametric feature weight: $\pi^{\text{Z}}_{l,[s,t]} \sim \mathcal{U}[-10^{-2}, 10^{-2}], \forall \, l \leq n_{\text{Z}},\, s+t \leq d_{\text{Z}}$
    \\ Estimate noise level, $\hat{\sigma}_i \leftarrow \text{MAD noise estimation} (\bar{I}(x_i, y_i))$
    \\ Generate Zernike polynomial maps $S_{l}^{\text{Z}}, \forall l$ and telescope obscuration $P$
    
    \bigskip
    \STATE \underline{Alternate optimization}:
    \FOR{$m=1$ to $m_{\max}$}
        \STATE Solve \autoref{eq:optim_problem} for $\pi^{\text{Z}}_{l,[s,t]}$, \hfill (parametric part)
        \\ $\quad\quad\quad\quad\quad \forall \, l \leq n_{\text{Z}},\, s+t \leq d_{\text{Z}}$ using $\eta_{m}^{\text{Z}}$, $N_{\text{ep}, m}^{\text{Z}}$
        \STATE Solve \autoref{eq:optim_problem} for $\pi^{\text{DD}}_{[l,m]},\, A_{\text{mix}},\, S_{k}^{\text{DD}}$, \hfill (data-driven part)
        \\ $\quad\quad\quad\quad\quad \forall \, k \leq n_{\text{DD}},\, l+m \leq d_{\text{DD}}$ using $\eta_{m}^{\text{DD}}$, $N_{\text{ep}, m}^{\text{DD}}$
    \ENDFOR
    
    \bigskip
    \STATE Return PSF model: $\Phi_{\theta}$
         
    \end{algorithmic}
\end{algorithm}

\subsection{PSF recovery}
\label{sc:inference}
PSF recovery consists of using the trained PSF model to estimate the PSF at other positions in the FOV where star observations are unavailable. The recovery requires specifying how the feature weight functions can extrapolate to different positions in the FOV. Then, it is possible to reuse the learned and fixed WFE features from \autoref{eq:wfe_psf_model} to calculate the WFE at the FOV position of interest. 

The procedure is straightforward when the weight functions are defined as FOV polynomials, as in the parts involving Zernike polynomials and WaveDiff data-driven part. It consists in evaluating the polynomial on a new position and obtaining the new feature weights.

\section{Numerical experiments}
\label{sc:numerical_experiments}
In this section, we start by defining the objectives of the numerical experiments. Then, we introduce the experiment setup and the data set we use to accomplish our goals. Finally, we present and discuss the results we find.

\subsection{Objectives}
To evaluate the novel framework and show its utility, we consider a scenario where unexpected physical phenomena introduced some complexity into the PSF field. This choice means that we are not explicitly accounting for the increase in complexity in our model. Another interpretation is that we underestimated the PSF field's complexity in our model. The practical way we set up the experiment is to simulate the WFE of a PSF field that is more complex than what the models can reproduce in the WFE space. In this setting, none of the models can have a zero reconstruction error over the ground truth WFE, even in a noiseless scenario. After simulating the complex WFE field, we can generate the required pixel PSFs using the optical forward model. Note that the added complexity is represented in the spatial variations of the WFE field and not in a change of the optical forward model.

This chosen setup allows comparing different model performances and observing to which degree the models can build an approximation of a complex WFE PSF field. The proposed setup is also helpful to study at which point the built approximation is good in terms of pixel PSF reconstruction error. We use this setting to study the following items:
\begin{itemize}
    \item[\textit{a)}] Pixel reconstruction performance at the observation resolution and at $3$ times the observation resolution (super-resolution) for polychromatic images.
 
    \item[\textit{b)}] Pixel reconstruction performance at the observation resolution as a function of wavelength for the best performing model of \textit{a)}.
    \item[\textit{c)}] Pixel reconstruction performance as a function of the total number of observed stars in FOV that are used to constrain the PSF models.
    \item[\textit{d)}] Error in the recovered WFE with respect to the ground truth WFE of the PSF field.
    \item[\textit{e)}] Errors using weak-lensing metrics.
\end{itemize}

\subsubsection{Reconstruction metrics}
The primary metric to compare the performance is the residual root mean square error (RMSE). We have access to the ground truth PSF field as we work with simulations. Therefore, we can calculate noiseless PSFs at any resolution and FOV position, making the performance comparison possible. We define the root mean square (RMS) of a matrix $A_i \in \mathbb{K}^{K \times K}$, or a vector in $\mathbb{K}^{K^2}$, as
\begin{equation}
 \text{RMS}(A_i) = \left( \frac{1}{K^2} \sum_{j=1}^{K^2} \left|A_{i;j}\right|^{2} \right)^{1/2} ,
\end{equation}
where we use a single index that corresponds to a flattened matrix. Let us consider a set of $n$ noiseless $K \times K$ dimensional images $\{ A_{i} \}_{i=1}^{n}$, and the corresponding reconstructions $\{ \hat{A}_{i} \}_{i=1}^{n}$. The main pixel reconstruction errors we use are the absolute error, $Err_{\text{abs}}$, and the relative error, $Err_{\text{rel}}$, which are defined as
\begin{equation}
 Err_{\text{abs}} = \frac{1}{n} \sum_{i=1}^{n} \text{RMS}(A_{i} - \hat{A}_{i}) \;,\quad Err_{\text{rel}} = \frac{1}{n} \sum_{i=1}^{n} \frac{\text{RMS}(A_{i} - \hat{A}_{i})}{\text{RMS}(A_{i})} \times 100\% \;.
\end{equation}
A constant value over the WFE matrix does not affect the pixel PSF. Therefore, we remove each WFE matrix's mean before computing the residual's RMS in the WFE error calculation. We only use the non-obscured elements of the WFE matrix to compute the RMS.

\subsubsection{Weak-lensing metrics}
Weak lensing sets the most stringent requirements for the errors of the PSF models. Cosmologists have developed shape and size metrics \cite{massey2012} that can be related to the errors in the cosmological parameters \cite{cropper2013}. Therefore, the PSF model requirements are usually set up regarding these metrics. These are defined in terms of the moments of the polychromatic observation $\bar{I}[u, v]$, following \cite{hirata2003}, we define
\begin{align}
    \label{eq:psf_moments_1}
    \bar{\mu} & = \frac{\int \mu \; \bar{I}[u, v] \; w[u, v] \; \text{d}u \, \text{d}v }{\int \bar{I}[u, v] \; w[u, v] \; \text{d}u \, \text{d}v}, \\
    M_{\mu \nu} &= \frac{\int \bar{I}[u, v] \; (\mu - \bar{\mu}) \; (\nu - \bar{\nu}) \; w[u, v] \; \text{d}u \, \text{d}v}{\int \bar{I}[u, v] w[u, v] \; \text{d}u \, \text{d}v},
    \label{eq:psf_moments_2}
\end{align}
where $\mu , \nu \in \{u, v\}$ and $w[u, v]$ is a weight window used to avoid the divergence of the integrals due to noise. \autoref{eq:psf_moments_1} defines the first-order moments, while \autoref{eq:psf_moments_2} defines the second-order moments. The size metric is defined as
\begin{equation}
    R^{2} = T = M_{u u} + M_{v v},
\end{equation}
and the shape metrics, or ellipticities, are defined as
\begin{equation}
    e = e_1 + {\rm i} e_2 = \frac{(M_{u u} - M_{v v}) + {\rm i} \, 2 M_{u v}}{T} .
\end{equation}
The method used to estimate these metrics is the widely-used adaptive moment algorithm from \texttt{Galsim}'s HSM module\footnote{\url{https://github.com/GalSim-developers/GalSim}} \cite{hirata2003,mandelbaum2005}. The adaptive moment algorithm measurement provides $\sigma$ as size, which relates to the above-defined size metric as $R^2 = 2\sigma^2$. The measurements are carried out on super-resolved polychromatic images.

To calculate the error metrics, we first have to measure the ellipticity and size parameters using the adaptive moments' algorithm over each image from the sets of ground truth images $\{ A_{i} \}_{i=1}^{n}$, and reconstructions $\{ \hat{A}_{i} \}_{i=1}^{n}$. The measurements are then used to build vectors as, for example, $\mathbf{e}_{1,2} = [e_{1,2; 1}, \ldots, e_{1,2; n}]$. Finally, the error metrics used are the RMSE over each metric that can be written as
\begin{equation}
    \text{RMSE}(\mathbf{e}_{1,2}) = \text{RMS}(\mathbf{e}_{1,2} - \hat{\mathbf{e}}_{1,2})\;, \quad \text{RMSE}(\mathbf{R}^{2}) = \frac{\text{RMS}(\mathbf{R}^{2} - \hat{\mathbf{R}}^{2})}{\langle \mathbf{R}^{2} \rangle},
\end{equation}
where the size metric is divided by its mean, as is usually done in the literature.

\subsection{Experiment setup}
\label{sc:experiment_set_up}
To demonstrate our novel framework, we simulate a simplified FOV with $2000$ star observations for training and $400$ noiseless target stars for testing. All positions are randomly distributed in the FOV. To evaluate the impact of the number of stars in the PSF field on the performance, we build three subsets $\mathcal{S}_1$, $\mathcal{S}_2$, and $\mathcal{S}_3$ containing $200$, $500$, and $1000$ stars, respectively. The stars are assigned to the subsets so that they verify $\mathcal{S}_1 \subset \mathcal{S}_2 \subset \mathcal{S}_3 \subset \mathcal{S}$, where $\mathcal{S}$ is the data set with the entire training stars. This assignment provides a fair comparison and is less sensitive to the randomness of the positions. The observations have a variable signal-to-noise ratio (SNR) that is uniformly distributed in the range $[10, 110]$. The SNR definition we use is
\begin{equation}
 \text{SNR} = \frac{\left\| \bar{I} \right\|_{F}^{2}}{\sigma_{\bar{I}}^2 N^2}, \quad \left\| \bar{I} \right\|_{F}^{2} = \sum_{i,j = 1}^{N,N} | \bar{I}_{i,j} |^{2},
\end{equation}
where $\bar{I} \in \mathbb{R}^{N \times N}$ is an observation corrupted with white Gaussian noise with standard deviation $\sigma_{\bar{I}}$, and the operation $\| \cdot \|_{F}^{2}$ denotes the Frobenius matrix norm.

We use the parameters of an optical model close to the \textit{Euclid}'s VIS instrument model \cite{laureijs2011, venancio2020}. We consider a broad passband of $[550, 900]$nm, a telescope's focal length of $f_{\text{L}}=24.5$m, a telescope pupil diameter of $M_{\text{D}}=1.2$m, a squared pixel size of $\Delta=12\mu$m\footnote{This corresponds to $0.1$ arcsecond.}, and an oversampling factor of $Q=3$. The obscuration used is fixed for every position in the FOV and is built as a superposition of simple geometrical shapes, as seen in \cite[Fig. 7]{venancio2020}. A moving average filter is applied to the obscuration to smooth the transitions between the obscured and illuminated parts to avoid aliasing. We set $n_{\lambda}=20$ wavelength bins for the approximation in \autoref{eq:spectral_approx}. The dimensions used are $K=256$ for the WFE space, $M=64$ for the high-resolution pixel space, and $N=32$ for the observation, or low-resolution, pixel space.

\begin{figure}
    \centering
    \includegraphics[width=\textwidth]{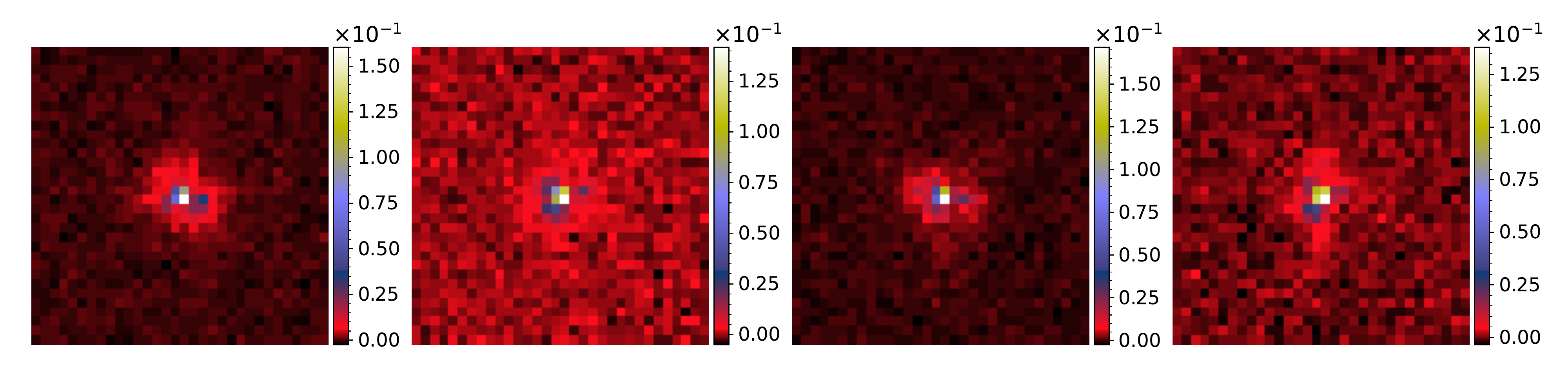}
    \includegraphics[width=\textwidth]{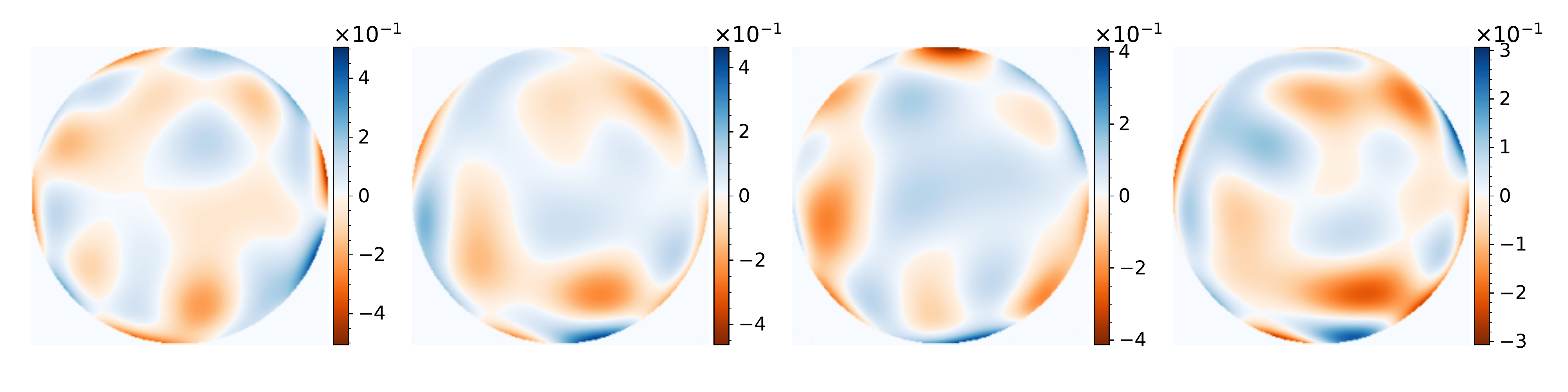}
    \caption{Example of PSFs from the simulated data set. The first row contains images of observed stars, while the second row contains the corresponding WFE map used to generate each one. \nblink{DD-features-dataset-examples}}
    \label{fi:dataset_example}
\end{figure}

SED information is required to build polychromatic observations. We randomly choose for each star one of $13$ stellar SED templates from \cite{pickles1998} following the work in \cite{kuntzer2016}. More information about the SEDs can be found in \ref{sc:stellar_SEDs}. The SED data is considered a perfectly-known input to the PSF model. The ground truth (GT) PSF field is built using $45$ Zernike polynomials with a $d_{\text{Z}}$ of $2$ for each mode. The polynomial coefficients are randomly chosen, maintaining the total amount of aberration at any position in the FOV close to a nominal value. The total amount of aberrations can be characterized by the RMS value of the WFE map, and the nominal value is defined as $100$nm. No further detector effects are considered in this work. The instrument's focal plane is constituted of a squared matrix of squared CCD chips, or CCD mosaic \cite{cropper2016}. This work does not consider any discontinuity between the different CCD chips and assumes that the CCDs are perfectly aligned in the focal plane and share the same properties. The diffraction phenomena are the only sources of chromatic variation in the PSF field. \autoref{fi:dataset_example} presents four PSF examples from the simulated data set, where the observed stars and their corresponding WFE is shown.

The following models are compared:
\begin{itemize}
    \item[\textit{i)}] \textit{Zernike 15}: a model using the proposed optical forward model but without any data-driven features. Only the fixed features, Zernike polynomials, are being used. We use $n_{\text{Z}}=15$, and $d_{\text{Z}}=2$. In this study, this model represents a parametric approach where its parameters were badly specified. As a consequence, the complexity of the model is far away from the ground truth PSF field.
    \item[\textit{ii)}] \textit{Zernike 40}: a model similar to \textit{i)}, with $d_{\text{Z}}=2$, but using $n_{\text{Z}}=40$. In this study, this model represents a parametric model where its parameters were slightly misspecified. This model is close to the ground truth PSF field. Er, it still needs more complexity as the number of Zernike polynomials is lower than the ground truth.
    \item[\textit{iii)}] \textit{WaveDiff}: the model described in \autoref{sc:data_driven_features} using $n_{\text{Z}}=15$, $d_{\text{Z}}=2$, and $d_{\text{DD}}=5$ that corresponds to $n_{\text{DD}}=21$. This model uses the same badly specified parameters of the fixed features as the model \textit{i)}.
    %
    \item[\textit{iv)}] \texttt{PSFEx}: The choice of the model parameters is based on the numerical experiment done in \cite{schmitz2020}. The model is built independently on each CCD. The spatial variations of the PSF, \texttt{PSFVAR\_KEYS}, are a function of the position parameters, \texttt{XWIN\_IMAGE, YWIN\_IMAGE}. The sampling and the image dimension are adjusted for our experiments. We use a \texttt{PSF\_SAMPLING} of \texttt{1.} and \texttt{0.33}, and a \texttt{PSF\_SIZE} of \texttt{32,32} and \texttt{64,64}, for reconstructions at the observation resolution and at $3$ times the observation resolution, respectively. We use a \texttt{PSFVAR\_DEGREES} of \texttt{3}, the degree of the polynomial used for the spatial variation function, as we found that it gave better results than other values. See \autoref{sc:related_work} for a detailed description.
    \item[\textit{v)}] \texttt{RCA}: the current state-of-the-art data-driven PSF model \cite{schmitz2020} which has been specially designed for the \textit{Euclid} mission. The model is built independently on each CCD. We verified that the parameters used in \cite{schmitz2020} were the best-performing ones. Therefore, we use \texttt{4} features, or EigenPSFs, with a denoising parameter \texttt{K} of \texttt{3}. We adjust the upsampling parameter \texttt{D} to match the two different reconstruction resolutions desired, that in our case is \texttt{1} and \texttt{0.33}.
    \item[\textit{vi)}] \texttt{MCCD}: a recently developed data-driven PSF model \cite{liaudat2020} that combines ideas from the two previous models. \texttt{PSFEx} and \texttt{RCA} build a PSF model per CCD chip using fewer stars. The \texttt{MCCD} model can build a PSF model in the full FOV taking into account the instrument's geometry. We compared models with different parameters and selected the one that was the best performing. The global polynomial degree is set to \texttt{6} with a global denoising parameter, \texttt{K\_GLOB}, set to \texttt{3}. For the local part, the hybrid model is being used with \texttt{4} graph features and a degree of \texttt{2} for the polynomial features. The local denoising parameter, \texttt{K\_LOC}, is also set to \texttt{3}. 
\end{itemize}

Our setup for testing \texttt{PSFEx} and \texttt{RCA} is different from \cite{ngole2016}, and \cite{schmitz2020}. Here, the observed PSF field has a spatially varying SNR, and we simulate more realistic polychromatic PSFs. The observed PSF positions are uniformly distributed in the FOV. The studies \cite{ngole2016, schmitz2020} used monochromatic PSFs sampled at $600$ nm, requiring a super-resolution factor of $2$ to be Nyquist sampled. However, in the current experiment, when working with polychromatic PSFs, we consider monochromatic PSFs from $550$ nm to $900$ nm, requiring an integer super-resolution factor of $3$ to be Nyquist sampled. As a consequence of the different experimental setups, the performance results of \texttt{PSFEx} and \texttt{RCA} are expected to be different compared to \cite{ngole2016, schmitz2020}. \ref{sc:dataset_adaptations} describes the adaptations the pixel-based PSF models required to handle the aforementioned simulations.

\autoref{tb:parameters} presents the hyperparameters used to train the different PSF models that use the optical forward model. These have been tuned to optimize performance rather than computation time. All the models use a batch size of $32$. 

\begin{table}
\centering
\caption{Hyperparameters of the different PSF models that use the optical forward model.}
\resizebox{\textwidth}{!}{%
    \begin{tabular}{lccccccccc}
        \toprule
        &&\multicolumn{4}{c}{$m_1$} &  \multicolumn{4}{c}{$m_2$} \\
        \cmidrule(r){3-6} \cmidrule(r){7-10}
        PSF model & Data set   & $\eta_{1}^{\text{Z}}[\times 10^{-3}]$ & $\eta_{1}^{\text{DD}}$ & $N_{\text{ep},1}^{\text{Z}}$ & $N_{\text{ep},1}^{\text{DD}}$  & $\eta_{2}^{\text{Z}}[\times 10^{-3}]$ & $\eta_{2}^{\text{DD}}$ & $N_{\text{ep},2}^{\text{Z}}$ & $N_{\text{ep},2}^{\text{DD}}$ \\
        \midrule
        \multirow{3}{*}{Zernike $15$}& $\mathcal{S}_{1}$  & \multirow{3}{*}{$5.0$}   & \multirow{3}{*}{-} & $40$ & \multirow{3}{*}{-} & \multirow{3}{*}{$1.0$} & \multirow{3}{*}{-} & $40$ & \multirow{3}{*}{-} \\
        &$\mathcal{S}_{2}$                 &  &  & $30$ &  & & & $30$ & \\
        &$\mathcal{S}_{3}$                 &  &  & $30$ &  & & & $30$ & \\
        &$\mathcal{S}$                     &  &  & $20$ &  & & & $20$ & \\
        \midrule
        \multirow{3}{*}{Zernike $40$}& $\mathcal{S}_{1}$  & \multirow{3}{*}{$5.0$}   & \multirow{3}{*}{-} & $40$ & \multirow{3}{*}{-} & \multirow{3}{*}{$1.0$} & \multirow{3}{*}{-} & $40$ & \multirow{3}{*}{-} \\
        &$\mathcal{S}_{2}$                 &  &  & $30$ &  & & & $30$ & \\
        &$\mathcal{S}_{3}$                 &  &  & $30$ &  & & & $30$ & \\
        &$\mathcal{S}$                     &  &  & $20$ &  & & & $20$ & \\
        \midrule
        \multirow{3}{*}{\shortstack[l]{WaveDiff}}& $\mathcal{S}_{1}$  & \multirow{3}{*}{$10.0$}   & \multirow{3}{*}{$0.1$} & $30$ & $300$ & \multirow{3}{*}{$4.0$} & \multirow{3}{*}{$0.06$} & $30$ & $300$ \\
        &$\mathcal{S}_{2}$                 &  &  & $30$ & $200$ & & & $30$ & $150$ \\
        &$\mathcal{S}_{3}$                 &  &  & $20$ & $150$ & & & $20$ & $100$ \\
        &$\mathcal{S}$                     &  &  & $15$ & $100$ & & & $15$ & $50$ \\
        \bottomrule
    \end{tabular}}
\label{tb:parameters}
\end{table}

\subsection{Results}
The results presented here were obtained from the set of testing stars that were not used to optimize the different PSF models. The WaveDiff model used an Nvidia Tesla V100 GPU of 32 GB. The optimization time scales with the number of observations in the training set. However, the number of epochs used for the optimization also varies. An epoch is defined as a complete cycle over the training dataset. Optimizing the WaveDiff model with the $2000$-star data set took $11.5$ hours, while the $200$-star data set took $8.2$ hours. Note that the hyperparameters, especially the number of epochs, were set to maximize the performance, not the computing time.

\subsubsection{Results: (a), Polychromatic errors}
\label{sc:result_objective_a}

\autoref{tb:results_a} summarizes the pixel reconstruction performance results on polychromatic images. There is a significant improvement of WaveDiff, compared to the pixel-based models, \texttt{PSFEx}, \texttt{RCA}, and \texttt{MCCD}. Considering the reconstruction at the observation resolution (x$1$), the WaveDiff model reaches $6$ to almost $11$ times lower absolute errors. The errors are decreased by a factor between $34$ and $51$ when considering the super-resolution task (x$3$).

\begin{table}
    \centering
    \caption{Polychromatic test star reconstruction $Err_{\text{abs}}$ and $Err_{\text{rel}}$ at the observation resolution (x1) and at super-resolution (x3). The presented results were obtained with models being trained on the largest training star set $\mathcal{S}$. \nblink{table-pixel-errors-metrics}}
        \begin{tabular}{lcc}
            \toprule
            &\multicolumn{2}{c}{$Err_{\text{abs}}$ [$\times 10^{-5}$] ($Err_{\text{rel}}$)}   \\
            \cmidrule(r){2-3}
            PSF model   & Resolution x1 & Resolution x3 \\
            \midrule
            \textit{i)} Zernike 15                   &  $77.5$ ($10.6\%$)    & $19.0$ ($12.8\%$)     \\
            \textit{ii)} Zernike 40                  &  $32.6$ ($4.4\%$)     & $8.7$ ($5.9\%$)      \\
            \textbf{\textit{iii)} WaveDiff} &  $\mathbf{6.4}$ ($\mathbf{0.86\%}$)     & $\mathbf{1.9}$ ($\mathbf{1.3\%}$)      \\
            \textit{iv)} \texttt{PSFEx}             &  $69.2$ ($9.5\%$)    & $66.3$ ($43.0\%$)     \\
            \textit{v)} \texttt{RCA}              &  $39.6$ ($5.4\%$)    & $85.3$ ($55.5\%$)     \\
            \textit{vi)} \texttt{MCCD}            &  $43.5$ ($6.0\%$)    & $97.7$ ($63.4\%$)     \\
            \bottomrule
        \end{tabular}
    \label{tb:results_a}
\end{table}

The Zernike $15$ model, or \textit{i)}, underperforms with respect to the WFE-based models at the observation resolution. It is even poorer than the state-of-the-art models, thus highlighting the lack of representation of a reduced number of Zernike polynomials. If we remove the data-driven features from WaveDiff, we obtain a model equivalent to model \textit{i)}. This difference in the performance between model \textit{i)} and the WaveDiff model reflects the importance of the data-driven features in the WFE-based PSF model. It also shows the effectiveness of the data-driven features in adapting to the observations and generalizing to the target positions.
Concerning the super-resolution task, the model \textit{i)} still outperforms the pixel-based models. There is a huge performance gap between the models using a differentiable forward model and those building their models directly in the pixel space. The super-resolution is taken into account naturally when using the optical forward model. This handling of the super-resolution is done in the image formation process encoded in the forward model with the subsequent downsampling to obtain a PSF at the observation resolution. The observed performance gap underlines the importance of adding prior physical information to the inverse problem to solve a challenging task such as super-resolution.

Model \textit{ii)} is performing poorly with respect to WaveDiff, even though the number of Zernike polynomials, $d_{\text{Z}}=40$, is close to the ground truth. This slight mismatch leads to significant PSF reconstruction errors, demonstrating parametric models' rigidity. The WaveDiff-original model with only $15$ Zernike polynomials does not suffer from this restriction, showing the flexibility of the data-driven component of the model.

\begin{figure}
    \centering
    \includegraphics[width=\textwidth]{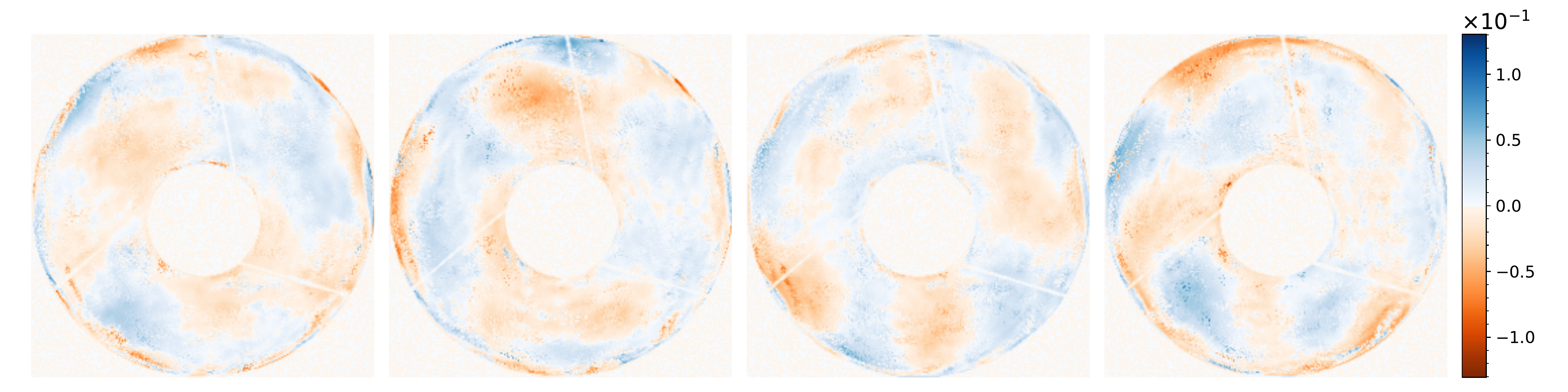}
    \caption{Examples of learned data-driven eigenWFEs, or WFE features, of the WaveDiff model. These features correspond to polynomial feature weight functions. Units are in $\mu$m. \nblink{DD-features-dataset-examples}}
    \label{fi:eigen_psfs}
\end{figure}

\autoref{fi:eigen_psfs} presents some examples of learned data-driven WFE features for the WaveDiff model. The features show that structure was learned, indicating that we are estimating a WFE manifold. The learning of the WFE manifold results from the optimization through the differentiable optical model. Note that no constraint has been applied to the data-driven WFE features.


\begin{figure}
    \centering
    \includegraphics[width=0.95\textwidth]{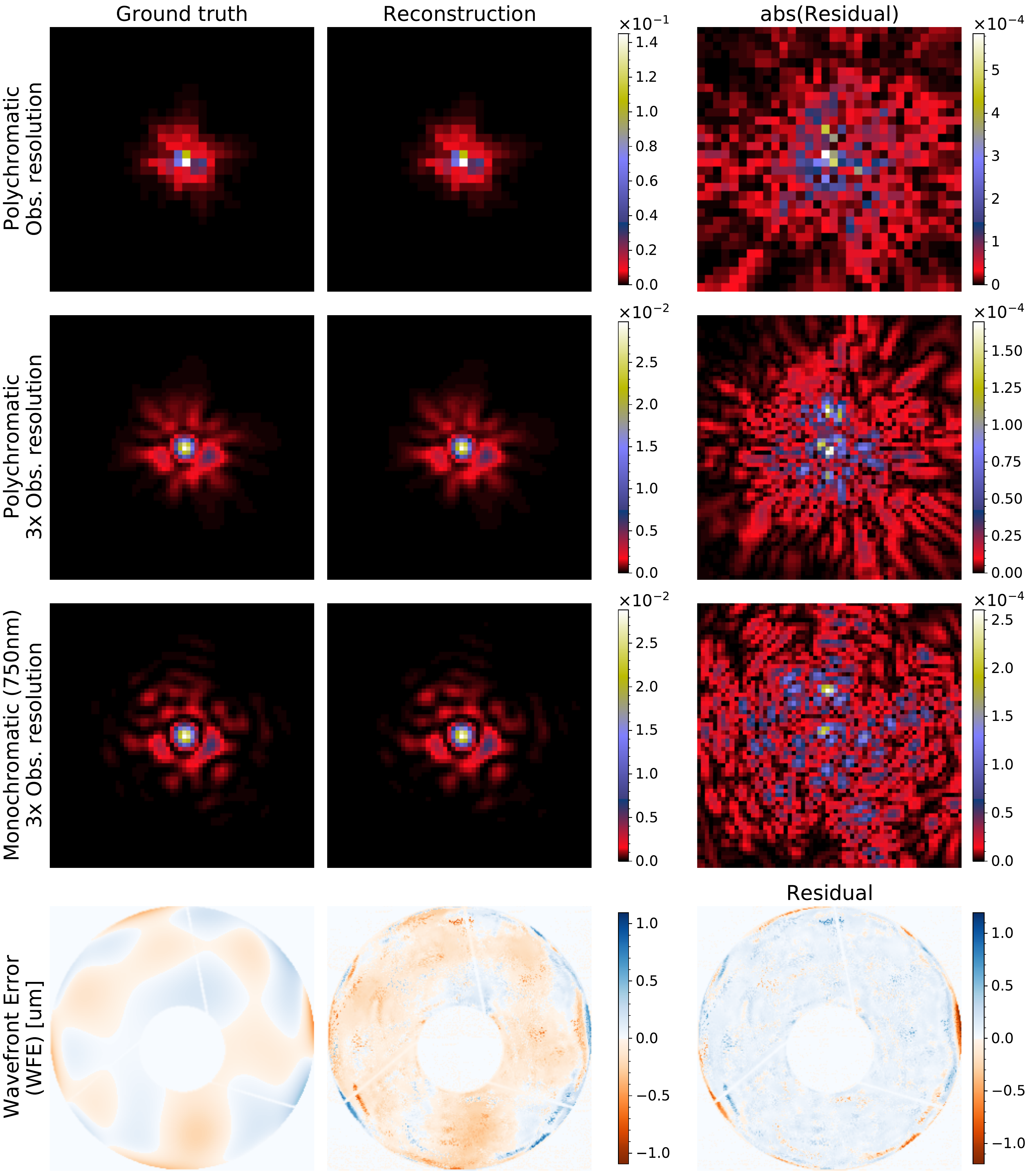}
    \caption{Visual reconstruction of a test PSF with the WaveDiff model. The first column shows the ground truth noiseless PSF, the second column the reconstruction of the PSF model, and the third column the residual between the ground truth and the model. 
    \textbf{First row:} Polychromatic pixel representation at the observation resolution. \textbf{Second row:} Polychromatic representation at $3$ times the observation resolution. \textbf{Third row:} Monochromatic pixel representation at $750$ nm and at $3$ times the observation resolution. \textbf{Fourth row:} WFE representation of the PSF, where units are in $\mu$m. We have removed the mean of the WFE and applied the obscurations.
    The first three rows show the absolute value of the residual, while the last shows the residual.
    \nblink{DD-features-dataset-examples}}
    \label{fi:visual_reconstruction_psf}
\end{figure}

\autoref{fi:visual_reconstruction_psf} presents the reconstruction of a test PSF done by the WaveDiff model. The first three rows show the excellent reconstruction quality of the WaveDiff model for the different pixel scenarios. The fourth row shows important errors in the WFE reconstruction; nonetheless, the pixel reconstructions are very similar. This fact shows the richness of the WFE space and how two very different WFE maps can recreate remarkably similar pixel PSF images.

\subsubsection{Results: (b), Monochromatic errors}
\label{sc:result_objective_b}

\begin{figure}
    \centering
    \includegraphics[width=\textwidth]{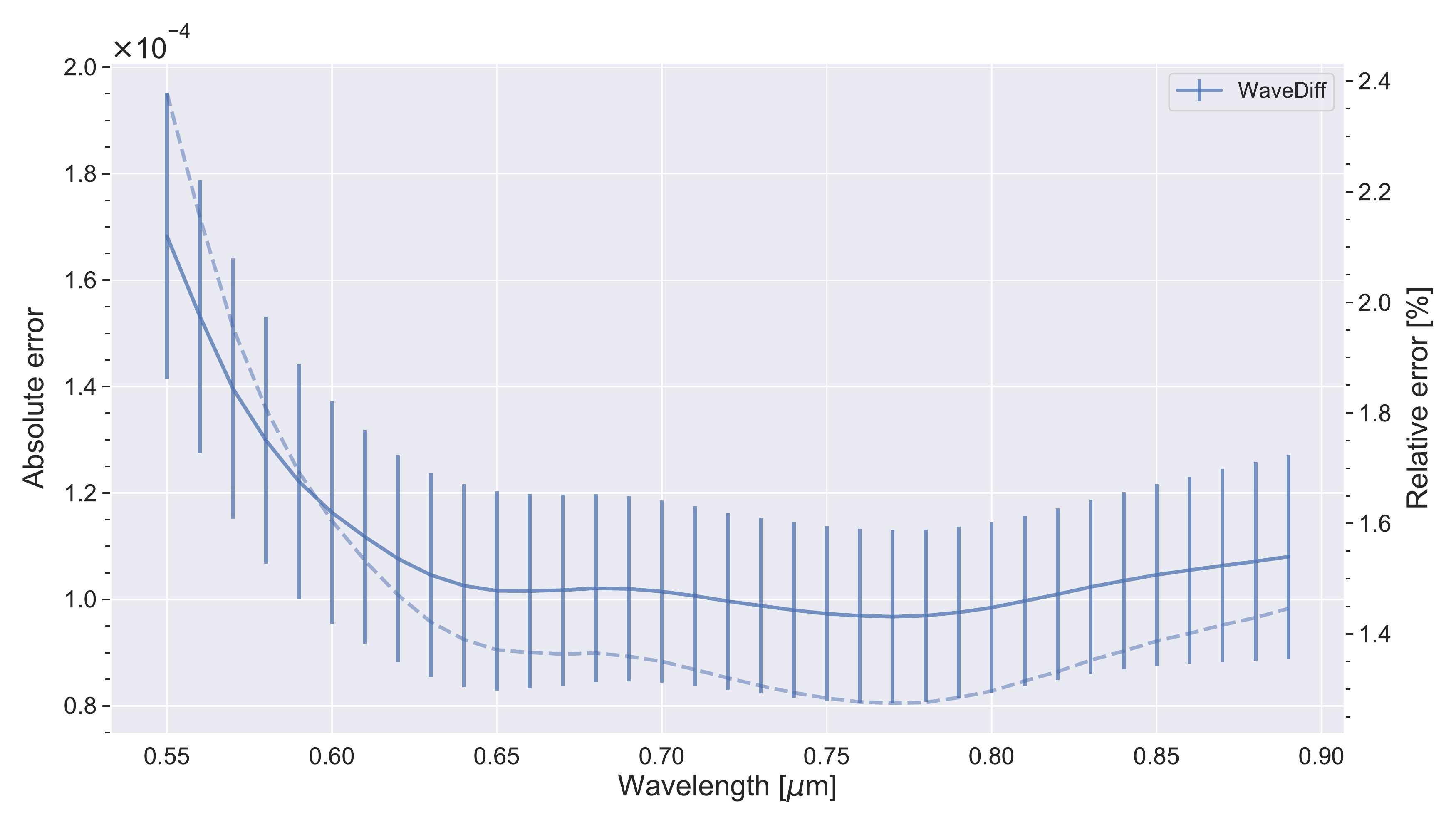}
    \caption{Test star reconstruction error as a function of wavelength for the WaveDiff model at three times the observation resolution.  
    \textbf{Left axis:} Absolute error, or $Err_{\text{abs}}$, and is plotted with a solid line. The error bars represent the variance over the RMS errors of the different test stars used for calculating $Err_{\text{abs}}$.
    \textbf{Right axis:} Relative error, or $Err_{\text{rel}}$, and is plotted with a dashed line.
    \nblink{WaveDiff-original-monochromatic-rmse}}
    \label{fi:result_b}
\end{figure}

\autoref{fi:result_b} presents the pixel reconstruction errors as a function of wavelength for the WaveDiff model.
One can see that the errors are kept low over all the instrument's passband. The model successfully captures the chromatic variations of the ground truth PSF field. The mean relative error over the passband is $1.46\%$. We highlight this result, as the data used to constrain the PSF model are noisy polychromatic PSFs. The loss function we use to optimize our model is built on the reconstruction of polychromatic PSFs. Estimating chromatic variations from polychromatic observations is a complicated ill-posed problem. Several sets of monochromatic PSFs can represent the same polychromatic PSF, as information is lost in the spectral integration of \autoref{eq:poly_star}. Ensuring that the monochromatic error is kept low is essential in validating the proposed PSF model. The error bars in \autoref{fi:result_b} represent the standard deviation of the reconstruction error within the set of test stars. The variance is kept considerably low with respect to $Err_{\text{abs}}$, the mean RMS error on the test set. A visual reconstruction example of a monochromatic PSF can be seen in the third row of \autoref{fi:visual_reconstruction_psf}.

\subsubsection{Results: (c), Errors as a function of the total number of training stars}
\label{sc:result_objective_c}

\begin{figure}
    \centering
    \includegraphics[width=\textwidth]{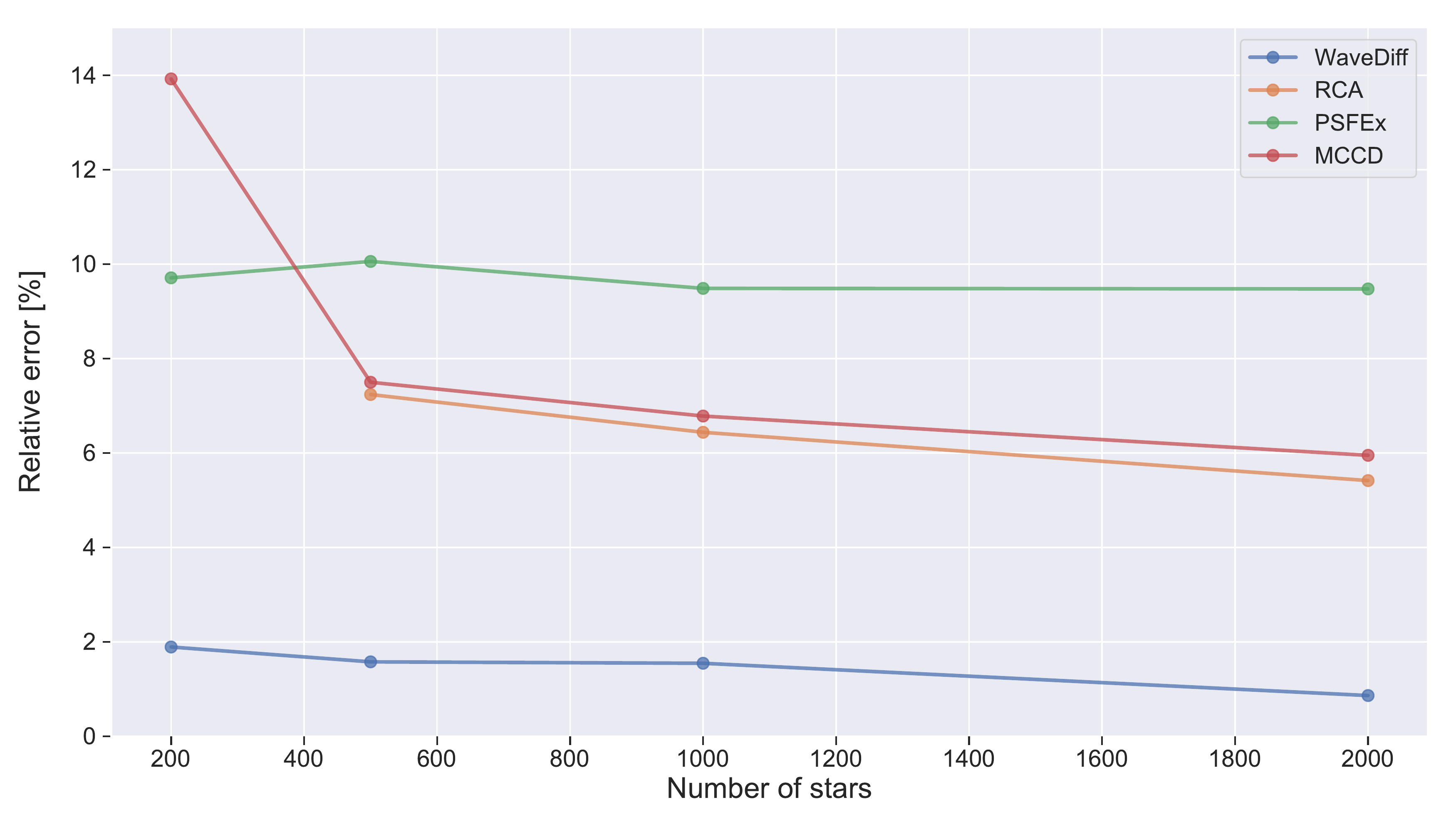}
    \caption{Polychromatic PSF relative reconstruction error, $Err_{\text{rel}}$, at observation resolution as a function of the total number of training stars in the FOV. The training data sets correspond, in ascending order, to $\mathcal{S}_1 \subset \mathcal{S}_2 \subset \mathcal{S}_3 \subset \mathcal{S}$, described in \autoref{sc:experiment_set_up}. The results are computed with respect to the reconstruction of test stars. Test stars are maintained throughout the models and training sets, allowing a truthful comparison. The result of \texttt{RCA} for the $\mathcal{S}_1$ ($200$ stars) data set is not shown as it is poor.
    \nblink{WaveDiff-original-rmse-vs-star-nb}}
    \label{fi:result_c_x1_res}
\end{figure}

The PSF modeling task becomes more arduous as we reduce the number of training stars used to constrain the PSF model, as less information is available to constrain the PSF models. \autoref{fi:result_c_x1_res} presents the pixel reconstruction relative errors, $Err_{\text{rel}}$, as a function of the number of training stars. WaveDiff maintains excellent performances even with the smallest training set with $10\%$ of the stars in $\mathcal{S}$. These results show the robustness of the proposed models with a low number of observed stars. Both models can handle a variable number of training stars with different SNRs. WaveDiff can incorporate the information present in the stars as the reconstruction errors monotonically decrease when we augment the number of training stars. 

\texttt{PSFEx} shows a relatively stable performance with the variation of the number of training stars. This result can be explained by the fact that it is the most constrained model. Consequently, it has difficulties incorporating more information into the model, but, in turn, it is more robust to a variation in the number of training stars. \texttt{RCA} exhibits somewhat similar behavior to \texttt{MCCD}, with the performance improving as the number of stars increases. \texttt{RCA} is not able to produce a result with the smallest data set $\mathcal{S}_{1}$. However, \texttt{MCCD} can still work with such a low number of training stars. The results from the largest data set, $\mathcal{S}$, in \autoref{fi:result_c_x1_res} coincide with the ones from the left column of \autoref{tb:results_a}. 

\begin{figure}
    \centering
    \includegraphics[width=\textwidth]{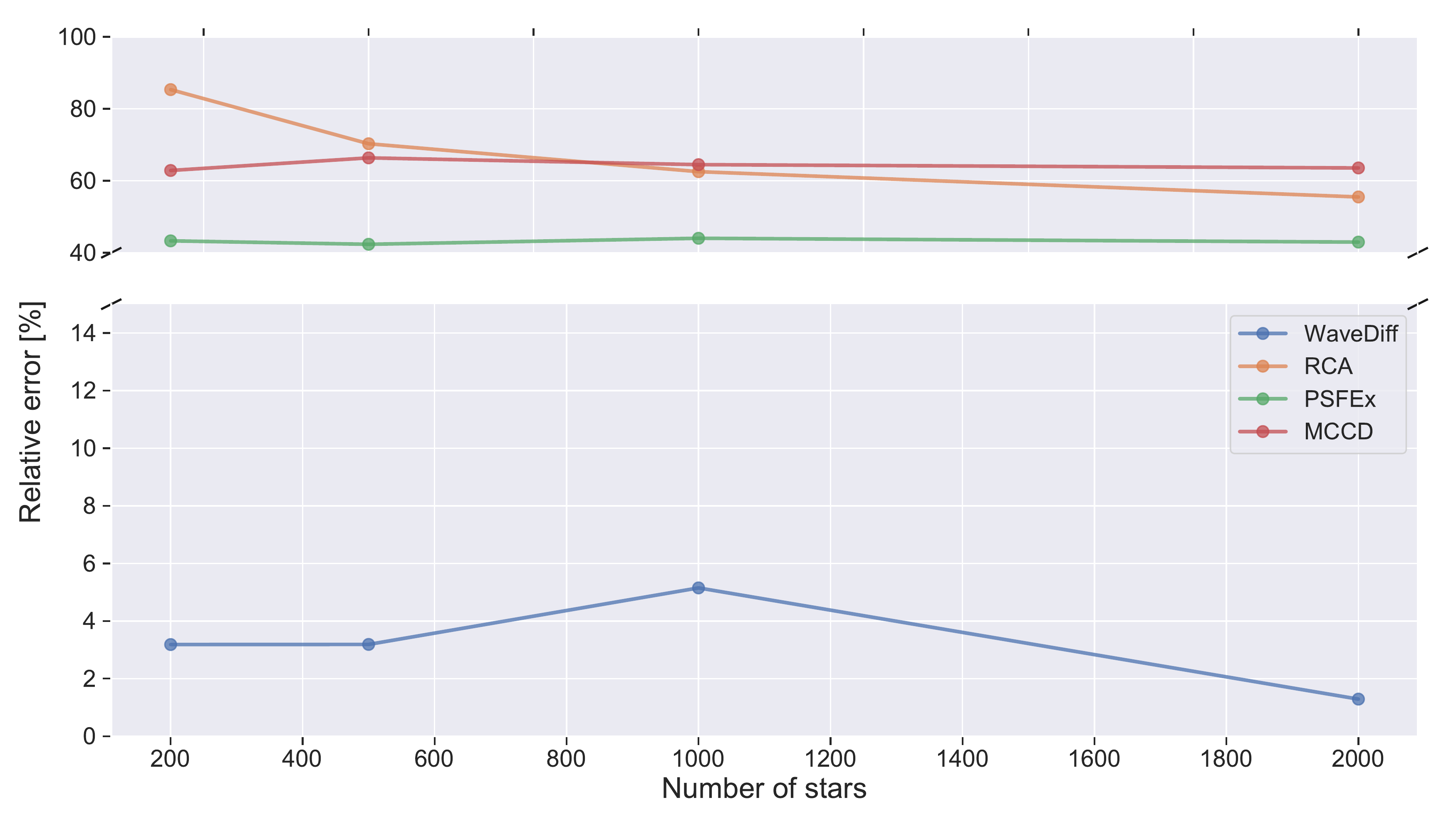}
    \caption{Polychromatic PSF relative reconstruction error, $Err_{\text{rel}}$, at three times (x$3$) the observation resolution as a function of the total number of training stars in the FOV. This experiment represents the super-resolution task.
    \nblink{WaveDiff-original-rmse-vs-star-nb}}
    \label{fi:result_c_x3_res}
\end{figure}

We repeat the previous experiment evaluating the reconstruction error at super-resolution, and we obtain the results shown in \autoref{fi:result_c_x3_res}. The performance in the $2000$-star training data set, $\mathcal{S}$, coincides with the right column of \autoref{tb:results_a}. 
We observe the previous trend, with a huge performance gap between WaveDiff, which uses an optical forward model, and the state-of-the-art pixel models. Considering the super-resolution metric, the errors of WaveDiff is not monotonically decreasing as we increase the number of stars. Nevertheless, the errors are consistently low, never over the relative $6\%$ error. This result shows that, for a variable number of training stars, the proposed models are satisfactorily generalizing to test positions and capturing the complexity of the high-resolution PSF field. Between the state-of-the-art pixel-based models, \texttt{PSFEx} is the best one performing in the super-resolution reconstruction task. In the reconstruction at the observation resolution \texttt{PSFEx} is underperforming with respect to \texttt{RCA} and \texttt{MCCD}. One explanation of this behavior is that in our experimental setup, we perform a super-resolution with a factor of $3$, which differs from the factor of $2$ used in \cite{ngole2016, schmitz2020}. Complex models tend to degenerate if they are not well-regularized in challenging super-resolution settings. However, simpler models like \texttt{PSFEx} tend to be more robust and give consistent results as their limited capacity regularizes them. This fact illustrates the importance of validating the reconstruction results at both resolutions.

\subsubsection{Results: (d), WFE reconstruction errors}
\label{sc:result_objective_d}

\autoref{tb:results_d_WFE} presents the performance results on the WFE recovery for the models based on WFE. It is interesting to observe that all the models obtain high errors in the WFE recovery. This result illustrates the difficulty of recovering the WFE from degraded in-focus observations. The ground truth WFE field is more complex than what the PSF models can represent, as mentioned in \autoref{sc:experiment_set_up}, so a zero WFE recovery error was impossible to achieve. 

Even though WaveDiff achieves an outstanding pixel reconstruction performance, they have significant errors in WFE recovery. These results indicate that the WFE manifold estimated by this model is far from the ground truth WFE manifold. However, the pixel PSF fields the different WFE manifolds represent are very close, which matters to our PSF model. The WFE manifold estimated by our model in these experiments is an auxiliary product allowing our models to incorporate physical information, capture chromatic variations, and achieve good super-resolution performance. \autoref{fi:visual_reconstruction_psf} presents an example of two similar PSFs being generated by two very different WFEs. The WFE recovery can be found in the fourth row.

\begin{table}
    \centering
    \caption{Target stars WFE recovery errors with the PSF models estimated with the $\mathcal{S}$ data set. \nblink{table-WFE-error-metrics}}
        \begin{tabular}{lc}
            \toprule
            PSF model                 & WFE $Err_{\text{abs}}$ [nm] ($Err_{\text{rel}}$) \\
            \midrule
            \textit{i)} Zernike 15             &  $106$ ($136\%$)       \\
            \textit{ii)} Zernike 40   &  $102$  ($130\%$)        \\
            \textbf{\textit{iii)} WaveDiff}    &  $\mathbf{101}$ ($\mathbf{129\%}$)        \\
            \bottomrule
        \end{tabular}
    \label{tb:results_d_WFE}
\end{table}

\subsubsection{Results: (e), Weak-lensing metrics}
\label{sc:result_objective_e}

\autoref{tb:results_e} presents the performance results of the different PSF models using weak-lensing metrics. The results obtained are coherent with the pixel errors in the super-resolution column; see \autoref{tb:results_a}. This result was expected as the weak-lensing metrics are computed on the polychromatic super-resolved reconstructed PSFs. The error reduction achieved by the WaveDiff model with respect to \texttt{RCA} constitutes a \textit{breakthrough in performance} for data-driven PSF modeling. The $e_1$ and $e_2$ errors are $20$ and $26$ times lower, and the size error, $R^{2}/\langle R^{2} \rangle $, is $276$ times lower. These results highlight the importance of the proposed model for current and future weak-lensing studies.

\begin{table}
    \centering
    \caption{Weak-lensing metrics of the different PSF models estimated with the $\mathcal{S}$ data set.
    \nblink{table-shape-size-error-metrics}}
        \begin{tabular}{lccc}
            \toprule
            &\multicolumn{3}{c}{RMSE}   \\
            \cmidrule(r){2-4}
            PSF model   & $e_1$[$\times 10^{-2}$] & $e_2$[$\times 10^{-2}$] & $R^{2}/\langle R^{2} \rangle $[$\times 10^{-1}$] \\
            \midrule
            \textit{i)} Zernike 15                   &  $3.92$    & $2.69$  & $0.92$    \\
            \textit{ii)} Zernike 40                  &  $2.12$    & $1.38$  & $0.97$    \\
            \textbf{\textit{iii)} WaveDiff} &  $\mathbf{0.23}$     & $\mathbf{0.16}$  & $\mathbf{0.13}$    \\
            \textit{iv)} \texttt{PSFEx}              &  $4.38$    & $4.25$  & $14.8$    \\
            \textit{v)} \texttt{RCA}               &  $4.61$    & $4.17$  & $36.0$   \\
            \textit{vi)} \texttt{MCCD}             &  $9.79$    & $7.32$  & $47.4$   \\
            \bottomrule
        \end{tabular}
    \label{tb:results_e}
    \end{table}

\section{Discussion}
\label{sc:discussions}

In this section, we discuss several scientific choices, the utility of the present work, possible improvements, and extensions. 

In \autoref{sc:training}, we did not introduce a regularizer term in the loss function. Consequently, the best-performing models, WaveDiff, does not include an explicit WFE regularization. We tried adding a $\nu \ell_2(\Phi_{\theta})$ regularization term to \autoref{eq:optim_problem} that constrained the total amount of aberration that is represented by the energy of the WFE. The physical motivation was to favor less aberrated models, as we saw that the WaveDiff model converged towards a WFE manifold with more aberration than the ground truth. This result is illustrated by the WFE reconstruction in the fourth row of \autoref{fi:visual_reconstruction_psf}. After tuning the $\nu$ hyperparameter accompanying the regularizer and running the experiments, we noticed that all the pixel performance results were degraded with respect to the original model. However, the WFE reconstruction error decreased considerably. Even though the estimated WFE manifold got closer to the ground truth WFE manifold of the PSF field, the pixel performance worsened. 
This experience showcases that regularizers in the WFE space should be used with care. We identify two regimes of the proposed WFE-based PSF modeling. The first regime aims to reconstruct the \textit{ground truth} WFE manifold, e.g., the case of parametric models. The second regime aims to reconstruct a \textit{useful} WFE manifold. In the latter, \textit{useful} refers to a WFE manifold with a pixel representation similar to the ground truth WFE manifold. The WaveDiff model we are proposing is an example of the second regime. 
The first regime makes sense if \textit{a priori} information about the ground truth WFE manifold is available and could be used to build the regularizer. This situation could be the case if complementary measurements of the optical system are available. The second regime should be taken with more care as the WFE regularizer can bias the estimated useful WFE manifold. This bias represents a more significant deviation in pixel space even if both manifolds become closer in the WFE space.

The proposed data-driven PSF model presents advantages with respect to parametric PSF models. We observe the advantages when comparing the Zernike $15$ and Zernike $40$ models with the WaveDiff model. However, it does not mean that parametric and data-driven models cannot be complementary. The WFE output of the parametric model can be used as input in the proposed framework. The model would shift from the first to the second regime mentioned above, trying to reconstruct the ground-truth WFE rather than a useful representation. The parametric model output could be introduced to the data-driven model using a WFE regularizer. The task of the data-driven optimization would be to estimate the difference between the ground truth WFE and the WFE estimated by the parametric model. 

%

%
Even though the simulations used in this work are representative of a space telescope, there are several simplifications we have assumed. For example, we have considered that all the CCD chips are perfectly aligned in the focal plane and have not included any detector-level effects. These and other effects could be included in the proposed framework. Detector-level effects, such as guiding errors or charge transfer inefficiency \cite{rhodes2010}, could be included in the optical forward model. The impact of the aforementioned effects should be considered by the model before handling real observations. An interesting study, but out of scope for this article, is to evaluate how errors in the input SED data propagate to the estimated PSF model.

%

%
New flavours, i.e., versions, of the WaveDiff PSF model can be developed by taking advantage of our framework and redefining the function $\Phi_{\theta}$ in \autoref{fi:psf_model_diagram} or adding a regularizer term $\mathcal{R}$ to our optimization objective in \autoref{eq:optim_problem}. Previous works that tried to leverage the power of modern neural networks into PSF models were blind to the physics of the inverse problem and built the model in pixel space \cite{jia2020a,jia2020b,jia2020c}. Our framework uses an end-to-end differentiable optical forward model, runs on GPU, and is entirely coded in TensorFlow \cite{tensorflow2015}. This construction makes it a promising candidate to develop \textit{physically-motivated} neural network approaches to boost the PSF model.

\section{Conclusion}
\label{sc:conclusions}

This work introduces a novel data-driven modeling method of a point spread function field coined WaveDiff, which represents a paradigm shift with respect to current state-of-the-art data-driven methods. We propose to include a differentiable physics-based forward model that allows changing the modeling space from pixels to the wavefront. This change transfers most of the modeling complexity into the forward model and simplifies the building of the point spread function model. We exploit the automatic differentiation capabilities of the model and use modern optimization techniques to estimate the model parameters in the wavefront space. The WaveDiff model is built using the TensorFlow framework and runs on GPU.
We have verified that our model can learn from in-focus observations a \textit{useful} wavefront-based manifold representation of the PSF field. We have not used special calibration data or wavefront information. The WaveDiff model is only constrained from noisy in-focus polychromatic under-sampled observations using the spectral energy distribution of each observed object. The proposed WaveDiff model can super-resolve the PSF field and is, to the best of our knowledge, the \textit{first data-driven PSF model} to effectively capture the wavelength dependence of the point spread function.

The proposed method was tested using a set of simulations of a space telescope which included a broad passband, different spectral energy distribution for the observed objects, variable signal-to-noise ratios, and the undersampling of the observations. We compared our WaveDiff model against three state-of-the-art data-driven point spread function models, including \texttt{PSFEx}, \texttt{MCCD}, and \texttt{RCA}, the current data-driven approach for the \textit{Euclid} satellite. The WaveDiff model represents a \textit{breakthrough in performance} obtaining $6$ and $44$ \textit{times} lower reconstruction error than \texttt{RCA} at observation resolution and super-resolution (x$3$), respectively. Our proposed model obtains $0.86\%$ and $1.3\%$ relative reconstruction error at x$1$ and x$3$ the observation resolution, respectively. We validated that the WaveDiff model captures the chromatic variations of the point spread function with the monochromatic reconstruction error. The WaveDiff model achieves a mean relative error of $1.46\%$ over the broad instrument's passband. We have further tested the WaveDiff model varying the number of observed stars used to constrain the model. Our model maintains good performance even when using the $10\%$ of the original observed stars.

Using simulations to validate the model allows accessing and comparing the estimated and the ground truth wavefronts. We have found that the wavefront manifold estimated by the WaveDiff model and the ground truth are far from each other. However, interestingly, both wavefront manifolds represent very similar pixel representations, as was shown in the pixel errors above. These results support using the wavefront space as an auxiliary space to represent the pixel point spread function field. It has not necessarily to be the objective to find the ground truth wavefront manifold, but a manifold that is \textit{useful} for our objectives. These are defined in pixel space rather than in the wavefront space.

Finally, we have compared the different point spread function models using weak-lensing metrics such as ellipticity and size. Compared to the \texttt{RCA} model, the WaveDiff model reduces ellipticity errors by a factor of $20$ and $26$ for each component, respectively. In addition, the size error obtained is $276$ times lower than the one obtained by \texttt{RCA}. These results highlight the utility and confirm the breakthrough in performance achieved by the proposed WaveDiff model.

\bigskip

In the spirit of reproducible and reusable research, all software, scripts, and analysis notebooks are publicly available at \url{\gitrepo}.

\section*{Acknowledgments}
This work was granted access to the HPC resources of IDRIS under the allocations 2021-AD011011554 and 2021-AD011012983 made by GENCI. This work has made use of the CANDIDE Cluster at the Institut d'Astrophysique de Paris and made possible by grants from the PNCG and the DIM-ACAV.
\textit{Software used in this work: Astropy \cite{Robitaille2013, PriceWhelan2018}, IPython \cite{Perez2007}, GalSim \cite{rowe2015}, Jupyter \cite{Kluyver2016}, Matplotlib \cite{Hunter2007}, Seaborn \cite{seaborn2021}, Numpy \cite{numpy2020}, TensorFlow \cite{tensorflow2015}.}

\section*{References}
\bibliography{PSF_prop}

\providecommand{\newblock}{}
\begin{thebibliography}{10}
\expandafter\ifx\csname url\endcsname\relax
  \def\url#1{{\tt #1}}\fi
\expandafter\ifx\csname urlprefix\endcsname\relax\def\urlprefix{URL }\fi
\providecommand{\eprint}[2][]{\url{#2}}

\bibitem{kilbinger2015}
Kilbinger M 2015 {\em Reports on Progress in Physics\/} {\bf 78} 086901

\bibitem{mandelbaum2018_bis3}
Mandelbaum R 2018 {\em Annual Review of Astronomy and Astrophysics\/} {\bf 56}
  393--433 ISSN 00664146 (\textit{Preprint} \eprint{1710.03235})

\bibitem{laureijs2011}
Laureijs R, Amiaux J, Arduini S, Augueres J~L, Brinchmann J, Cole R, Cropper M,
  Dabin C, Duvet L, Ealet A {\em et~al.\/} 2011 {\em ArXiv e-prints\/}
  (\textit{Preprint} \eprint{1110.3193})

\bibitem{wfirst}
Spergel D, Gehrels N, Baltay C, Bennett D, Breckinridge J, Donahue M, Dressler
  A, Gaudi B~S, Greene T, Guyon O, Hirata C, Kalirai J, Kasdin N~J, Macintosh
  B, Moos W, Perlmutter S, Postman M, Rauscher B, Rhodes J, Wang Y, Weinberg D,
  Benford D, Hudson M, Jeong W~S, Mellier Y, Traub W, Yamada T, Capak P,
  Colbert J, Masters D, Penny M, Savransky D, Stern D, Zimmerman N, Barry R,
  Bartusek L, Carpenter K, Cheng E, Content D, Dekens F, Demers R, Grady K,
  Jackson C, Kuan G, Kruk J, Melton M, Nemati B, Parvin B, Poberezhskiy I,
  Peddie C, Ruffa J, Wallace J~K, Whipple A, Wollack E and Zhao F 2015
  Wide-field infrarred survey telescope-astrophysics focused telescope assets
  wfirst-afta 2015 report (\textit{Preprint} \eprint{1503.03757})

\bibitem{LSST2009}
{LSST Science Collaboration}, {Abell} P~A, {Allison} J, {Anderson} S~F,
  {Andrew} J~R, {Angel} J~R~P, {Armus} L, {Arnett} D, {Asztalos} S~J, {Axelrod}
  T~S and et~al 2009 {\em arXiv e-prints\/} (\textit{Preprint}
  \eprint{0912.0201})

\bibitem{moffat1988}
{Bendinelli} O, {Parmeggiani} G and {Zavatti} F 1988 {\em Journal of
  Astrophysics and Astronomy\/} {\bf 9} 17--24

\bibitem{piotrowski2013}
{Piotrowski, L W}, {Batsch, T}, {Czyrkowski, H}, {Cwiok, M}, {Dabrowski, R},
  {Kasprowicz, G}, {Majcher, A}, {Majczyna, A}, {Malek, K}, {Mankiewicz, L},
  {Nawrocki, K}, {Opiela, R}, {Siudek, M}, {Sokolowski, M}, {Wawrzaszek, R},
  {Wrochna, G}, {Zaremba, M} and {Zarnecki, A F} 2013 {\em A\&A\/} {\bf 551}
  A119 \urlprefix\url{https://doi.org/10.1051/0004-6361/201219230}

\bibitem{bertin2011}
{Bertin} E 2011 {Automated Morphometry with SExtractor and PSFEx} {\em
  Astronomical Data Analysis Software and Systems XX\/} ({\em Astronomical
  Society of the Pacific Conference Series\/} vol 442) ed {Evans} I~N,
  {Accomazzi} A, {Mink} D~J and {Rots} A~H p 435

\bibitem{miller2013}
Miller L, Heymans C, Kitching T, Van~Waerbeke L, Erben T, Hildebrandt H,
  Hoekstra H, Mellier Y, Rowe B, Coupon J {\em et~al.\/} 2013 {\em MNRAS\/}
  {\bf 429} 2858--2880

\bibitem{jarvis2020}
Jarvis M, Bernstein G~M, Amon A, Davis C, L{\'{e}}get P~F, Bechtol K, Harrison
  I, Gatti M, Roodman A, Chang C, Chen R, Choi A, Desai S, Drlica-Wagner A,
  Gruen D, Gruendl R~A, Hernandez A, MacCrann N, Meyers J, Navarro-Alsina A,
  Pandey S, Plazas A~A, Secco L~F, Sheldon E, Troxel M~A, Vorperian S, Wei K,
  Zuntz J, Abbott T~M~C, Aguena M, Allam S, Avila S, Bhargava S, Bridle S~L,
  Brooks D, {Carnero Rosell} A, {Carrasco Kind} M, Carretero J, Costanzi M,
  da~Costa L~N, {De Vicente} J, Diehl H~T, Doel P, Everett S, Flaugher B,
  Fosalba P, Frieman J, Garc{\'{i}}a-Bellido J, Gaztanaga E, Gerdes D~W,
  Gutierrez G, Hinton S~R, Hollowood D~L, Honscheid K, James D~J, Kent S, Kuehn
  K, Kuropatkin N, Lahav O, Maia M~A~G, March M, Marshall J~L, Melchior P,
  Menanteau F, Miquel R, Ogando R~L~C, Paz-Chinch{\'{o}}n F, Rykoff E~S,
  Sanchez E, Scarpine V, Schubnell M, Serrano S, Sevilla-Noarbe I, Smith M,
  Suchyta E, Swanson M~E~C, Tarle G, Varga T~N, Walker A~R, Wester W and
  Wilkinson R~D 2020 {\em Monthly Notices of the Royal Astronomical Society\/}
  {\bf 501} 1282--1299 ISSN 0035-8711 (\textit{Preprint} \eprint{2011.03409})
  \urlprefix\url{https://academic.oup.com/mnras/article/501/1/1282/6006887}

\bibitem{jee2007}
Jee M~J, Blakeslee J~P, Sirianni M, Martel A~R, White R~L and Ford H~C 2007
  {\em Publications of the Astronomical Society of the Pacific\/} {\bf 119}
  1403--1419 \urlprefix\url{https://doi.org/10.1086/524849}

\bibitem{schrabback2010}
{Schrabback, T}, {Hartlap, J}, {Joachimi, B}, {Kilbinger, M}, {Simon, P},
  {Benabed, K}, {Bradac, M}, {Eifler, T}, {Erben, T}, {Fassnacht, C D}, {High,
  F William}, {Hilbert, S}, {Hildebrandt, H}, {Hoekstra, H}, {Kuijken, K},
  {Marshall, P J}, {Mellier, Y}, {Morganson, E}, {Schneider, P}, {Semboloni,
  E}, {Van Waerbeke, L} and {Velander, M} 2010 {\em A\&A\/} {\bf 516} A63
  \urlprefix\url{https://doi.org/10.1051/0004-6361/200913577}

\bibitem{gentile2013}
Gentile M, Courbin F and Meylan G 2013 {\em A\&A\/} {\bf 549} A1

\bibitem{ngole2015}
Ngol{\`e} F, Starck J~L, Ronayette S, Okumura K and Amiaux J 2015 {\em A\&A\/}
  {\bf 575} A86

\bibitem{ngole2016}
Ngol{\`e} F, Starck J~L, Okumura K, Amiaux J and Hudelot P 2016 {\em Inverse
  Problems\/} {\bf 32} 124001

\bibitem{schmitz2020}
Schmitz M~A, Starck J~L, Mboula F~N, Auricchio N, Brinchmann J, Capobianco
  R~I~V, Cl{\'e}dassou R, Conversi L, Corcione L, Fourmanoit N, Frailis M,
  Garilli B, Hormuth F, Hu D, Israel H, Kermiche S, Kitching T~D, Kubik B, Kunz
  M, Ligori S, Lilje P~B, Lloro I, Mansutti O, Marggraf O and Massey R~J 2020
  {\em A\&A\/} {\bf 636} A78 (\textit{Preprint} \eprint{1906.07676})
  \urlprefix\url{https://arxiv.org/pdf/1906.07676.pdf}

\bibitem{refregier2003a}
Refregier A 2003 {\em Monthly Notices of the Royal Astronomical Society\/} {\bf
  338} 35--47 ISSN 0035-8711 (\textit{Preprint}
  \eprint{https://academic.oup.com/mnras/article-pdf/338/1/35/3832113/338-1-35.pdf})
  \urlprefix\url{https://doi.org/10.1046/j.1365-8711.2003.05901.x}

\bibitem{massey2005}
Massey R and Refregier A 2005 {\em Monthly Notices of the Royal Astronomical
  Society\/} {\bf 363} 197--210 ISSN 0035-8711 (\textit{Preprint}
  \eprint{https://academic.oup.com/mnras/article-pdf/363/1/197/4125992/363-1-197.pdf})
  \urlprefix\url{https://doi.org/10.1111/j.1365-2966.2005.09453.x}

\bibitem{ngole2017}
Ngol{\`e} F and Starck J~L 2017 {\em SIAM Journal on Imaging Sciences\/} {\bf
  10} 1549--1578

\bibitem{herbel2018}
Herbel J, Kacprzak T, Amara A, Refregier A and Lucchi A 2018 {\em Journal of
  Cosmology and Astroparticle Physics\/} {\bf 2018} 054--054
  \urlprefix\url{https://doi.org/10.1088/1475-7516/2018/07/054}

\bibitem{jia2020a}
Jia P, Wu X, Li Z, Li B, Wang W, Liu Q and Popowicz A 2020 Modelling the point
  spread function of wide field small aperture telescopes with deep neural
  networks -- applications in point spread function estimation
  (\textit{Preprint} \eprint{2011.10243})

\bibitem{jia2020b}
Jia P, Li X, Li Z, Wang W and Cai D 2020 {\em Monthly Notices of the Royal
  Astronomical Society\/} {\bf 493} 651--660 ISSN 0035-8711 (\textit{Preprint}
  \eprint{https://academic.oup.com/mnras/article-pdf/493/1/651/32513220/staa319.pdf})
  \urlprefix\url{https://doi.org/10.1093/mnras/staa319}

\bibitem{jia2020c}
{Jia} P, {Wu} X, {Yi} H, {Cai} B and {Cai} D 2020 {\em AJ\/} {\bf 159} 183
  (\textit{Preprint} \eprint{2003.00615})

\bibitem{krist1993}
{Krist} J 1993 {Tiny Tim : an HST PSF Simulator} {\em Astronomical Data
  Analysis Software and Systems II\/} ({\em Astronomical Society of the Pacific
  Conference Series\/} vol~52) ed {Hanisch} R~J, {Brissenden} R~J~V and
  {Barnes} J p 536

\bibitem{krist1995}
{Krist} J 1995 {Simulation of HST PSFs using Tiny Tim} {\em Astronomical Data
  Analysis Software and Systems IV\/} ({\em Astronomical Society of the Pacific
  Conference Series\/} vol~77) ed {Shaw} R~A, {Payne} H~E and {Hayes} J~J~E p
  349

\bibitem{krist2011}
Krist J~E, Hook R~N and Stoehr F 2011 {20 years of Hubble Space Telescope
  optical modeling using Tiny Tim} {\em Optical Modeling and Performance
  Predictions V\/} vol 8127 ed Kahan M~A International Society for Optics and
  Photonics (SPIE) pp 166 -- 181
  \urlprefix\url{https://doi.org/10.1117/12.892762}

\bibitem{cropper2016}
Cropper M, Pottinger S, Niemi S, Azzollini R, Denniston J, Szafraniec M, Awan
  S, Mellier Y, Berthe M, Martignac J, Cara C, Giorgio A~M~D, Sciortino A,
  Bozzo E, Genolet L, Cole R, Philippon A, Hailey M, Hunt T, Swindells I,
  Holland A, Gow J, Murray N, Hall D, Skottfelt J, Amiaux J, Laureijs R, Racca
  G, Salvignol J~C, Short A, Alvarez J~L, Kitching T, Hoekstra H, Massey R and
  Israel H 2016 {VIS: the visible imager for Euclid} {\em Space Telescopes and
  Instrumentation 2016: Optical, Infrared, and Millimeter Wave\/} vol 9904 ed
  MacEwen H~A, Fazio G~G, Lystrup M, Batalha N, Siegler N and Tong E~C
  International Society for Optics and Photonics (SPIE) pp 269 -- 284
  \urlprefix\url{https://doi.org/10.1117/12.2234739}

\bibitem{krist1995b}
Krist J~E and Burrows C~J 1995 {\em Appl. Opt.\/} {\bf 34} 4951--4964
  \urlprefix\url{http://ao.osa.org/abstract.cfm?URI=ao-34-22-4951}

\bibitem{HST_PSF2017}
{Hoffmann} S~L and {Anderson} J 2017 {A Study of PSF Models for ACS/WFC}
  Instrument Science Report ACS 2017-8

\bibitem{liaudat2020}
Liaudat T, Bonnin J, Starck J~L, Schmitz M~A, Guinot A, Kilbinger M and Gwyn
  S~D~J 2021 {\em A\&A\/} {\bf 646} A27
  \urlprefix\url{https://doi.org/10.1051/0004-6361/202039584}

\bibitem{noll1976}
Noll R~J 1976 {\em J. Opt. Soc. Am.\/} {\bf 66} 207--211
  \urlprefix\url{http://www.osapublishing.org/abstract.cfm?URI=josa-66-3-207}

\bibitem{tensorflow2015}
Abadi M, Agarwal A, Barham P, Brevdo E, Chen Z, Citro C, Corrado G~S, Davis A,
  Dean J, Devin M, Ghemawat S, Goodfellow I, Harp A, Irving G, Isard M, Jia Y,
  Jozefowicz R, Kaiser L, Kudlur M, Levenberg J, Man\'{e} D, Monga R, Moore S,
  Murray D, Olah C, Schuster M, Shlens J, Steiner B, Sutskever I, Talwar K,
  Tucker P, Vanhoucke V, Vasudevan V, Vi\'{e}gas F, Vinyals O, Warden P,
  Wattenberg M, Wicke M, Yu Y and Zheng X 2015 {TensorFlow}: Large-scale
  machine learning on heterogeneous systems software available from
  tensorflow.org \urlprefix\url{https://www.tensorflow.org/}

\bibitem{liaudat2021}
Liaudat T, Starck J~L, Kilbinger M and Frugier P~A 2021 {Rethinking the
  modeling of the instrumental response of telescopes with a differentiable
  optical model} {\em NeurIPS 2021 Machine Learning for Physical sciences
  workshop\/} (\textit{Preprint} \eprint{2111.12541})
  \urlprefix\url{http://arxiv.org/abs/2111.12541}

\bibitem{liaudat2021_b}
Liaudat T, Starck J~L and Kilbinger M 2021 {Semi-Parametric Wavefront Modelling
  for the Point Spread Function} {\em {52{\`e}me Journ{\'e}es de Statistique de
  la Soci{\'e}t{\'e} Fran{\c c}aise de Statistique (SFdS)}\/} (Nice, France)
  \urlprefix\url{https://hal.archives-ouvertes.fr/hal-03444576}

\bibitem{li2016}
Li B~S, Li G~L, Cheng J, Peterson J and Cui W 2016 {\em Research in Astronomy
  and Astrophysics\/} {\bf 16} 007 ISSN 1674-4527 (\textit{Preprint}
  \eprint{1604.07126})
  \urlprefix\url{https://iopscience.iop.org/article/10.1088/1674-4527/16/9/139}

\bibitem{starck2011}
Starck J~L, Murtagh F and Bertero M 2011 Starlet transform in astronomical data
  processing {\em Handbook of Mathematical Methods in Imaging\/} (Springer) pp
  1489--1531

\bibitem{goodman2005}
Goodman J~W 2005 {\em Introduction to Fourier optics, 3rd ed., by JW Goodman.
  Englewood, CO: Roberts \& Co. Publishers, 2005\/} {\bf 1}

\bibitem{schmitz2018}
Schmitz M~A, Heitz M, Bonneel N, Ngole F, Coeurjolly D, Cuturi M, Peyr{\'e} G
  and Starck J~L 2018 {\em SIAM Journal on Imaging Sciences\/} {\bf 11}
  643--678

\bibitem{schmitz2019}
Schmitz M~A 2019 {\em {Euclid weak lensing : PSF field estimation}\/} Theses
  {Universit{\'e} Paris Saclay (COmUE)}
  \urlprefix\url{https://tel.archives-ouvertes.fr/tel-02462793}

\bibitem{mccann1997}
McCann R~J 1997 {\em Advances in Mathematics\/} {\bf 128} 153--179 ISSN
  0001-8708
  \urlprefix\url{https://www.sciencedirect.com/science/article/pii/S0001870897916340}

\bibitem{agueh2011}
Agueh M and Carlier G 2011 {\em SIAM Journal on Mathematical Analysis\/} {\bf
  43} 904--924 (\textit{Preprint} \eprint{https://doi.org/10.1137/100805741})
  \urlprefix\url{https://doi.org/10.1137/100805741}

\bibitem{soulez2016}
Soulez F, Courbin F and Unser M 2016 {\em Proc. SPIE 9912, Advances in Optical
  and Mechanical Technologies for Telescopes and Instrumentation II, 991277
  (July 22, 2016)\/} (\textit{Preprint} \eprint{1608.01816})
  \urlprefix\url{https://arxiv.org/pdf/1608.01816.pdf}

\bibitem{Wang2020}
Wang F, Bian Y, Wang H, Lyu M, Pedrini G, Osten W, Barbastathis G and Situ G
  2020 {\em Light: Science and Applications\/} {\bf 9} ISSN 20477538
  \urlprefix\url{http://dx.doi.org/10.1038/s41377-020-0302-3}

\bibitem{Wong2021}
Wong A, Pope B, Desdoigts L, Tuthill P, Norris B and Betters C 2021 {\em
  Journal of the Optical Society of America B\/} {\bf 38} 2465 ISSN 0740-3224
  (\textit{Preprint} \eprint{arXiv:2107.00952v1})

\bibitem{Jurling2014}
Jurling A~S and Fienup J~R 2014 {\em Journal of the Optical Society of America
  A\/} {\bf 31} 1348 ISSN 1084-7529

\bibitem{schechtman2015}
Shechtman Y, Eldar Y~C, Cohen O, Chapman H~N, Miao J and Segev M 2015 {\em IEEE
  Signal Processing Magazine\/} {\bf 32} 87--109

\bibitem{baydin2017}
Baydin A~G, Pearlmutter B~A, Radul A~A and Siskind J~M 2017 {\em J. Mach.
  Learn. Res.\/} {\bf 18} 5595--5637 ISSN 1532-4435

\bibitem{tuthill2018}
Tuthill P, Bendek E, Guyon O, Horton A, Jeffries B, Jovanovic N, Klupar P,
  Larkin K, Norris B, Pope B and Shao M 2018 {The TOLIMAN space telescope} {\em
  Optical and Infrared Interferometry and Imaging VI\/} vol 10701 ed
  Creech-Eakman M~J, Tuthill P~G and M{\'e}rand A International Society for
  Optics and Photonics (SPIE) p 107011J
  \urlprefix\url{https://doi.org/10.1117/12.2313269}

\bibitem{zuntz2018}
Zuntz J, Sheldon E, Samuroff S, Troxel M~A, Jarvis M, MacCrann N, Gruen D, Prat
  J, S{\'a}nchez C, Choi A {\em et~al.\/} 2018 {\em MNRAS\/} {\bf 481}
  1149--1182

\bibitem{bertin1996}
{Bertin, E} and {Arnouts, S} 1996 {\em Astron. Astrophys. Suppl. Ser.\/} {\bf
  117} 393--404 \urlprefix\url{https://doi.org/10.1051/aas:1996164}

\bibitem{beck2009}
Beck A and Teboulle M 2009 {\em SIAM J. Img. Sci.\/} {\bf 2} 183--202
  \urlprefix\url{https://doi.org/10.1137/080716542}

\bibitem{condat2013}
Condat L 2013 {\em Journal of Optimization Theory and Applications\/} {\bf 158}
  460--479

\bibitem{starck2015}
Starck J~L, Murtagh F and Fadili J 2015 {\em Sparse Image and Signal
  Processing: Wavelets and Related Geometric Multiscale Analysis\/} 2nd ed
  (Cambridge University Press)

\bibitem{racca2016}
Racca G~D, Laureijs R, Stagnaro L, Salvignol J~C, Alvarez J~L, Criado G~S,
  Venancio L~G, Short A, Strada P, B{\"o}nke T, Colombo C, Calvi A, Maiorano E,
  Piersanti O, Prezelus S, Rosato P, Pinel J, Rozemeijer H, Lesna V, Musi P,
  Sias M, Anselmi A, Cazaubiel V, Vaillon L, Mellier Y, Amiaux J, Berth{\'e} M,
  Sauvage M, Azzollini R, Cropper M, Pottinger S, Jahnke K, Ealet A, Maciaszek
  T, Pasian F, Zacchei A, Scaramella R, Hoar J, Kohley R, Vavrek R, Rudolph A
  and Schmidt M 2016 {The Euclid mission design} {\em Space Telescopes and
  Instrumentation 2016: Optical, Infrared, and Millimeter Wave\/} vol 9904 ed
  MacEwen H~A, Fazio G~G, Lystrup M, Batalha N, Siegler N and Tong E~C
  International Society for Optics and Photonics (SPIE) pp 235 -- 257
  \urlprefix\url{https://doi.org/10.1117/12.2230762}

\bibitem{fourier1978}
Bracewell R 1978 {\em The Fourier Transform and its Applications\/} 2nd ed
  (Tokyo: McGraw-Hill Kogakusha, Ltd.)

\bibitem{baron2022}
Baron M, Sassolas B, Frugier P~A, Venancio L~M~G, Amiaux J, Castelnau M, Keller
  F, Dovillaire G, Treimany P, Juv{\'e}nal R, Miller L, Pinard L and Ealet A
  2022 {Measurement and modelling of the chromatic dependence of a reflected
  wavefront on the Euclid space telescope dichroic mirror} {\em Space
  Telescopes and Instrumentation 2022: Optical, Infrared, and Millimeter
  Wave\/} vol 12180 ed Coyle L~E, Matsuura S and Perrin M~D International
  Society for Optics and Photonics (SPIE) p 121804V
  \urlprefix\url{https://doi.org/10.1117/12.2630072}

\bibitem{schmidt2010}
Schmidt J~D 2010 {\em {Numerical Simulation of Optical Wave Propagation with
  Examples in MATLAB}\/} (SPIE) ISBN 9780819483270
  \urlprefix\url{https://spiedigitallibrary.org/ebooks/PM/Numerical-Simulation-of-Optical-Wave-Propagation-with-Examples-in-MATLAB/eISBN-9780819483270/10.1117/3.866274}

\bibitem{parikh2014}
Parikh N and Boyd S 2014 {\em Found. Trends Optim.\/} {\bf 1} 127--239 ISSN
  2167-3888 \urlprefix\url{https://doi.org/10.1561/2400000003}

\bibitem{xu2013}
Xu Y and Yin W 2013 {\em SIAM Journal on Imaging Sciences\/} {\bf 6} 1758--1789
  ISSN 1936-4954

\bibitem{liu2020}
Liu L, Jiang H, He P, Chen W, Liu X, Gao J and Han J 2019 {On the Variance of
  the Adaptive Learning Rate and Beyond} (\textit{Preprint}
  \eprint{1908.03265}) \urlprefix\url{http://arxiv.org/abs/1908.03265}

\bibitem{kingma2014}
Kingma D~P and Ba J 2014  (\textit{Preprint} \eprint{1412.6980})

\bibitem{massey2012}
Massey R, Hoekstra H, Kitching T, Rhodes J, Cropper M, Amiaux J, Harvey D,
  Mellier Y, Meneghetti M, Miller L {\em et~al.\/} 2012 {\em MNRAS\/} {\bf 429}
  661--678

\bibitem{cropper2013}
Cropper M, Hoekstra H, Kitching T, Massey R, Amiaux J, Miller L, Mellier Y,
  Rhodes J, Rowe B, Pires S {\em et~al.\/} 2013 {\em MNRAS\/} {\bf 431}
  3103--3126

\bibitem{hirata2003}
Hirata C and Seljak U 2003 {\em MNRAS\/} {\bf 343} 459--480

\bibitem{mandelbaum2005}
Mandelbaum R, Hirata C~M, Seljak U, Guzik J, Padmanabhan N, Blake C, Blanton
  M~R, Lupton R and Brinkmann J 2005 {\em MNRAS\/} {\bf 361} 1287--1322

\bibitem{venancio2020}
Venancio L~M~G, Carminati L, Amiaux J, Bonino L, Racca G, Vavrek R, Laureijs R,
  Short A, Boenke T and Strada P 2020 {Status of the performance of the Euclid
  spacecraft} {\em Space Telescopes and Instrumentation 2020: Optical,
  Infrared, and Millimeter Wave\/} vol 11443 ed Lystrup M, Perrin M~D, Batalha
  N, Siegler N and Tong E~C International Society for Optics and Photonics
  (SPIE) pp 45 -- 60 \urlprefix\url{https://doi.org/10.1117/12.2562490}

\bibitem{pickles1998}
Pickles A~J 1998 {\em Publications of the Astronomical Society of the
  Pacific\/} {\bf 110} 863--878 \urlprefix\url{https://doi.org/10.1086/316197}

\bibitem{kuntzer2016}
Kuntzer T, Tewes M and Courbin F 2016 {\em A\&A\/} {\bf 591} A54

\bibitem{rhodes2010}
Rhodes J, Leauthaud A, Stoughton C, Massey R, Dawson K, Kolbe W and Roe N 2010
  {\em Publications of the Astronomical Society of the Pacific\/} {\bf 122}
  439--450 \urlprefix\url{https://doi.org/10.1086/651675}

\bibitem{Robitaille2013}
Robitaille T~P, Tollerud E~J, Greenfield P, Droettboom M, Bray E, Aldcroft T,
  Davis M, Ginsburg A, Price-Whelan A~M, Kerzendorf W~E, Conley A, Crighton N,
  Barbary K, Muna D, Ferguson H, Grollier F, Parikh M~M, Nair P~H,
  G{\"{u}}nther H~M, Deil C, Woillez J, Conseil S, Kramer R, Turner J~E~H,
  Singer L, Fox R, Weaver B~A, Zabalza V, Edwards Z~I, Azalee~Bostroem K, Burke
  D~J, Casey A~R, Crawford S~M, Dencheva N, Ely J, Jenness T, Labrie K, Lim
  P~L, Pierfederici F, Pontzen A, Ptak A, Refsdal B, Servillat M and Streicher
  O 2013 {\em Astronomy {\&} Astrophysics\/} {\bf 558} A33 ISSN 0004-6361
  \urlprefix\url{http://www.aanda.org/10.1051/0004-6361/201322068}

\bibitem{PriceWhelan2018}
Price-Whelan A~M, Sip{\H{o}}cz B~M, G{\"{u}}nther H~M, Lim P~L, Crawford S~M,
  Conseil S, Shupe D~L, Craig M~W, Dencheva N, Ginsburg A, VanderPlas J~T,
  Bradley L~D, P{\'{e}}rez-Su{\'{a}}rez D, de~Val-Borro M, Aldcroft T~L, Cruz
  K~L, Robitaille T~P, Tollerud E~J, Ardelean C, Babej T, Bach Y~P, Bachetti M,
  Bakanov A~V, Bamford S~P, Barentsen G, Barmby P, Baumbach A, Berry K~L,
  Biscani F, Boquien M, Bostroem K~A, Bouma L~G, Brammer G~B, Bray E~M,
  Breytenbach H, Buddelmeijer H, Burke D~J, Calderone G, Rodr{\'{i}}guez J~L~C,
  Cara M, Cardoso J~V~M, Cheedella S, Copin Y, Corrales L, Crichton D, D'Avella
  D, Deil C, Depagne {\'E}, Dietrich J~P, Donath A, Droettboom M, Earl N, Erben
  T, Fabbro S, Ferreira L~A, Finethy T, Fox R~T, Garrison L~H, Gibbons S~L~J,
  Goldstein D~A, Gommers R, Greco J~P, Greenfield P, Groener A~M, Grollier F,
  Hagen A, Hirst P, Homeier D, Horton A~J, Hosseinzadeh G, Hu L, Hunkeler J~S,
  Ivezi{\'{c}} {\v Z}, Jain A, Jenness T, Kanarek G, Kendrew S, Kern N~S,
  Kerzendorf W~E, Khvalko A, King J, Kirkby D, Kulkarni A~M, Kumar A, Lee A,
  Lenz D, Littlefair S~P, Ma Z, Macleod D~M, Mastropietro M, McCully C,
  Montagnac S, Morris B~M, Mueller M, Mumford S~J, Muna D, Murphy N~A, Nelson
  S, Nguyen G~H, Ninan J~P, N{\"{o}}the M, Ogaz S, Oh S, Parejko J~K, Parley N,
  Pascual S, Patil R, Patil A~A, Plunkett A~L, Prochaska J~X, Rastogi T, Janga
  V~R, Sabater J, Sakurikar P, Seifert M, Sherbert L~E, Sherwood-Taylor H, Shih
  A~Y, Sick J, Silbiger M~T, Singanamalla S, Singer L~P, Sladen P~H, Sooley
  K~A, Sornarajah S, Streicher O, Teuben P, Thomas S~W, Tremblay G~R, Turner
  J~E~H, Terr{\'{o}}n V, Kerkwijk M~H~v, de~la Vega A, Watkins L~L, Weaver B~A,
  Whitmore J~B, Woillez J and Zabalza V 2018 {\em The Astronomical Journal\/}
  {\bf 156} 123 ISSN 1538-3881
  \urlprefix\url{https://iopscience.iop.org/article/10.3847/1538-3881/aabc4f}

\bibitem{Perez2007}
Perez F and Granger B~E 2007 {\em Computing in Science {\&} Engineering\/} {\bf
  9} 21--29 ISSN 1521-9615
  \urlprefix\url{http://ieeexplore.ieee.org/document/4160251/}

\bibitem{rowe2015}
Rowe B, Jarvis M, Mandelbaum R, Bernstein G~M, Bosch J, Simet M, Meyers J~E,
  Kacprzak T, Nakajima R, Zuntz J {\em et~al.\/} 2015 {\em Astronomy and
  Computing\/} {\bf 10} 121--150

\bibitem{Kluyver2016}
Kluyver T, Ragan-Kelley B, P{\'{e}}rez F, Granger B~E, Bussonnier M, Frederic
  J, Kelley K, Hamrick J~B, Grout J, Corlay S, Ivanov P, Avila D, Abdalla S,
  Willing C and {et al} 2016 {Jupyter Notebooks - a publishing format for
  reproducible computational workflows} {\em ELPUB\/}

\bibitem{Hunter2007}
Hunter J~D 2007 {\em Computing In Science {\&} Engineering\/} {\bf 9} 90--95

\bibitem{seaborn2021}
Waskom M~L 2021 {\em Journal of Open Source Software\/} {\bf 6} 3021
  \urlprefix\url{https://doi.org/10.21105/joss.03021}

\bibitem{numpy2020}
Harris C~R, Millman K~J, van~der Walt S~J, Gommers R, Virtanen P, Cournapeau D,
  Wieser E, Taylor J, Berg S, Smith N~J, Kern R, Picus M, Hoyer S, van Kerkwijk
  M~H, Brett M, Haldane A, del R{\'{i}}o J~F, Wiebe M, Peterson P,
  G{\'{e}}rard-Marchant P, Sheppard K, Reddy T, Weckesser W, Abbasi H, Gohlke C
  and Oliphant T~E 2020 {\em Nature\/} {\bf 585} 357--362
  \urlprefix\url{https://doi.org/10.1038/s41586-020-2649-2}

\bibitem{chung1997}
Chung F~R~K 1997 {\em Spectral Graph Theory\/} (American Mathematical Society)

\bibitem{ricaud2019}
Ricaud B, Borgnat P, Tremblay N, Gon{\c c}alves P and Vandergheynst P 2019 {\em
  Comptes Rendus Physique\/} {\bf 20} 474--488 ISSN 1631-0705 fourier and the
  science of today / Fourier et la science d'aujourd'hui
  \urlprefix\url{https://www.sciencedirect.com/science/article/pii/S1631070519301094}

\end{thebibliography}

\appendix

\section{WaveDiff model's flavours}
\label{ap:wavediff_graph_flavours}

In this appendix, we include two additional PSF models, WaveDiff-graph and WaveDiff-polygraph. These models are extensions, or flavours, of the WaveDiff model presented in the article. We detail the differences of these model with respect to the original WaveDiff and repeat the numerical experiments described in the article's body.

\subsection{New data-driven models}
\label{ap:graph_DD_models}

\paragraph{WaveDiff-graph} 
We want to add constraints to the feature weights from \autoref{eq:wfe_psf_model} to capture more localized spatial variations of the PSF field. To achieve this goal, we follow the approach presented in \cite{ngole2016} and already applied in \cite{schmitz2020, liaudat2020}. We aim to capture variations that occur at different spatial frequencies. Unfortunately, the fact that the PSF positions in the FOV are randomly distributed makes it hard to define the different types of variations. To tackle this issue, we define a fully-connected undirected weighted graph where each node corresponds to the position of a star. The weights are built as a parametric function of parameters $\Lambda$ of the distance between the stars. Once the graph is defined, we compute its Laplacian matrix $L_{\text{p}, \Lambda} \in \mathbb{R}^{n_{\text{stars}} \times n_{\text{stars}}}$ \cite{chung1997}, that encodes all the weights and connections in the graph. This matrix is real and symmetric, and we can decompose it as $L_{\text{p}, \Lambda} = V_{\Lambda} \Sigma_{\Lambda} V_{\Lambda}^{\top}$, with $\Sigma_{\Lambda}$ being a diagonal matrix containing the eigenvalues. The eigenvectors of the Laplacian matrix, $V_{\Lambda}$, represent the graph harmonics or a basis for defining the graph Fourier transform \cite{ricaud2019}. 
From these, a dictionary is built to factorize the feature weights, as follows. In order to capture more spatial variations, we use $r$ different parameters, $\{\Lambda_1, \ldots, \Lambda_r\}$ for the graph's weight function. These parameters define different graphs and, subsequently, different harmonics. We concatenate the graph-harmonics, or Laplacian eigenvectors, of all the graphs to build the graph dictionary that we define as $V = [V_{\Lambda_1}, \ldots, V_{\Lambda_r}]$.
We write $\mathbf{f}_{k}^{\text{DD}} = [f_{k}^{\text{DD}}(x_1,y_1),\, \ldots\,, f_{k}^{\text{DD}}(x_{n_{\text{stars}}},y_{n_{\text{stars}}})] \in \mathbb{R}^{1 \times n_{\text{stars}}}$, where the weight function of the DD feature $k$ is expressed as a vector of all star positions.
Then, we factorize the feature weight vector as $\mathbf{f}_{k}^{\text{DD}} = \bm{\alpha}_k V^{\top}$, where $\bm{\alpha}_k \in \mathbb{R}^{1 \times r n_{\text{stars}}}$. The DD feature weights are built as a linear combination of graph-harmonics that are represented as row vectors in $V^{\top}$. The vector $\bm{\alpha}_k$ defines this combination's weights.
We impose sparsity on the representation vector so that the learned feature can specialize on a specific or a small subset of graph harmonics. The sparsity in the $\bm{\alpha}_k$ vector can be achieved by adding an $\ell_1$ constraint in the final loss function. We call it the \textit{graph constraint}. The optimization algorithm will select the best graph-harmonics for a specific feature that allows obtaining a good representation of our observations. 
For a more detailed description of how we define the graph weights and the choice of the $\Lambda$ parameters, see \cite[Sections~5.2,~5.5.3]{ngole2016}.

\paragraph{WaveDiff-polygraph} 
The feature weights simultaneously use the previously mentioned constraints and the original WaveDiff constraint from \autoref{eq_06:data_driven_model}. It contains two contributions one based on FOV position polynomials and the other based on the graph constraint. The motivation is to be able to capture smooth as well as localized PSF spatial variations.

\subsection{Model optimization}
The optimization objective in \autoref{eq:optim_problem} needs to be modified to cope with the new model's flavours, which require a regularization term. For the WaveDiff-graph model, we use a $\mu \| \bm{\alpha}_k \|_{1}$ regularization to impose sparsity on the different $\bm{\alpha}_k$ vectors that were defined in \autoref{sc:data_driven_features}, and are used to build the DD feature weights. The hyperparameter $\mu$ allows us to control the strength of the regularization. We use an identical regularization for the graph part in the WaveDiff-polygraph model. 

The two new models reuse the optimization procedure described in Algorithm \ref{al:wavediff_training}. However, they require to add the graph regularization strengh $\mu$ to the hyper-parameters, and to initialize the graph constraint dictionary following \cite[\S5.2,\S5.5.3]{ngole2016}.

\subsection{PSF recovery}
In the case of the new WaveDiff models with graph flavors, the PSF recovery is not direct. During the optimization, it is not only the DD features being learned but also their corresponding spatial variations that depend on the graph constraint, described in \ref{ap:graph_DD_models}. We want to recover the PSF at any position in the FOV while respecting the learned spatial variations. To accomplish this, we consider the different weight functions $f_{k}^{\text{DD}}$, or the vector $\mathbf{f}_{k}^{\text{DD}}$, learned for each feature $S_{k}^{\text{DD}}$. These encompass the learned spatial variations that contain a selection of useful graph harmonics. We follow the strategy from \cite{liaudat2020} and use the learned $f_{k}^{\text{DD}}$ at different FOV positions to build a local interpolant. Then, we use it to interpolate the feature weights to the desired position. We use a Radial Basis Function (RBF) interpolation scheme. For a new position $(x_j, y_j)$, we select the closest $N_{\text{RBF}}$ training star positions $\{ (x_i, y_i) \}_{i=1, \ldots, N_{\text{RBF}}}$. The interpolant function of the weights corresponding to feature $k$ is built using the following equation
\begin{equation}
    g_k(x,y) = \sum_{i=1}^{N_{\text{RBF}}} \lambda_{k,i} \, K\left(\left\| [x - x_i \,,\, y - y_i]^{\top} \right\|_{2}\right),
    \label{eq:rbf_interpolation function}
\end{equation}
where $K: \mathbb{R}_{+} \to \mathbb{R}$ is the interpolation kernel or RBF that takes as input the distance between the desired and the training position. The weights that define the interpolant $g_k$ are $\{ \lambda_{k,i} \}_{i=1, \ldots, N_{\text{RBF}}}$. We learn these weights by using a set of exact interpolation constraints in the training positions that writes $\{g_k(x_i,y_i) = f_{k}^{\text{DD}}(x_i,y_i) \}_{i=1, \ldots, N_{\text{RBF}}}$. We follow \cite[\S3.7]{liaudat2020} and use a thin plate kernel that is defined as $K(r) = r^2 \ln(r)$, and set $N_{\text{RBF}}$ to $20$. To resume, we have to build one interpolant function for each feature $k$ and each new position $(x_j, y_j)$. Even though it may seem that it is a costly procedure in terms of time and resources, the fact that the interpolation is done using $1$-D functions makes it fast. 

\subsection{Experiment setup}
The reuse the experiment setup from \autoref{sc:experiment_set_up}. The details of the two models are presented as follows:

\begin{itemize}
    \item[\textit{vii)}] \textit{WaveDiff-graph}: a similar model as \textit{iii)}, WaveDiff, with a badly specified number of Zernike polynomials, using $n_{\text{Z}}=15$ and $d_{\text{Z}}=2$. However, this model uses the previously described graph constraint for the data-driven features. In order to perform a fair comparison, we use $n_{\text{DD}}=21$.
    \item[\textit{viii)}] \textit{WaveDiff-polygraph}: a similar model as \textit{iii)}, WaveDiff, and \textit{vii)}, using $n_{\text{Z}}=15$ and $d_{\text{Z}}=2$. For the data-driven features, this model uses $d_{\text{DD}}=3$ that corresponds to $10$ features and $10$ more features corresponding to the graph constraint. 
\end{itemize}

The WaveDiff-graph and the WaveDiff-polygraph use a regularization parameter $\mu$ of $1\times 10^{-8}$. The other hyper-parameters used by the model can be found in \autoref{tb:parameters_graph}.

\begin{table}
    \centering
    \caption{Hyperparameters of the WaveDiff flavours.}
    \resizebox{\textwidth}{!}{%
        \begin{tabular}{lccccccccc}
            \toprule
            &&\multicolumn{4}{c}{$m_1$} &  \multicolumn{4}{c}{$m_2$} \\
            \cmidrule(r){3-6} \cmidrule(r){7-10}
            PSF model & Data set   & $\eta_{1}^{\text{Z}}[\times 10^{-3}]$ & $\eta_{1}^{\text{DD}}$ & $N_{\text{ep},1}^{\text{Z}}$ & $N_{\text{ep},1}^{\text{DD}}$  & $\eta_{2}^{\text{Z}}[\times 10^{-3}]$ & $\eta_{2}^{\text{DD}}$ & $N_{\text{ep},2}^{\text{Z}}$ & $N_{\text{ep},2}^{\text{DD}}$ \\
            \midrule
            \multirow{3}{*}{\shortstack[l]{WaveDiff-polygraph}}& $\mathcal{S}_{1}$  & \multirow{3}{*}{$10.0$}   & \multirow{3}{*}{$0.1$} & $30$ & $300$ & \multirow{3}{*}{$4.0$} & \multirow{3}{*}{$0.06$} & $30$ & $300$ \\
            &$\mathcal{S}_{2}$                 &  &  & $30$ & $200$ & & & $30$ & $150$ \\
            &$\mathcal{S}_{3}$                 &  &  & $20$ & $150$ & & & $20$ & $100$ \\
            &$\mathcal{S}$                     &  &  & $15$ & $100$ & & & $15$ & $50$ \\
            \midrule
            \multirow{3}{*}{WaveDiff-graph}& $\mathcal{S}_{1}$  & \multirow{3}{*}{$10.0$}   & \multirow{3}{*}{$0.4$} & $30$ & $300$ & \multirow{3}{*}{$4.0$} & \multirow{3}{*}{$0.2$} & $30$ & $300$ \\
            &$\mathcal{S}_{2}$                 &  &  & $30$ & $200$ & & & $30$ & $150$ \\
            &$\mathcal{S}_{3}$                 &  &  & $20$ & $150$ & & & $20$ & $100$ \\
            &$\mathcal{S}$                     &  &  & $15$ & $100$ & & & $15$ & $50$ \\
            \bottomrule
        \end{tabular}}
    \label{tb:parameters_graph}
    \end{table}

\subsection{Results}

The results of the WaveDiff-graph and WaveDiff-polygraph models are presented in comparison to the original WaveDiff model.

\subsubsection{Results: (a), Polychromatic errors}

\begin{table}
    \centering
    \caption{Polychromatic test star reconstruction $Err_{\text{abs}}$ and $Err_{\text{rel}}$ at the observation resolution (x1) and at super-resolution (x3). The presented results were obtained with models being trained on the largest training star set $\mathcal{S}$. \nblink{table-pixel-errors-metrics}}
        \begin{tabular}{lcc}
            \toprule
            &\multicolumn{2}{c}{$Err_{\text{abs}}$ [$\times 10^{-5}$] ($Err_{\text{rel}}$)}   \\
            \cmidrule(r){2-3}
            PSF model   & Resolution x1 & Resolution x3 \\
            \midrule
            \textbf{\textit{iii)} WaveDiff} &  $\mathbf{6.4}$ ($\mathbf{0.86\%}$)     & $\mathbf{1.9}$ ($\mathbf{1.3\%}$)      \\
            \textit{vii)} WaveDiff-graph             &  $71.8$ ($9.8\%$)    & $19.0$ ($12.7\%$)      \\
            \textit{viii)} WaveDiff-polygraph          &  $6.9$ ($0.94\%$)    & $1.9$ ($1.3\%$)      \\
            \bottomrule
        \end{tabular}
    \label{tb:results_a_graph}
\end{table}

%
We compare the performance of the different WaveDiff flavors. There is a remarkable difference, around an order of magnitude, between the WaveDiff-graph and the original and polygraph flavors. This fact shows that \textit{only} using the graph constraint for the weight function is not effective in building an appropriate WFE manifold for the PSF field, in contrast with the polynomial weight functions. 

\autoref{fi:eigen_psfs} presented examples of learned data-driven WFE features for the original WaveDiff model. We found that the polynomial weight functions, which are smoother and more constrained than their graph counterparts, allow learning more structure into the data-driven features.

\subsubsection{Results: (c), Errors as a function of the total number of training stars}

\begin{figure}
    \centering
    \includegraphics[width=\textwidth]{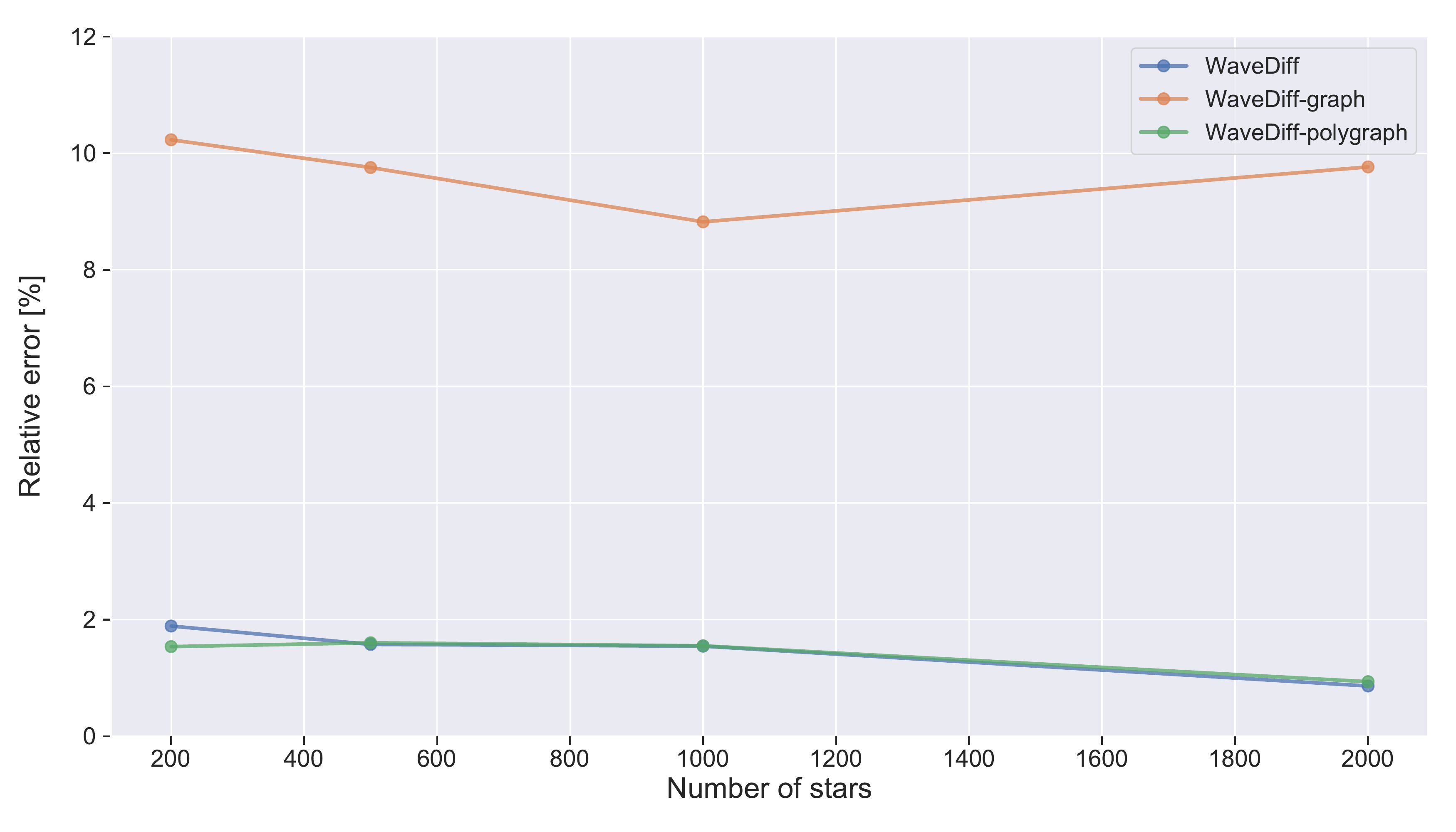}
    \caption{Polychromatic PSF relative reconstruction error, $Err_{\text{rel}}$, at observation resolution as a function of the total number of training stars in the FOV. The same conditions of \autoref{fi:result_c_x1_res} were used.
    \nblink{WaveDiff-original-rmse-vs-star-nb}}
    \label{fi:result_c_x1_res_graph}
\end{figure}

\begin{figure}
    \centering
    \includegraphics[width=\textwidth]{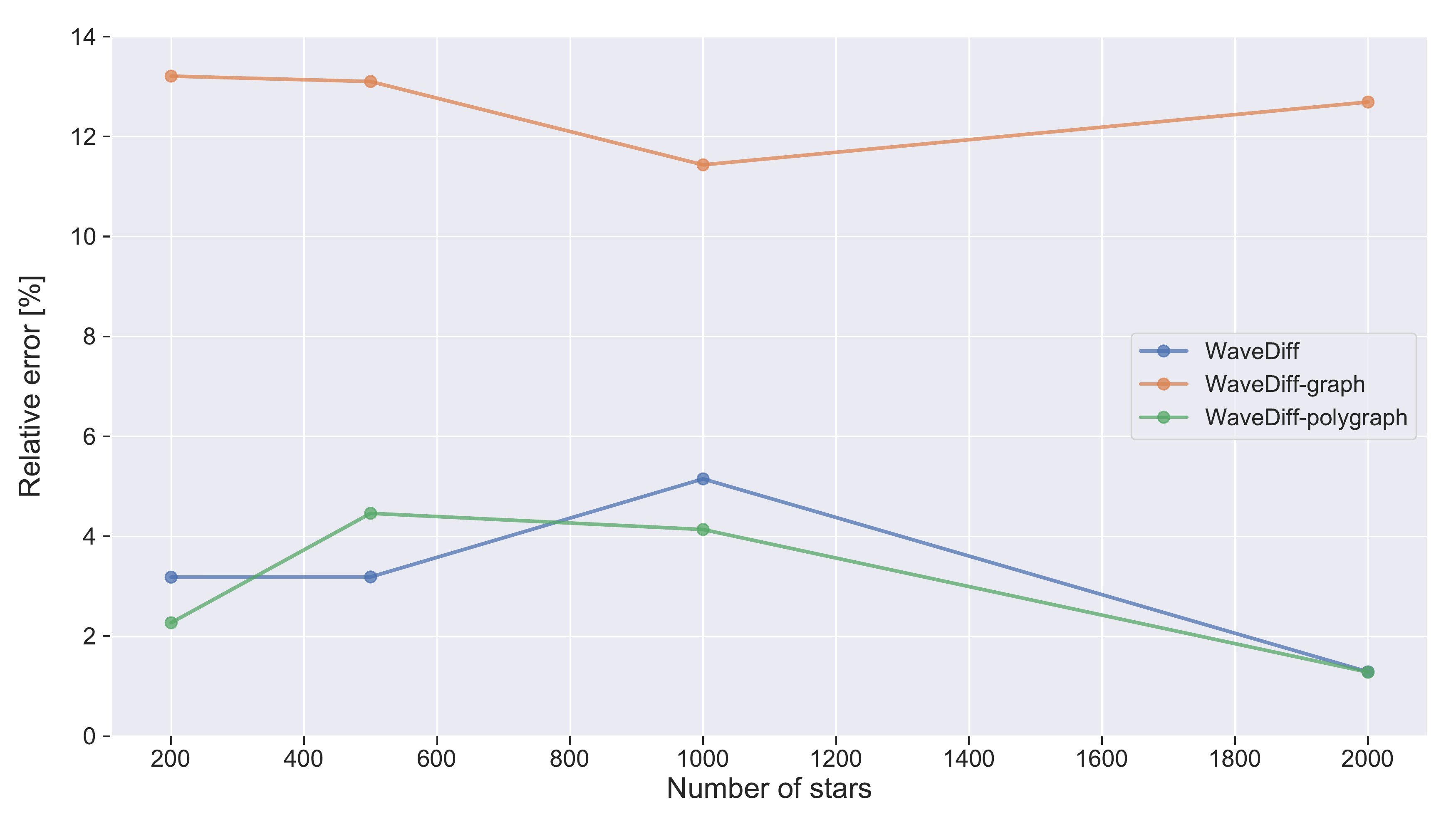}
    \caption{Polychromatic PSF relative reconstruction error, $Err_{\text{rel}}$, at three times (x$3$) the observation resolution as a function of the total number of training stars in the FOV. The same conditions of \autoref{fi:result_c_x3_res} were used.
    \nblink{WaveDiff-original-rmse-vs-star-nb}}
    \label{fi:result_c_x3_res_graph}
\end{figure}

\autoref{fi:result_c_x1_res_graph} and \autoref{fi:result_c_x3_res_graph} present the relative pixel reconstruction errors as a function of the number of training stars at one and three times the observation resolution. The previous trend is observed in these results with the WaveDiff-graph performing poorly while the WaveDiff-polygraph performing similar to the original WaveDiff model.

\subsubsection{Results: (d), WFE reconstruction errors}

\autoref{tb:results_d_WFE_graph} presents the WFE reconstruction results comparing the three WaveDiff models. All of them have important errors in the WFE recovery, indicating that the graph constraint in not helping to get a closer WFE representation with respect to the ground truth WFE.

\begin{table}
    \centering
    \caption{Target stars WFE recovery errors with the PSF models estimated with the $\mathcal{S}$ data set. \nblink{table-WFE-error-metrics}}
        \begin{tabular}{lc}
            \toprule
            PSF model                 & WFE $Err_{\text{abs}}$ [nm] ($Err_{\text{rel}}$) \\
            \midrule
            \textbf{\textit{iii)} WaveDiff}    &  $\mathbf{101}$ ($\mathbf{129\%}$)        \\
            \textit{vii)} WaveDiff-graph       &  $108$ ($138\%$)       \\
            \textit{viii)} WaveDiff-polygraph    &  $111$ ($142\%$)       \\
            \bottomrule
        \end{tabular}
    \label{tb:results_d_WFE_graph}
\end{table}

\subsubsection{Results: (e), Weak-lensing metrics}
The weak-lensing metrics from the three WaveDiff models are presented in \autoref{tb:results_e_graph}. We observe the same behaviour as in the super-resolution pixel error. The WaveDiff-polygraph performing close, and even better, to the original WaveDiff model while the WaveDiff-graph performs poorly.

\begin{table}
    \centering
    \caption{Weak-lensing metrics of the different PSF models estimated with the $\mathcal{S}$ data set.  \nblink{table-shape-size-error-metrics}}
        \begin{tabular}{lccc}
            \toprule
            &\multicolumn{3}{c}{RMSE}   \\
            \cmidrule(r){2-4}
            PSF model   & $e_1$[$\times 10^{-2}$] & $e_2$[$\times 10^{-2}$] & $R^{2}/\langle R^{2} \rangle $[$\times 10^{-1}$] \\
            \midrule
            \textbf{\textit{iii)} WaveDiff} &  $\mathbf{0.23}$     & $0.16$  & $\mathbf{0.13}$    \\
            \textit{vii)} WaveDiff-graph              &  $4.19$    & $1.64$  & $1.39$    \\
            \textit{viii)} WaveDiff-polygraph           &  $0.24$    & $\mathbf{0.14}$   & $0.16$    \\
            \bottomrule
        \end{tabular}
    \label{tb:results_e_graph}
\end{table}

\subsection{Discussion}

We found similar performances for the original WaveDiff and WaveDiff-polygraph models. The results presented in \autoref{tb:results_a_graph} showed a slightly better performance of the original WaveDiff model. 
It is, however, hard to conclude which model is better with the different metrics used. The structure imposed into the $\Phi_{\theta}$ function constrains the WFE manifold learned from the observations. Which $\Phi_{\theta}$ function allows learning a better WFE manifold depends on the ground-truth WFE, which in turn depends on the telescope we are modeling. In the current setting, both models achieve great performances. A follow-up study would evaluate the models under high-fidelity optical simulations that generally involve costly ray-tracing techniques.

\section{Derivation of the wavelength dependent zero padding formula}
\label{sc:app_oversamp_formula}
\begin{table}[h]
    \centering
    \caption{Optic variables used for the optical forward model.}
        \begin{tabular}{cl}
            \toprule
            Variable        & Description \\
            \midrule
            $\lambda$       &  Wavelength [$\mu$m]    \\
            $K$             &  Matrix dimension of the pupil  \\
            $f_L$           &  Telescope's focal length [m]   \\
            $M_{\text{D}}$  &  Pupil diameter [m]   \\
            $\Delta$        &  Pixel size [$\mu$m]   \\
            $Q$             &  Oversampling factor [Dimensionless]   \\
            $p(\lambda)$    &  Matrix dimension including pupil and external obscuration \\
            \bottomrule
        \end{tabular}
    \label{tb:optic_variables}
\end{table}
We use the optical variables from \autoref{tb:optic_variables} to derive \autoref{eq:wavelength_dep_zero_pad}. The diameter of the pupil aperture is by definition $M_{\text{D}}$, so the grid spacing in the pupil plane is $ \delta_{\xi} = M_{\text{D}} \,/\, K$. The length of the simulated pupil plane is $M_{\text{D}}\, p(\lambda) \,/\, K$, as it has a total of $p(\lambda)$ grid elements and $p(\lambda) \geq K$. The length of the simulated focal plane part is $\Delta \, p(\lambda)$. This result is calculated as the length of one pixel times the matrix dimension we will apply the \texttt{FFT}, or the number of grid elements per dimension. Recall that we are modeling the propagation of a wave from the pupil plane, variables $[\xi, \eta]$, to the focal plane, variables $[u, v]$, under Fraunhofer diffraction.

\autoref{eq:mono_PSF_model_theory_2} shows that the frequencies from the pupil plane coordinates $[\xi, \eta]$ are directly mapped to the focal plane's spatial coordinates $[u, v]$. The equations write
\begin{equation}
    f_{\xi} = \frac{u}{\lambda f_{L}} \; , \quad f_{\eta} = \frac{v}{\lambda f_{L}} \,.
\end{equation}
we consider one of the two equations above thanks to the symmetry of the problem. Replacing then the maximum frequency in the pupil plane, $1 / \delta_{\xi}$, and maximum position in the focal plane we obtain
\begin{equation}
    \frac{K}{M_{\text{D}}} = \frac{p(\lambda) \Delta}{\lambda f_L} \,.
\end{equation}
We can adjust the image's resolution by dividing the pixel size $\Delta$ by the oversampling factor $Q$. In other words, with $Q>1$, we are reducing the pixel size and increasing the system resolution. We include the oversampling factor, and we obtain
\begin{equation}
    \frac{K}{M_{\text{D}}} = \frac{p(\lambda) \Delta}{\lambda f_L Q} \,.
\end{equation}
Finally, if we solve for $p(\lambda)$, we obtain \autoref{eq:wavelength_dep_zero_pad}. We refer the reader to \cite[\S4]{schmidt2010} for more information on optical simulations.

\section{Templates of the stellar spectral energy distributions}
\label{sc:stellar_SEDs}

\begin{figure}
    \centering
    \includegraphics[width=\textwidth]{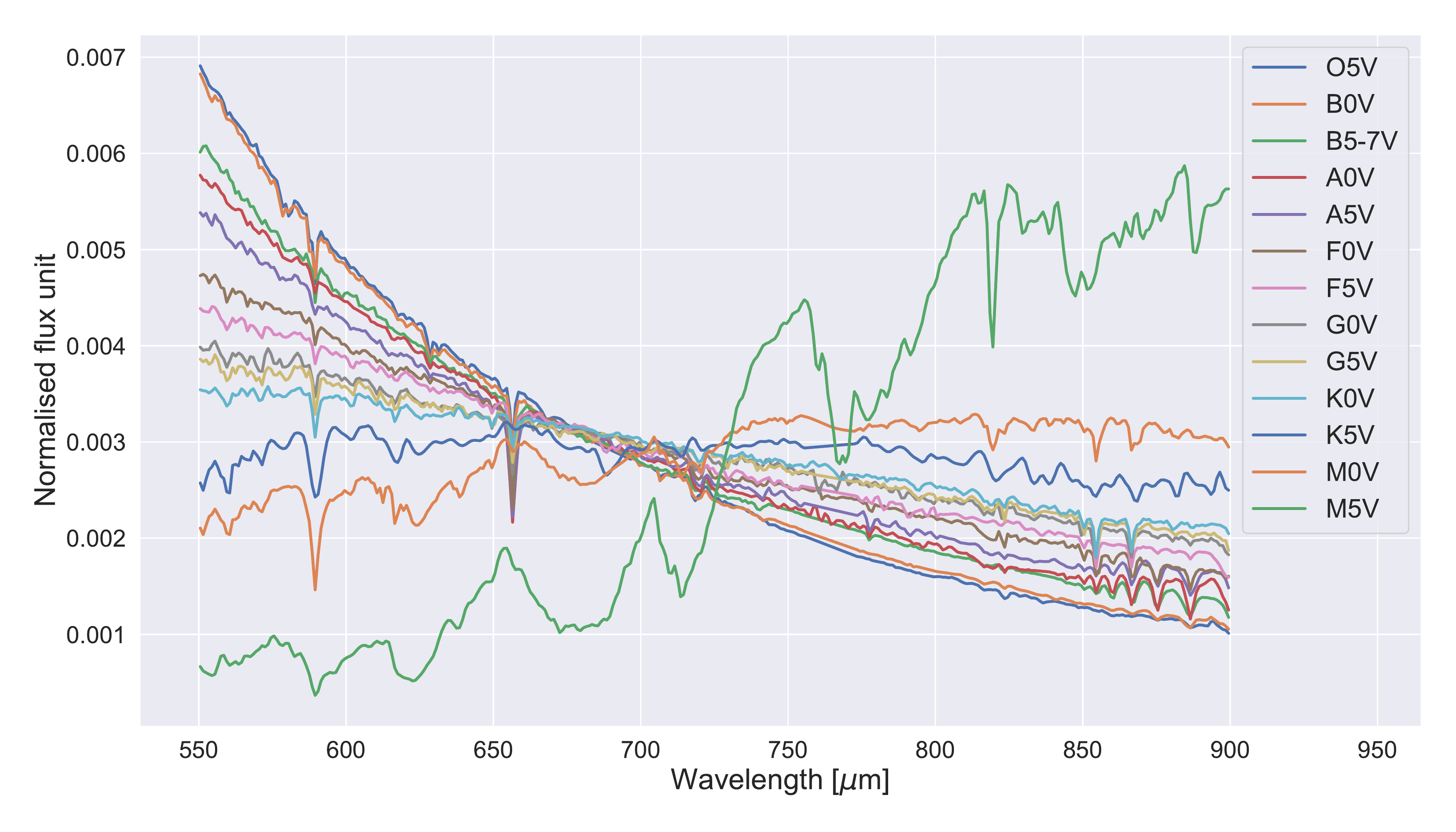}
    \caption{Templates for the stellar SEDs used to build the polychromatic PSFs.
    \nblink{SED_plot}}
    \label{fi:SED_plots}
\end{figure}

\autoref{fi:SED_plots} presents the stellar SEDs used in our experiments to build our polychromatic PSF observations. We follow the work from \cite{kuntzer2016} on his study of stellar SED classification for \textit{Euclid} and use $13$ different types of stellar SEDs from \cite{pickles1998}.

\section{Data set adaptation for \texttt{PSFEx}, \texttt{RCA}, and \texttt{MCCD}}
\label{sc:dataset_adaptations}

The \texttt{PSFEx} and \texttt{RCA} models build independent PSF models for each CCD chip. However, we have simultaneously simulated the PSF field in the whole FOV. Therefore, dividing the FOV respecting the instrument's geometry is necessary. In this case, we follow \textit{Euclid}'s VIS instrument geometry \cite{cropper2016}. The focal plane consists of a $6\times6$ CCD matrix, where each CCD is a $4096$ pixel square. Depending on the star positions, we divide the FOV into $36$ subsets. As the star positions are uniformly distributed over the FOV, the star density is approximately constant on each CCD. The input of the \texttt{RCA} model is shared with \texttt{MCCD}. The latter model continues with internal processing, where all the CCDs are merged into a single data set with global coordinates. We have included the geometry and the coordinate transformations required by the \texttt{MCCD} model.

To create the PSF model with \texttt{PSFEx}, we need to process the simulations with \texttt{SExtractor} \cite{bertin2011}. For that, we create a full CCD image of $4096 \times 4096$ pixels with the postage stamps of the stars at their corresponding positions. Each star has its already defined varying SNR. Choosing the proper parameters to ensure that \textit{all} the stars are detected and extracted by \texttt{SExtractor} can be tedious, particularly for low SNR stars. We build a second mock CCD image that we only use for detection. In this new mock image with low-valued pixels, we place high-valued pixels on the positions of the stars, thus forcing the detection of the stars. The extraction of the stars is then done using the first mock CCD image. Following this procedure, we ensure that all the stars are correctly detected and extracted by \texttt{SExtractor}. Finally, the star observations can be used to build PSF models with \texttt{PSFEx}.

\end{document}